\documentclass[aps,prd,twocolumn,superscriptaddress,nofootinbib]{revtex4-1}

\usepackage{latexsym}
\usepackage{amsmath}
\usepackage{amssymb}
\usepackage{amsfonts}
\usepackage{bm}

\usepackage{color}
\definecolor{purple}{rgb}{0.5,0,0.5}
\definecolor{blue}{rgb}{0.0,0,0.9}
\definecolor{prdblue}{rgb}{0.133,0.118,0.498}
\usepackage[colorlinks=true, pdfstartview=FitV, linkcolor=prdblue, citecolor= prdblue, urlcolor=prdblue]{hyperref}

\usepackage{supertabular} 
\usepackage{placeins}
\usepackage{epsfig}
\usepackage{graphicx}

\begin{document}

%\preprint{}

\title{Triply heavy tetraquarks $\bar{b}c\bar{q}c$ and $\bar{c}b\bar{q}b$ in a constituent quark model}


\author{Gang Yang}
\email[]{yanggang@zjnu.edu.cn}
\affiliation{Department of Physics, Zhejiang Normal University, Jinhua 321004, China}
%\thanks{}

\author{Jialun Ping}
\email[]{jlping@njnu.edu.cn}
\affiliation{Department of Physics and Jiangsu Key Laboratory for Numerical Simulation of Large Scale Complex Systems, Nanjing Normal University, Nanjing 210023, P. R. China}

\author{Jorge Segovia}
\email[]{jsegovia@upo.es}
\affiliation{Departamento de Sistemas F\'isicos, Qu\'imicos y Naturales, Universidad Pablo de Olavide, E-41013 Sevilla, Spain}

%Collaboration name if desired (requires use of superscript address

%%%%%%%%%\date{\today}

\begin{abstract}
A systematic investigation of $S$-wave triply heavy tetraquark systems with quark content $\bar{b}c\bar{q}c$ and $\bar{c}b\bar{q}b$ $(q = u,\,d,\,s)$, spin-parity quantum numbers $J^P=0^+$, $1^+$, $2^+$ and isospin $I=0,\,\frac{1}{2}$, is carried out within the constituent quark model framework.
The four-body bound and resonance states are determined by solving the Schrödinger equation employing the high-precision and efficient Gaussian Expansion Method (GEM) in conjunction with the powerful Complex Scaling Method (CSM). Besides, a comprehensive coupled-channel analysis of the $S$-wave tetraquark systems is performed, taking into account meson-meson, diquark-antidiquark and K-type configurations, as well as all allowed color structures.
Several narrow resonant states are identified in each $I(J^P)$ channel. In particular, resonances for the $\bar{b}c\bar{q}c$ system are found in the mass range of $8.87-9.36$ GeV, while those for the $\bar{c}b\bar{q}b$ system lie between $12.13-12.45$ GeV. Most of the predicted resonances are compact tetraquark states with sizes smaller than $1.0$ fm, although four states exhibit more extended structures with sizes around $1.3$ fm, suggesting a more loosely bound nature.
Magnetic moments and dominant wave function components of these exotic states are also analyzed. The results indicate that K-type configurations play a major role in the structure of the observed resonances. Finally, possible strong decay channels (golden modes) for these states are theoretically proposed.
\end{abstract}

\pacs{
12.38.-t \and % Quantum Chromodynamics
12.39.-x      % Potential Models
}
\keywords{
Quantum Chromodynamics \and
Quark models
}

\maketitle

%%%%%%%%%%%%%%%%%%%%%%%%%%%%%%%%%%%%%%%%%%%%%%%%%%%%%%%

\section{Introduction}
\label{sec:intro}

Over the past decades, a large number of exotic hadrons, \emph{viz.} states that lie beyond the conventional quark model of baryons and mesons, have continued to be reported by high-energy experiments around the world. In the charmed sector, focusing specifically on open-charm tetraquark candidates, the LHCb collaboration has reported several notable findings. Among them are the charm-strange tetraquark candidates $X_{0,(1)}(2900)$ and $T^{0(++)}_{c\bar{s}}(2900)$, observed in the study of three-body strong decays of $B$ mesons~\cite{LHCb:2020pxc, LHCb:2020bls, LHCb:2022sfr, LHCb:2022lzp}. Additionally, in 2021, LHCb identified a doubly charmed tetraquark candidate, $T^+_{cc}(3875)$~\cite{LHCb:2021vvq, LHCb:2021auc}.

In the hidden-charm sector, four exotic states with masses around 7.0 GeV were reported in succession by the LHCb, CMS, and ATLAS collaborations~\cite{LHCb:2020bwg, CMS:2023owd, ATLAS:2023bft}. These experimental discoveries have sparked extensive theoretical efforts employing a wide range of sophisticated approaches to explore their nature. For detailed discussions on the theoretical frameworks and properties of these exotic states, readers are referred to several comprehensive and representative reviews~\cite{Dong:2020hxe, Chen:2016qju, Chen:2016spr, Guo:2017jvc, Liu:2019zoy, Yang:2020atz, Dong:2021bvy, Chen:2021erj, Cao:2023rhu, Mai:2022eur, Meng:2022ozq, Chen:2022asf, Guo:2022kdi, Ortega:2020tng, Huang:2023jec, Lebed:2023vnd, Zou:2021sha, Du:2021fmf, Liu:2024uxn, Johnson:2024omq, Entem:2025bqt, Hanhart:2025bun, Wang:2025dur, Wang:2025sic, Francis:2024fwf, Chen:2024eaq, Husken:2024rdk, Liu:2024uxn, Johnson:2024omq}.

Nevertheless, no experimental signals of triply heavy tetraquarks have been observed to date, and only a limited number of theoretical investigations are available. Several conclusions can be drawn from the existing studies. First, the existence of bound states in the $QQ\bar{Q}\bar{q}$ tetraquark systems appears unlikely in various theoretical approaches, including effective field theory~\cite{Liu:2019mxw, Li:2025wod}, QCD sum rules~\cite{Jiang:2017tdc, Xu:2025zna, Zhang:2024jvv}, and a range of phenomenological models~\cite{Zhu:2023lbx, Weng:2021ngd, Lu:2021kut, Li:2025fmf, Yang:2025wqo, Yang:2024nyc}. However, a few studies have proposed the possible existence of bound states in triply charmed and triply bottomed tetraquark systems~\cite{Liu:2022jdl, Jiang:2017tdc}.

Additionally, theoretical calculations have been performed to estimate the mass spectra, magnetic moments, charge radii, and decay properties of stable excitations of $QQ\bar{Q}\bar{q}$ tetraquarks~\cite{Zhu:2023lbx, Jiang:2017tdc, Mutuk:2023yev, Xing:2019wil, Chen:2016ont, Xu:2025zna, Li:2025wod, Yang:2025wqo, Yang:2024nyc}. In particular, our previous study based on a constituent quark model~\cite{Yang:2024nyc} identified narrow resonant states in all allowed $I(J^P)$ quantum channels of $S$-wave triply heavy tetraquarks $cc\bar{c}\bar{q}$ and $bb\bar{b}\bar{q}$ $(q = u,\,d,\,s)$. These resonances are generally located in the energy ranges of $5.6-5.9$ GeV for the charm sector and $15.3-15.7$ GeV for the bottom sector.

A subsequent investigation of triply heavy tetraquark systems, specifically $\bar{b}c\bar{q}c$ and $\bar{c}b\bar{q}b$ configurations with $q = u,\,d,\,s$, is carried out within the same quark model that has demonstrated considerable success in predicting and describing various exotic hadrons. These include hidden-, single-, double-, and fully heavy tetraquarks~\cite{gy:2020dht, gy:2020dhts, Yang:2021hrb, Yang:2023mov, Yang:2021zhe, Yang:2022cut, Yang:2021izl, Yang:2023mov, Yang:2023wgu}, as well as pentaquarks~\cite{Yang:2015bmv, Yang:2018oqd, gy:2020dcp, Yang:2022bfu, Yang:2023dzb}, among others. Aimed at predicting possible bound and resonant states in the novel $QQ\bar{Q}\bar{q}$ tetraquark systems, the study comprehensively considers fully $S$-wave configurations, encompassing meson-meson, diquark-antidiquark, and K-type structures, along with all relevant color channels, for constructing the wave functions of these systems. Furthermore, quantum states with spin-parity $J^P = 0^+$, $1^+$, and $2^+$ and isospin $I = 0$ and $\frac{1}{2}$ are explored using a complex-range analysis within the framework of the powerful complex scaling method (CSM). In this theoretical approach, the complex-scaled four-body Schr\"odinger equation is solved with high accuracy and computational efficiency using the Gaussian Expansion Method (GEM).

The structure of this article is as follows. Section~\ref{sec:model} introduces the constituent quark model and the construction of wave functions for triply heavy tetraquarks. Theoretical results and corresponding discussions are presented in Section~\ref{sec:results}. Finally, a brief summary of the study is provided in Section~\ref{sec:summary}.

%%%%%%%%%%%%%%%%%%%%%%%%%%%%%%%%%%%%%%%%%%%%%%%%%%%%%%%%%%%%%%%%%%%%

\section{Theoretical framework}
\label{sec:model}

\subsection{The Hamiltonian}

A complex scaled Schr\"odinger equation to investigate possible four-body bound and resonance states is utilized, the equation reads as
\begin{equation}\label{CSMSE}
\left[ H(\theta)-E(\theta) \right] \Psi_{JM}(\theta)=0.
\end{equation}

Generally, the Hamiltonian of four-body system for a QCD-inspired constituent quark model is
\begin{equation}
H(\theta) = \sum_{i=1}^{4}\left( m_i+\frac{(\vec{p\,}_i e^{-i\theta})^2}{2m_i}\right) - T_{\text{CM}} + \sum_{j>i=1}^{4} V(\vec{r}_{ij} e^{i\theta}) \,,
\label{eq:Hamiltonian}
\end{equation}
where $m_{i}$ and $\vec{p}_i$ are, respectively, the constituent mass and momentum of a quark, $T_{\text{CM}}$ is the center-of-mass kinetic energy, and the last term is the two-body interaction potential. The complex-range analysis is performed by introducing an artificial parameter of rotated angle $\theta$ in the Hamiltonian, and thus three kinds of complex eigenvalues of Eq.~\eqref{CSMSE} (bound, resonance and scattering states) can be simultaneously studied. In particular, resonance and bound states are independent of the rotated angle $\theta$, with the first ones located above the threshold line with a total two-body strong decay width $\Gamma=-2\,\text{Im}(E)$, and the second ones always placed on the real-axis of the complex energy plane. The scattering states appear aligned along threshold lines and are unstable against different values of $\theta$.

The dynamics of triply heavy tetraquarks, $\bar{b}c\bar{q}c$ and $\bar{c}b\bar{q}b$, are driven by 2-body complex-scaled potentials,
\begin{equation}
\label{CQMV}
V(\vec{r}_{ij} e^{i\theta}) = V_{\text{CON}}(\vec{r}_{ij} e^{i\theta}) + V_{\text{OGE}}(\vec{r}_{ij} e^{i\theta})  \,.
\end{equation}
In particular, color confinement and perturbative one-gluon exchange interactions are considered the most relevant features of QCD in the heavy quark sector. Since this work focuses on low-lying $S$-wave, positive-parity, triply heavy tetraquark states, only the central and spin-spin components of the potential are taken into account. As in Ref.~\cite{Yang:2024nyc}, the interactions associated with dynamical chiral symmetry breaking, specifically those involving Goldstone boson exchange, are omitted in the present analysis of tetraquark systems.

Concerning the color confining interaction, lattice-regularized QCD has demonstrated that multi-gluon exchanges produce an attractive linearly rising potential proportional to the distance between infinite-heavy quarks~\cite{Bali:2005fu}. Nevertheless, the spontaneous creation of light-quark pairs from the QCD vacuum may give rise at the same scale to a breakup of the created color flux-tube~\cite{Bali:2005fu}. Therefore, we can phenomenologically describe the above two observations by
\begin{equation}
V_{\text{CON}}(\vec{r}_{ij} e^{i\theta})=\left[-a_{c}(1-e^{-\mu_{c}r_{ij} e^{i\theta}})+\Delta \right] 
(\lambda_{i}^{c}\cdot \lambda_{j}^{c}) \,,
\label{eq:conf}
\end{equation}
where $\lambda^c$ denote the SU(3) color Gell-Mann matrices, whereas $a_{c}$, $\mu_{c}$ and $\Delta$ are model parameters. For a real-range potential of Eq.~\eqref{eq:conf} $(\theta=0^\circ)$, one can find that it is linear at short inter-quark distances with an effective confinement strength $\sigma = -a_{c} \, \mu_{c} \, (\lambda^{c}_{i}\cdot \lambda^{c}_{j})$, while the potential becomes a constant at large distances, $V_{\text{thr.}} = (\Delta-a_c)(\lambda^{c}_{i}\cdot \lambda^{c}_{j})$.

Beyond the chiral symmetry breaking energy scale, one also expects the dynamics to be driven by perturbative effects of QCD. In particular, the one-gluon exchange potential, which includes the so-called Coulomb and color-magnetic interactions, is the leading order part:
\begin{align}
&
V_{\text{OGE}}(\vec{r}_{ij} e^{i\theta}) = \frac{1}{4} \alpha_{s} (\lambda_{i}^{c}\cdot \lambda_{j}^{c}) \Bigg[\frac{1}{r_{ij} e^{i\theta}} \nonumber \\ 
&
\hspace*{1.60cm} - \frac{1}{6m_{i}m_{j}} (\vec{\sigma}_{i}\cdot\vec{\sigma}_{j}) 
\frac{e^{-r_{ij} e^{i\theta} /r_{0}(\mu_{ij})}}{r_{ij} e^{i\theta} r_{0}^{2}(\mu_{ij})} \Bigg] \,,
\end{align}
where $\vec{\sigma}$ denote as the Pauli matrices, and $r_{0}(\mu_{ij})=\hat{r}_{0}/\mu_{ij}$ depends on the reduced mass of a $q\bar{q}$ pair. Besides, the regularized contact term is
\begin{equation}
\delta(\vec{r}_{ij} e^{i\theta}) \sim \frac{1}{4\pi r_{0}^{2}(\mu_{ij})}\frac{e^{-r_{ij} e^{i\theta} / r_{0}(\mu_{ij})}}{r_{ij} e^{i\theta} } \,.
\end{equation}

An effective scale-dependent strong coupling constant, $\alpha_s(\mu_{ij})$, provides a consistent description of mesons and baryons from light to heavy quark sectors. The frozen coupling constant is used of, for instance, Ref.~\cite{Segovia:2013wma},
\begin{equation}
\alpha_{s}(\mu_{ij})=\frac{\alpha_{0}}{\ln\left(\frac{\mu_{ij}^{2}+\mu_{0}^{2}}{\Lambda_{0}^{2}} \right)} \,,
\end{equation}
where $\alpha_{0}$, $\mu_{0}$ and $\Lambda_{0}$ are quark model parameters.

All of the above quark model parameters are summarized in Table~\ref{tab:model}. In particular, the quark model has already been successfully explaining the hadronic phenomenology: hadron spectra~\cite{Valcarce:1995dm, Vijande:2004he, Segovia:2008zza, Segovia:2008zz, Segovia:2011zza, Ortega:2016hde, Yang:2017xpp}, hadron-hadron interactions~\cite{Fernandez:1993hx, Valcarce:1994nr, Ortega:2009hj, Ortega:2016mms, Ortega:2016pgg, Ortega:2018cnm, Ortega:2020uvc} and multiquark systems~\cite{Vijande:2006jf, Yang:2015bmv, Yang:2017rpg}. Additionally, for later concern, theoretical and experimental (if available) masses of ground and first radial excitation states of $Q\bar{Q}$ and $Q\bar{q}$ $(q=u,\,d,\,s;\, Q=c,\,b)$ mesons are listed in Table~\ref{MesonMass}.

\begin{table}[!t]
\caption{\label{tab:model} The constituent quark model parameters.}
\begin{ruledtabular}
\begin{tabular}{llr}
Quark masses     & $m_q\,(q=u,\,d)$ (MeV) & 313 \\
                 & $m_s$ (MeV) &  555 \\
                 & $m_c$ (MeV) & 1752 \\
                 & $m_b$ (MeV) & 5100 \\[2ex]
OGE              & $\alpha_0$              & 2.118 \\
                 & $\Lambda_0~$(fm$^{-1}$) & 0.113 \\
                 & $\mu_0~$(MeV)           & 36.976 \\
                 & $\hat{r}_0~$(MeV~fm)    & 28.17 \\[2ex]
Confinement      & $a_c$ (MeV)         & 430 \\
                 & $\mu_c$ (fm$^{-1})$ & 0.70 \\
                 & $\Delta$ (MeV)      & 181.10 \\
\end{tabular}
\end{ruledtabular}
\end{table}

\begin{table}[!t]
\caption{\label{MesonMass} Theoretical and experimental (if available) masses of $1S$ and $2S$ states of $Q\bar{Q}$ and $Q\bar{q}\,(q=u, d, s;\, Q=c,\,b)$ mesons, unit in MeV.}
\begin{ruledtabular}
\begin{tabular}{lccc}
Meson & $nL$ & $M_{\text{The.}}$ & $M_{\text{Exp.}}$\\
\hline
$D$ & $1S$ &  $1897$ & $1870$ \\
        & $2S$ & $2648$ & $-$        \\[2ex]
$D^*$ & $1S$ &  $2017$ & $2007$ \\
            & $2S$ & $2704$ & $-$       \\[2ex]
$D_s$ & $1S$ &  $1989$ & $1968$ \\
            & $2S$ & $2705$ & $-$       \\[2ex]
$D^*_s$ & $1S$ &  $2115$ & $2112$ \\
            & $2S$ & $2769$ & $-$          \\[2ex]
            
$B$ & $1S$ &  $5278$ & $5280$ \\
        & $2S$ & $5984$ & $-$ \\[2ex]
$B^*$ & $1S$ &  $5319$ & $5325$ \\
           & $2S$ & $6005$ & $-$ \\[2ex]
$B_s$ & $1S$ &  $5355$ & $5367$ \\
            & $2S$ & $6017$ & $-$ \\[2ex]
$B^*_s$ & $1S$ &  $5400$ & $5415$ \\
               & $2S$ & $6042$ & $-$\\[2ex]
$B_c$ & $1S$  & $6276$ & $6275$ \\
            & $2S$ & $6859$ & $6872$ \\[2ex]
$B^*_c$ & $1S$  & $6330$ & $-$ \\
               & $2S$ & $6890$ & $-$
\end{tabular}
\end{ruledtabular}
\end{table}

\begin{figure}[ht]
\epsfxsize=3.4in \epsfbox{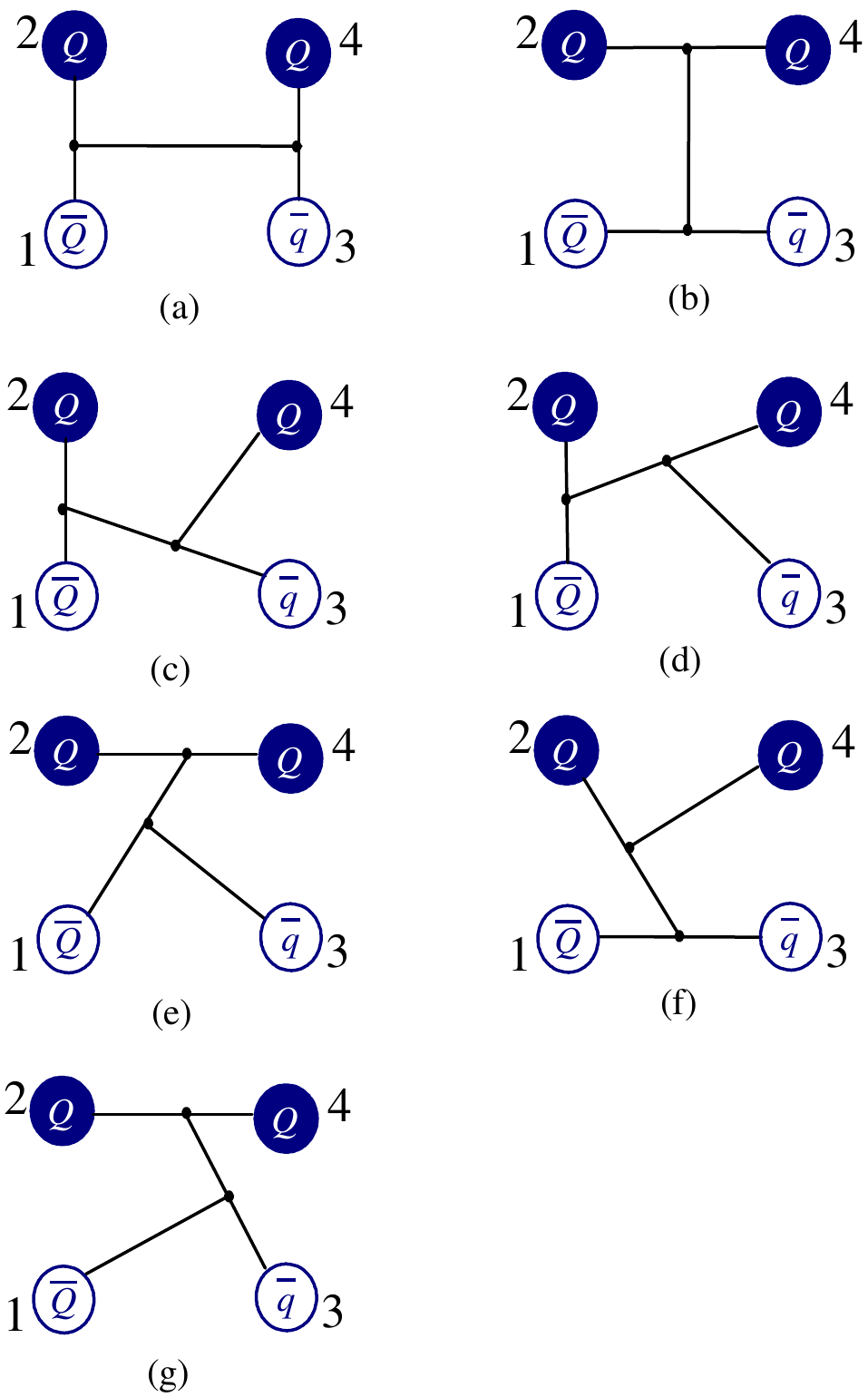}
\caption{\label{QQqq} Complete $S$-wave configurations are considered for the triply heavy tetrauqarks $\bar{b}c\bar{q}c$ and $\bar{c}b\bar{q}b$ $(q=u,\,d,\,s)$. Particularly, panel $(a)$ is meson-meson structure, panel $(b)$ is diquark-antidiquark arrangement, and the five K-type configurations are from panel $(c)$ to $(g)$.}
\end{figure}

\subsection{The tetraquark wave function}

All possible $S$-wave configurations of triply heavy tetraquarks, illustrated in Figure~\ref{QQqq}, are considered in this study. Specifically, the analysis includes one meson-meson structure shown in Fig.~\ref{QQqq}(a), one diquark-antidiquark arrangement displayed in Fig.~\ref{QQqq}(b), as well as five K-type configurations, which are rarely discussed in the existing literature on multiquark systems, and they are depicted in panels (c) to (g). Employing this diverse set of configurations offers a computationally efficient approach. Otherwise, if the calculation were restricted to a single structural type, it would be necessary to include a full set of radial and orbital excitations to achieve comparable accuracy.

The total wave function of a hadron is constructed as the internal product of its color, spin, flavor, and spatial components at the quark level. Starting with the color degree of freedom, in the meson–meson configuration, a color-singlet tetraquark wave function can be formed by coupling two color-singlet clusters, $1 \otimes 1$:
\begin{align}
\label{Color1}
\chi^c_1 &= \frac{1}{3}(\bar{r}r+\bar{g}g+\bar{b}b)\times (\bar{r}r+\bar{g}g+\bar{b}b) \,,
\end{align}
or two coupled color-octet clusters, $8\otimes 8$:
\begin{align}
\label{Color2}
\chi^c_2 &= \frac{\sqrt{2}}{12}(3\bar{b}r\bar{r}b+3\bar{g}r\bar{r}g+3\bar{b}g\bar{g}b+3\bar{g}b\bar{b}g+3\bar{r}g\bar{g}r
\nonumber\\
&+3\bar{r}b\bar{b}r+2\bar{r}r\bar{r}r+2\bar{g}g\bar{g}g+2\bar{b}b\bar{b}b-\bar{r}r\bar{g}g
\nonumber\\
&-\bar{g}g\bar{r}r-\bar{b}b\bar{g}g-\bar{b}b\bar{r}r-\bar{g}g\bar{b}b-\bar{r}r\bar{b}b) \,.
\end{align}
The first color state is the so-called color-singlet channel and the second one is the named hidden-color case.

The color wave functions associated to the diquark-antidiquark arrangement are the coupled color triplet-antitriplet clusters, $3\otimes \bar{3}$:
\begin{align}
\label{Color3}
\chi^c_3 &= \frac{\sqrt{3}}{6}(\bar{r}r\bar{g}g-\bar{g}r\bar{r}g+\bar{g}g\bar{r}r-\bar{r}g\bar{g}r+\bar{r}r\bar{b}b
\nonumber\\
&-\bar{b}r\bar{r}b+\bar{b}b\bar{r}r-\bar{r}b\bar{b}r+\bar{g}g\bar{b}b-\bar{b}g\bar{g}b
\nonumber\\
&+\bar{b}b\bar{g}g-\bar{g}b\bar{b}g) \,,
\end{align}
and the coupled color sextet-antisextet clusters, $6\otimes \bar{6}$:
\begin{align}
\label{Color4}
\chi^c_4 &= \frac{\sqrt{6}}{12}(2\bar{r}r\bar{r}r+2\bar{g}g\bar{g}g+2\bar{b}b\bar{b}b+\bar{r}r\bar{g}g+\bar{g}r\bar{r}g
\nonumber\\
&+\bar{g}g\bar{r}r+\bar{r}g\bar{g}r+\bar{r}r\bar{b}b+\bar{b}r\bar{r}b+\bar{b}b\bar{r}r
\nonumber\\
&+\bar{r}b\bar{b}r+\bar{g}g\bar{b}b+\bar{b}g\bar{g}b+\bar{b}b\bar{g}g+\bar{g}b\bar{b}g) \,.
\end{align}

Meanwhile, the color-singlet wave functions of the mentioned five K-type structures are given by
\begin{align}
\label{Color5}
\chi^c_5 &= \frac{1}{6\sqrt{2}}(\bar{r}r\bar{r}r+\bar{g}g\bar{g}g-2\bar{b}b\bar{b}b)+
\nonumber\\
&\frac{1}{2\sqrt{2}}(\bar{r}b\bar{b}r+\bar{r}g\bar{g}r+\bar{g}b\bar{b}g+\bar{g}r\bar{r}g+\bar{b}g\bar{g}b+\bar{b}r\bar{r}b)-
\nonumber\\
&\frac{1}{3\sqrt{2}}(\bar{g}g\bar{r}r+\bar{r}r\bar{g}g)+\frac{1}{6\sqrt{2}}(\bar{b}b\bar{r}r+\bar{b}b\bar{g}g+\bar{r}r\bar{b}b+\bar{g}g\bar{b}b) \,,
\end{align}
\begin{align}
\label{Color6}
\chi^c_6 &= \chi^c_1 \,,
\end{align}
\begin{align}
\label{Color7}
\chi^c_7 &= \chi^c_1 \,,
\end{align}
\begin{align}
\label{Color8}
\chi^c_8 &= \chi^c_2 \,,
\end{align}
\begin{align}
\label{Color9}
\chi^c_9 &= \frac{1}{2\sqrt{6}}(\bar{r}b\bar{b}r+\bar{r}r\bar{b}b+\bar{g}b\bar{b}g+\bar{g}g\bar{b}b+\bar{r}g\bar{g}r+\bar{r}r\bar{g}g+
\nonumber\\
&\bar{b}b\bar{g}g+\bar{b}g\bar{g}b+\bar{g}g\bar{r}r+\bar{g}r\bar{r}g+\bar{b}b\bar{r}r+\bar{b}r\bar{r}b)+
\nonumber\\
&\frac{1}{\sqrt{6}}(\bar{r}r\bar{r}r+\bar{g}g\bar{g}g+\bar{b}b\bar{b}b) \,,
\end{align}
\begin{align}
\label{Color10}
\chi^c_{10} &= \frac{1}{2\sqrt{3}}(\bar{r}b\bar{b}r-\bar{r}r\bar{b}b+\bar{g}b\bar{b}g-\bar{g}g\bar{b}b+\bar{r}g\bar{g}r-\bar{r}r\bar{g}g-
\nonumber\\
&\bar{b}b\bar{g}g+\bar{b}g\bar{g}b-\bar{g}g\bar{r}r+\bar{g}r\bar{r}g-\bar{b}b\bar{r}r+\bar{b}r\bar{r}b) \,,
\end{align}
\begin{align}
\label{Color11}
\chi^c_{11} &= \chi^c_9 \,,
\end{align}
\begin{align}
\label{Color12}
\chi^c_{12} &= -\chi^c_{10} \,.
\end{align}
\begin{align}
\label{Color13}
\chi^c_{13} &= \chi^c_9 \,,
\end{align}
\begin{align}
\label{Color14}
\chi^c_{14} &= \chi^c_{10} \,.
\end{align}

For the spin degree-of-freedom, the $S$-wave ground state can have total spin $S$ ranging from $0$ to $2$. Accordingly, the spin wave functions, denoted as $\chi^{\sigma_i}_{S, M_S}$, are constructed for each case. Without loss of generality, the spin projection $M_S$ can be taken to be equal to $S$:
\begin{align}
\label{SWF0}
\chi_{0,0}^{\sigma_{u_1}}(4) &= \chi^\sigma_{00}\chi^\sigma_{00} \,, \\
\chi_{0,0}^{\sigma_{u_2}}(4) &= \frac{1}{\sqrt{3}}(\chi^\sigma_{11}\chi^\sigma_{1,-1}-\chi^\sigma_{10}\chi^\sigma_{10}+\chi^\sigma_{1,-1}\chi^\sigma_{11}) \,, \\
\chi_{0,0}^{\sigma_{u_3}}(4) &= \frac{1}{\sqrt{2}}\big((\sqrt{\frac{2}{3}}\chi^\sigma_{11}\chi^\sigma_{\frac{1}{2}, -\frac{1}{2}}-\sqrt{\frac{1}{3}}\chi^\sigma_{10}\chi^\sigma_{\frac{1}{2}, \frac{1}{2}})\chi^\sigma_{\frac{1}{2}, -\frac{1}{2}} \nonumber \\ 
&-(\sqrt{\frac{1}{3}}\chi^\sigma_{10}\chi^\sigma_{\frac{1}{2}, -\frac{1}{2}}-\sqrt{\frac{2}{3}}\chi^\sigma_{1, -1}\chi^\sigma_{\frac{1}{2}, \frac{1}{2}})\chi^\sigma_{\frac{1}{2}, \frac{1}{2}}\big) \,, \\
\chi_{0,0}^{\sigma_{u_4}}(4) &= \frac{1}{\sqrt{2}}(\chi^\sigma_{00}\chi^\sigma_{\frac{1}{2}, \frac{1}{2}}\chi^\sigma_{\frac{1}{2}, -\frac{1}{2}}-\chi^\sigma_{00}\chi^\sigma_{\frac{1}{2}, -\frac{1}{2}}\chi^\sigma_{\frac{1}{2}, \frac{1}{2}}) \,,
\end{align}
for $(S,M_S)=(0,0)$, by 
\begin{align}
\label{SWF1}
\chi_{1,1}^{\sigma_{w_1}}(4) &= \chi^\sigma_{00}\chi^\sigma_{11} \,, \\ 
\chi_{1,1}^{\sigma_{w_2}}(4) &= \chi^\sigma_{11}\chi^\sigma_{00} \,, \\
\chi_{1,1}^{\sigma_{w_3}}(4) &= \frac{1}{\sqrt{2}} (\chi^\sigma_{11} \chi^\sigma_{10}-\chi^\sigma_{10} \chi^\sigma_{11}) \,, \\
\chi_{1,1}^{\sigma_{w_4}}(4) &= \sqrt{\frac{3}{4}}\chi^\sigma_{11}\chi^\sigma_{\frac{1}{2}, \frac{1}{2}}\chi^\sigma_{\frac{1}{2}, -\frac{1}{2}}-\sqrt{\frac{1}{12}}\chi^\sigma_{11}\chi^\sigma_{\frac{1}{2}, -\frac{1}{2}}\chi^\sigma_{\frac{1}{2}, \frac{1}{2}} \nonumber \\ 
&-\sqrt{\frac{1}{6}}\chi^\sigma_{10}\chi^\sigma_{\frac{1}{2}, \frac{1}{2}}\chi^\sigma_{\frac{1}{2}, \frac{1}{2}} \,, \\
\chi_{1,1}^{\sigma_{w_5}}(4) &= (\sqrt{\frac{2}{3}}\chi^\sigma_{11}\chi^\sigma_{\frac{1}{2}, -\frac{1}{2}}-\sqrt{\frac{1}{3}}\chi^\sigma_{10}\chi^\sigma_{\frac{1}{2}, \frac{1}{2}})\chi^\sigma_{\frac{1}{2}, \frac{1}{2}} \,, \\
\chi_{1,1}^{\sigma_{w_6}}(4) &= \chi^\sigma_{00}\chi^\sigma_{\frac{1}{2}, \frac{1}{2}}\chi^\sigma_{\frac{1}{2}, \frac{1}{2}} \,,
\end{align}
for $(S,M_S)=(1,1)$, and by 
\begin{align}
\label{SWF2}
\chi_{2,2}^{\sigma_{1}}(4) &= \chi^\sigma_{11}\chi^\sigma_{11} \,,
\end{align}
for $(S,M_S)=(2,2)$. Particularly, superscripts $u_1,\ldots,u_4$ and $w_1,\ldots,w_6$ determine the spin wave function for each configuration of tetraquark system, their values are listed in Table~\ref{SpinIndex}. Additionally, the expressions above are derived by considering the spin coupling between two sub-clusters, whose individual spin wave functions are constructed using standard SU(2) algebra. The required basis states are given by:
\begin{align}
\label{Spin}
&\chi^\sigma_{00} = \frac{1}{\sqrt{2}}(\chi^\sigma_{\frac{1}{2}, \frac{1}{2}} \chi^\sigma_{\frac{1}{2}, -\frac{1}{2}}-\chi^\sigma_{\frac{1}{2}, -\frac{1}{2}} \chi^\sigma_{\frac{1}{2}, \frac{1}{2}}) \,, \\
&\chi^\sigma_{11} = \chi^\sigma_{\frac{1}{2}, \frac{1}{2}} \chi^\sigma_{\frac{1}{2}, \frac{1}{2}} \,, \\
&\chi^\sigma_{1,-1} = \chi^\sigma_{\frac{1}{2}, -\frac{1}{2}} \chi^\sigma_{\frac{1}{2}, -\frac{1}{2}} \,, \\
&\chi^\sigma_{10} = \frac{1}{\sqrt{2}}(\chi^\sigma_{\frac{1}{2}, \frac{1}{2}} \chi^\sigma_{\frac{1}{2}, -\frac{1}{2}}+\chi^\sigma_{\frac{1}{2}, -\frac{1}{2}} \chi^\sigma_{\frac{1}{2}, \frac{1}{2}}) \,.
\end{align}

\begin{table}[!t]
\caption{\label{SpinIndex} The values of the superscripts $u_1,\ldots,u_4$ and $w_1,\ldots,w_6$ that determine the spin wave function for each configuration of the triply heavy tetraquark systems.}
\begin{ruledtabular}
\begin{tabular}{lccccccc}
& Di-meson & Diquark-antidiquark & $K_1$ & $K_2$ & $K_3$ & $K_4$ & $K_5$ \\
\hline
$u_1$ & 1 & 3 & & & & & \\
$u_2$ & 2 & 4 & & & & & \\
$u_3$ &   &   & 5 & 7 &  9 & 11 & 13 \\
$u_4$ &   &   & 6 & 8 & 10 & 12 & 14 \\[2ex]
$w_1$ & 1 & 4 & & & & & \\
$w_2$ & 2 & 5 & & & & & \\
$w_3$ & 3 & 6 & & & & & \\
$w_4$ &   &   & 7 & 10 & 13 & 16 & 19 \\
$w_5$ &   &   & 8 & 11 & 14 & 17 & 20\\
$w_6$ &   &   & 9 & 12 & 15 & 18 & 21
\end{tabular}
\end{ruledtabular}
\end{table}

The flavor wave functions $\chi^{f}_{I}$ of the two considered triply heavy tetraquarks are simply denoted as
\begin{align}
\chi^{f}_{I} = \bar{b}c\bar{q}c,\,\,\,  \chi^{f}_{I} = \bar{c}b\bar{q}b,\,(q=u,\,d,\,s) \,.
\end{align}

The Schr\"odinger-like four-body system equation is solved by means of the Rayleigh-Ritz variational principle. Generally, within a complex scaled theoretical framework, the spatial wave function is:
\begin{equation}
\label{eq:WFexp}
\psi_{LM_L}= \left[ \left[ \phi_{n_1l_1}(\vec{\rho}e^{i\theta}\,) \phi_{n_2l_2}(\vec{\lambda}e^{i\theta}\,)\right]_{l} \phi_{n_3l_3}(\vec{R}e^{i\theta}\,) \right]_{L M_L} \,,
\end{equation}
where the internal Jacobi coordinates are written as
\begin{align}
\vec{\rho} &= \vec{x}_1-\vec{x}_{2(4)} \,, \\
\vec{\lambda} &= \vec{x}_3 - \vec{x}_{4(2)} \,, \\
\vec{R} &= \frac{m_1 \vec{x}_1 + m_{2(4)} \vec{x}_{2(4)}}{m_1+m_{2(4)}}- \frac{m_3 \vec{x}_3 + m_{4(2)} \vec{x}_{4(2)}}{m_3+m_{4(2)}} \,,
\end{align}
for the meson-meson configuration of Fig.~\ref{QQqq} $(a)$, and as
\begin{align}
\vec{\rho} &= \vec{x}_1-\vec{x}_3 \,, \\
\vec{\lambda} &= \vec{x}_2 - \vec{x}_4 \,, \\
\vec{R} &= \frac{m_1 \vec{x}_1 + m_3 \vec{x}_3}{m_1+m_3}- \frac{m_2 \vec{x}_2 + m_4 \vec{x}_4}{m_2+m_4} \,,
\end{align}
for the diquark-antidiquark arrangement of Fig.~\ref{QQqq} $(b)$. The five K-type configurations presented in Fig.~\ref{QQqq} $(c)$ to $(g)$ are ($i, j, k, l$ take values according to the panels $(c)$ to $(g)$ of Fig.~\ref{QQqq}):
\begin{align}
\vec{\rho} &= \vec{x}_i-\vec{x}_j \,, \\
\vec{\lambda} &= \vec{x}_k- \frac{m_i \vec{x}_i + m_j \vec{x}_j}{m_i+m_j} \,, \\
\vec{R} &= \vec{x}_l- \frac{m_i \vec{x}_i + m_j \vec{x}_j+m_k \vec{x}_k}{m_i+m_j+m_k} \,.
\end{align}
Obviously, the center-of-mass kinetic term, $T_\text{CM}$, can be completely eliminated for a non-relativistic system defined in any of the above sets of relative motion coordinates.

In the Rayleigh-Ritz variational method, a key aspect lies in the basis expansion of the true wave function, as expressed in Eq.~\eqref{eq:WFexp}. To achieve this, the Gaussian Expansion Method (GEM)~\cite{Hiyama:2003cu} is employed. Within this framework, the spatial wave functions corresponding to the four relative motions are expanded using Gaussian basis functions, with size parameters chosen in geometric progression. Further details on the mathematical formulation of this method can be found, for example, in Ref.~\cite{Yang:2015bmv}. As a result, the orbital wave functions $\phi$ appearing in Eq.~\eqref{eq:WFexp} for an $S$-wave tetraquark system take the following form:
\begin{align}
&
\phi_{nlm}(\vec{r}e^{i\theta}\,) = \sqrt{1/4\pi} \, N_{nl} \, (re^{i\theta})^{l} \, e^{-\nu_{n} (re^{i\theta})^2} \,.
\end{align}

At last, the fully wave function, which fulfills the Pauli principle, is written as
\begin{align}
\label{TPs}
 \Psi_{J M_J, I} &= \sum_{i, j} c_{ij} \Psi_{J M_J, I, i, j} \nonumber \\
 &=\sum_{i, j} c_{ij} {\cal A} \left[ \left[ \psi_{L M_L} \chi^{\sigma_i}_{S M_S}(4) \right]_{J M_J} \chi^{f}_I \chi^{c}_j \right] \,,
\end{align}
where $\cal{A}$ is the antisymmetry operator of triply heavy tetraquark systems, which take into account the fact of involving two identical heavy quarks. The definition, according to Fig.~\ref{QQqq}, is 
\begin{equation}
\label{Antisym}
{\cal{A}} = 1-(24) \,.
\end{equation}
This is necessary in the present theoretical framework, since the complete wave function of a tetraquark system is constructed from the sub-clusters: meson-meson, diquark-antidiquark and K-type configurations.

As a further step toward understanding the nature of exotic hadrons in the triply heavy sector, the internal structure of possible resonant states is investigated through a quantitative analysis of interquark distances,
\begin{equation}
\label{quarkdistance}
{r_{q\bar{q}}} = Re(\sqrt{\langle \Psi_{J M_J, I} \vert  (r_{q\bar{q}} e^{i\theta})^2 \vert \Psi_{J M_J, I} \rangle}) \,,
\end{equation}
magnetic moment,
\begin{equation}
\label{MM}
{\mu_m} = Re(\langle \Psi_{J M_J, I} \vert  \sum_{i=1}^{4} \frac{\hat Q_i}{2m_i} \hat \sigma^z_i \vert \Psi_{J M_J, I} \rangle) \,,
\end{equation}
and a qualitative survey of dominant components,
\begin{equation}
\label{DCC}
C_p = Re(\sum_{i,j} \langle c^l_{ij} \Psi_{J M_J, I, i, j} \vert c^r_{ij} \Psi_{J M_J, I, i, j} \rangle) \,.
\end{equation}
%In particular, these complex observables of resonance with small width are analyzed by considering the real-parts. 
Furthermore, in Eq.~\eqref{MM}, $\hat Q_i$ denotes the electric charge operator of the $i$-th quark, while $\sigma^z_i$ represents the $z$-component of the Pauli spin matrix. The coefficients $c^l_{ij}$ and $c^r_{ij}$ correspond to the left and right generalized eigenvectors, respectively, of the fully antisymmetric complex wave function.

An important limitation of our theoretical framework must be emphasized: the various multiquark channels are not mutually orthogonal. Consequently, two types of calculations are carried out to evaluate the wave function components of the triply heavy tetraquark states. In the first approach, only the diagonal elements of the resulting matrix from Eq.~\eqref{DCC} are considered when determining the contribution from each channel. In the second approach, the off-diagonal elements are also taken into account by summing all elements in the $i$-th row and attributing them to the $i$-th diagonal element, thereby capturing interference effects between different configurations. A comparison on these two sets of results will enable us to significantly identify the dominant components of the either bound or resonant state, but also estimate if the contribution from off-diagonal elements is significant.
 
%%%%%%%%%%%%%%%%%%%%%%%%%%%%%%%%%%%%%%%%%%%%%%%%%%%%%%%%%%%%%%%%%%%

\section{Results}
\label{sec:results}

This section presents the results and discussion concerning triply heavy tetraquarks. As a starting point, the systems are analyzed on the real energy axis by setting the rotation angle $\theta$ to $0^\circ$. However, when performing a fully coupled-channel matrix computation, potential resonances are embedded within the continuum spectrum. These different types of states (bound, resonant and scattering ones) can be clearly distinguished in the complex energy plane by introducing a non-zero rotation angle $\theta$ through the Complex Scaling Method (CSM). To facilitate the extraction of well-defined eigenvalues and eigenvectors for the tetraquark systems, with all $S$-wave configurations included, the rotation angle is varied systematically from $0^\circ$ to $14^\circ$ as an artificial but effective parameter within the CSM framework.

In general, the calculated masses of the low-lying $\bar{Q}Q\bar{q}Q$ tetraquark states are summarized in Tables~\ref{GresultCC1} through~\ref{GresultCCT}. Initially, within the real-range analysis, the lowest masses for each tetraquark system, in the allowed $I(J^P)$ quantum numbers, are provided in Tables~\ref{GresultCC1}, ~\ref{GresultCC2}, ~\ref{GresultCC3}, ~\ref{GresultCC4}, ~\ref{GresultCC5}, ~\ref{GresultCC6}, ~\ref{GresultCC7}, ~\ref{GresultCC8}, ~\ref{GresultCC9}, ~\ref{GresultCC10}, ~\ref{GresultCC11}, and ~\ref{GresultCC12}, respectively. In these tables, the first column lists the considered configurations: meson-meson, diquark-antidiquark, and K-type. Where applicable, the corresponding non-interacting meson-meson threshold values from experimental data are noted in parentheses. The second column assigns an index to each channel, corresponding to a specific combination of spin ($\chi_J^{\sigma_i}$) and color ($\chi_j^c$) wave functions, which are explicitly shown in the third column. The fourth column reports the theoretical mass calculated for each channel, while the final column presents the result obtained from a coupled-channel calculation for the respective type of tetraquark configuration. The last row of each table gives the lowest-lying mass obtained from the fully coupled-channel calculation within the real-range framework.

\begin{figure}[!t]
\includegraphics[clip, trim={3.0cm 1.7cm 3.0cm 0.8cm}, width=0.45\textwidth]{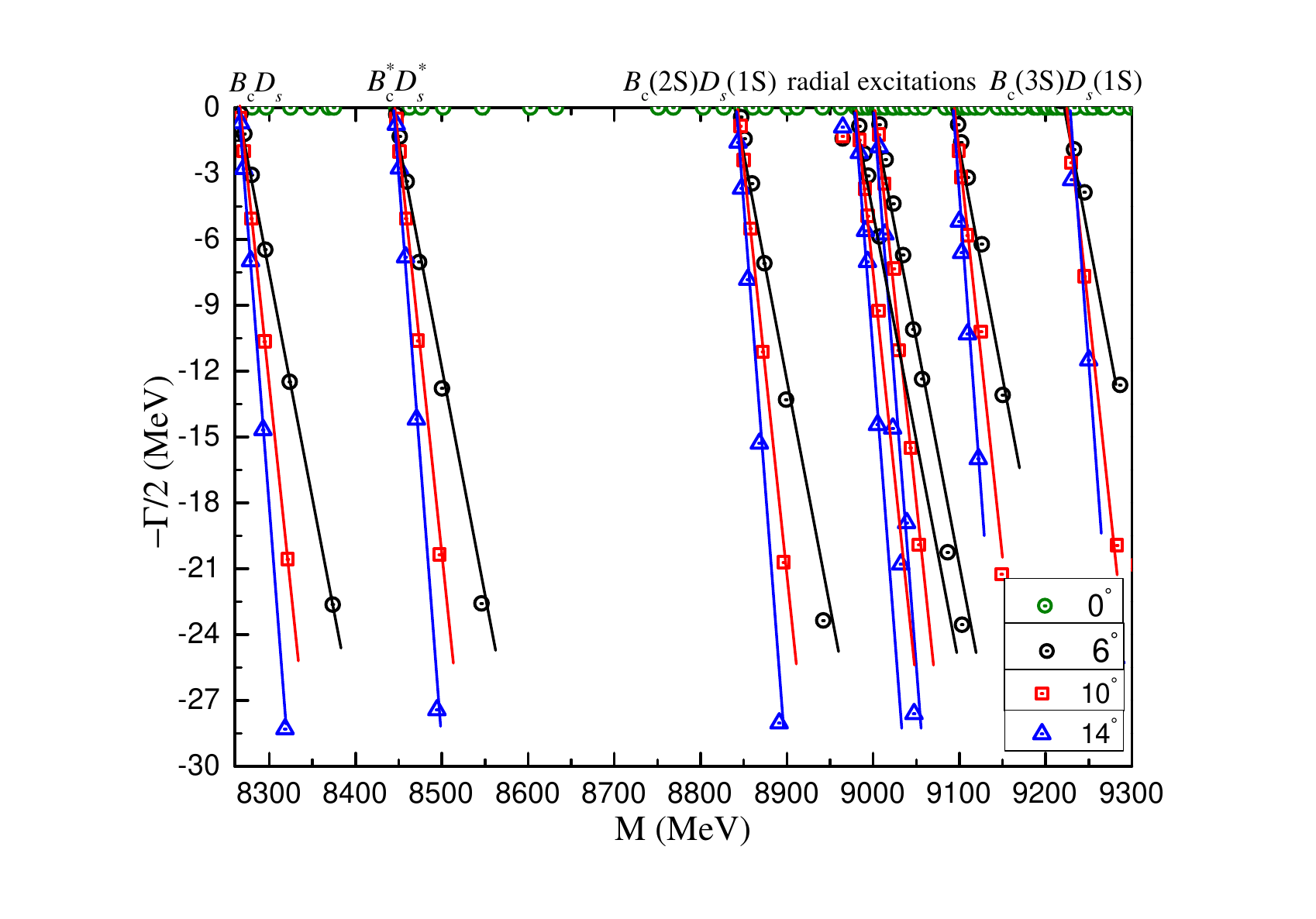} \\[1ex]
\includegraphics[clip, trim={3.0cm 1.9cm 3.0cm 1.0cm}, width=0.45\textwidth]{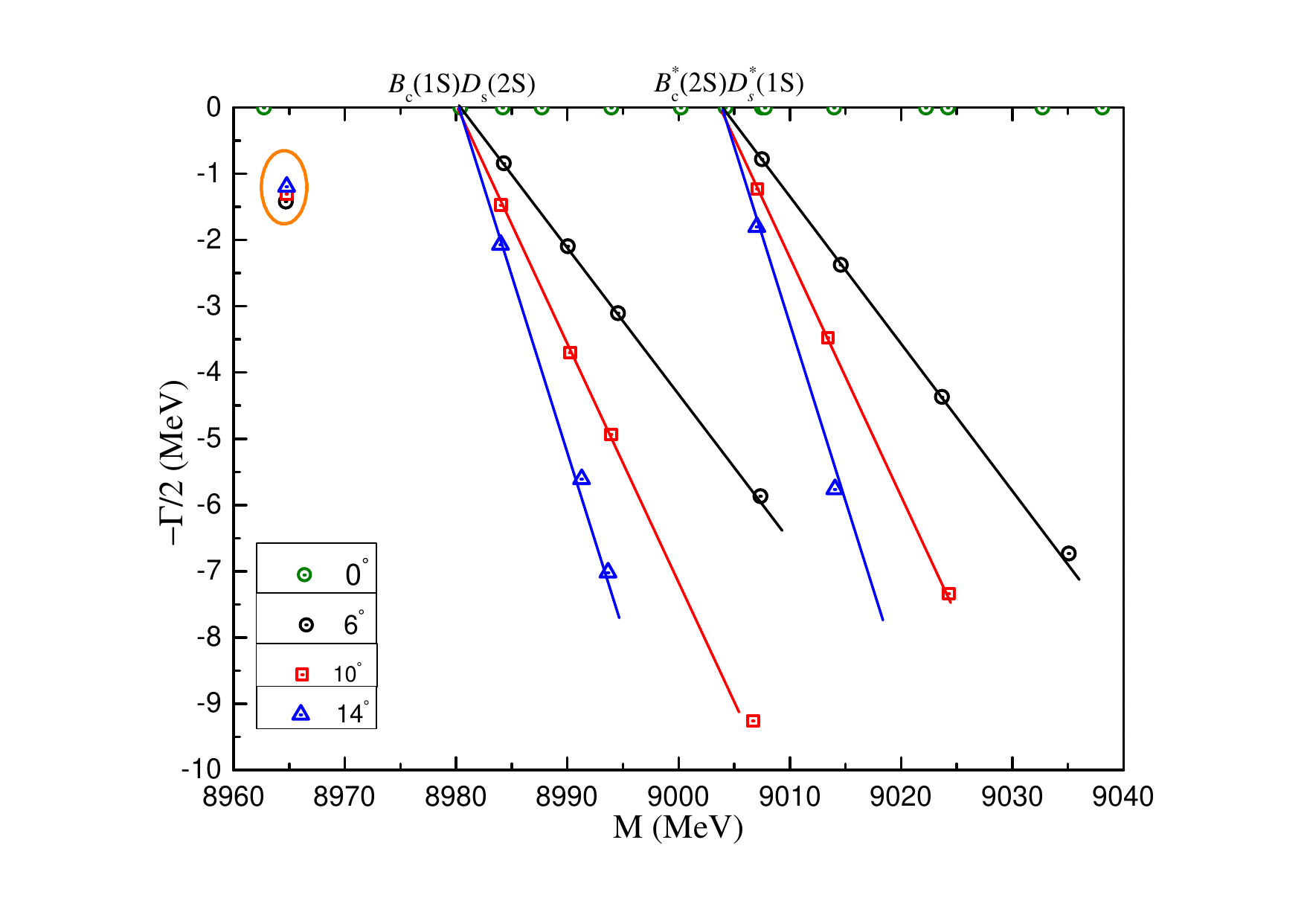}
\caption{\label{PP4} The complete coupled-channels calculation of $\bar{b}c\bar{s}c$ tetraquark system with $I(J^P)=0(0^+)$ quantum numbers. Particularly, the bottom panel is enlarged parts of dense energy region from $8.96\,\text{GeV}$ to $9.04\,\text{GeV}$.}
\end{figure}

Following the real-range investigation, fully coupled-channel calculations for each quantum state of the triply heavy tetraquark systems are performed using the Complex Scaling Method (CSM). Specifically, Figures~\ref{PP1} to~\ref{PP12} display the distribution of the complex eigenenergies, with the identified resonance states highlighted by circles. To gain further insight into the nature of these resonances, several properties are computed -- including root-mean-square (RMS) radii, magnetic moments, and dominant configuration components. The corresponding results are presented in Tables~\ref{GresultR1}, ~\ref{GresultR2}, ~\ref{GresultR3}, ~\ref{GresultR4}, ~\ref{GresultR5}, ~\ref{GresultR6}, ~\ref{GresultR7}, ~\ref{GresultR8}, ~\ref{GresultR9}, ~\ref{GresultR10}, ~\ref{GresultR11}, and ~\ref{GresultR12}. 

Finally, a summary of the most significant findings is provided in Table~\ref{GresultCCT}.

\subsection{The $\mathbf{\bar{b}c\bar{q}c}$ tetraquarks}

%%%%%%%%%%%%%%%%%%%%%%%%%%%%%%%%%%%%%%%%

\begin{table}[!t]
\caption{\label{GresultCC1} Lowest-lying $\bar{b}c\bar{q}c$ tetraquark states with $I(J^P)=\frac{1}{2}(0^+)$ calculated within the real range formulation of the constituent quark model.
The allowed meson-meson, diquark-antidiquark and K-type configurations are listed in the first column; when possible, the experimental value of the non-interacting meson-meson threshold is labeled in parentheses. Each channel is assigned an index in the 2nd column, it reflects a particular combination of spin ($\chi_J^{\sigma_i}$) and color ($\chi_j^c$) wave functions that are shown explicitly in the 3rd column. The theoretical mass obtained in each channel is shown in the 4th column and the coupled result for each kind of configuration is presented in the 5th column.
When a complete coupled-channels calculation is performed, last row of the table indicates the calculated lowest-lying mass (unit: MeV).}
\begin{ruledtabular}
\begin{tabular}{lcccc}
~~Channel   & Index & $\chi_J^{\sigma_i}$;~$\chi_j^c$ & $M$ & Mixed~~ \\
        &   &$[i; ~j]$ &  \\[2ex]
$(B_c D)^1 (8145)$  & 1  & [1;~1]  & $8173$ & \\
$(B^*_c D^*)^1$  & 2  & [2;~1]   & $8347$ & $8173$  \\[2ex]
$(B_c D)^8$          & 3  & [1;~2]  & $8592$ & \\
$(B^*_c D^*)^8$       & 4  & [2;~2]   & $8549$ & $8528$  \\[2ex]
$(cc)(\bar{b}\bar{q})$      & 5     & [3;~4]  & $8599$ & \\
$(cc)^*(\bar{b}\bar{q})^*$  & 6  & [4;~3]   & $8625$ & $8579$ \\[2ex]
$K_1$  & 7  & [5;~5]   & $8534$ & \\
            & 8  & [5;~6]   & $8580$ & \\
            & 9  & [6;~5]   & $8586$ & \\
            & 10  & [6;~6]   & $8521$ & $8433$ \\[2ex]
$K_2$  & 11  & [7;~7]   & $8548$ & \\
             & 12  & [7;~8]   & $8536$ & \\
             & 13  & [8;~7]   & $8482$ & \\
             & 14  & [8;~8]   & $8588$ & $8467$ \\[2ex]
$K_3$  & 15  & [9;~10]   & $8612$ & \\
             & 16  & [10;~9]   & $8591$ & $8566$ \\[2ex]
$K_4$  & 17  & [11;~12]   & $8626$ & \\
             & 18  & [12;~12]   & $8576$ & \\
             & 19  & [11;~11]   & $8967$ & \\
             & 20  & [12;~11]   & $8592$ & $8511$ \\[2ex]
$K_5$  & 21  & [13;~14]   & $8623$ & \\
             & 22  & [14;~13]   & $8567$ & $8547$ \\[2ex]
\multicolumn{4}{c}{Complete coupled-channels:} & $8173$
\end{tabular}
\end{ruledtabular}
\end{table}

\begin{figure}[!t]
\includegraphics[clip, trim={3.0cm 1.7cm 3.0cm 0.8cm}, width=0.45\textwidth]{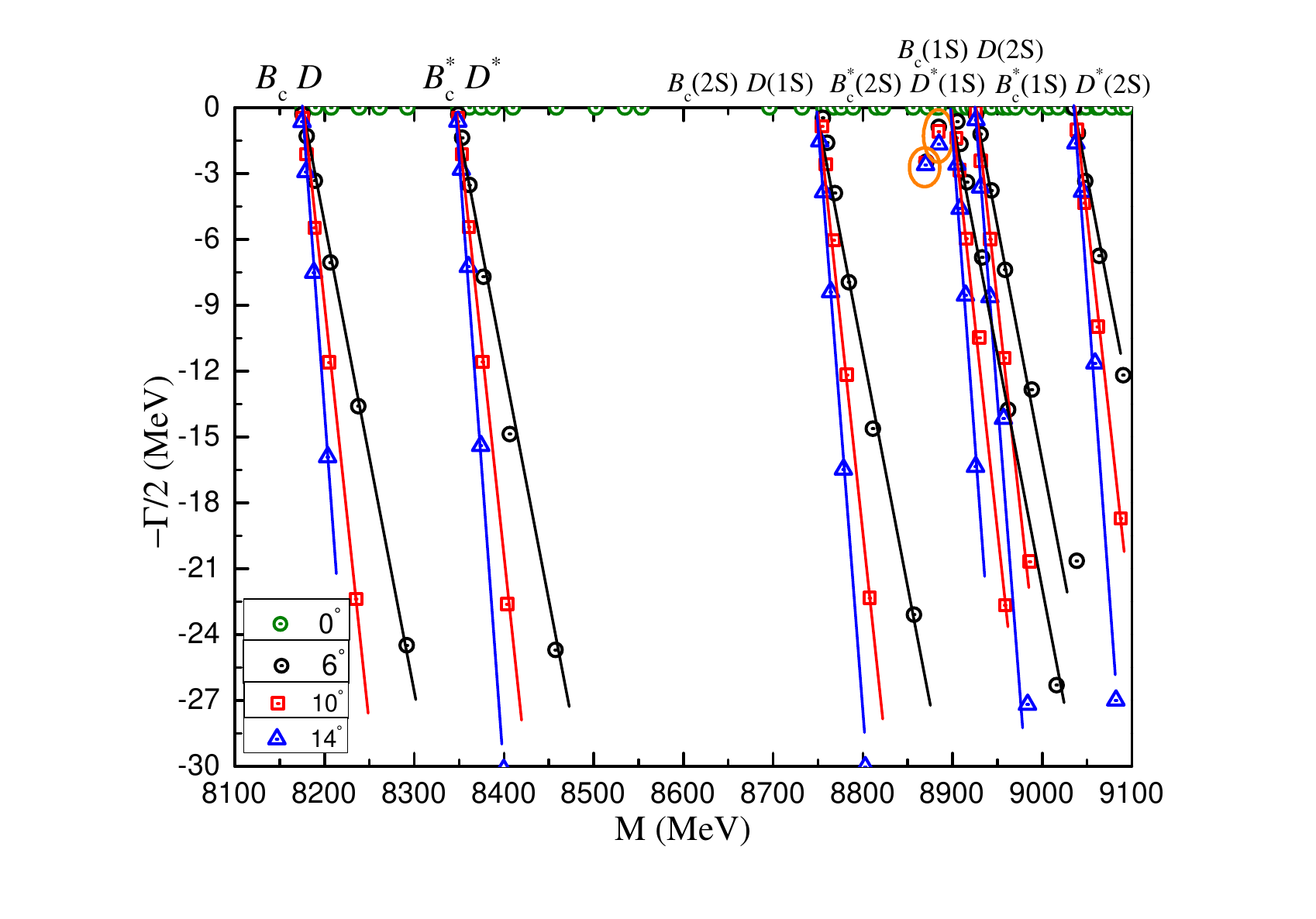} \\[1ex]
\includegraphics[clip, trim={3.0cm 1.7cm 3.0cm 0.8cm}, width=0.45\textwidth]{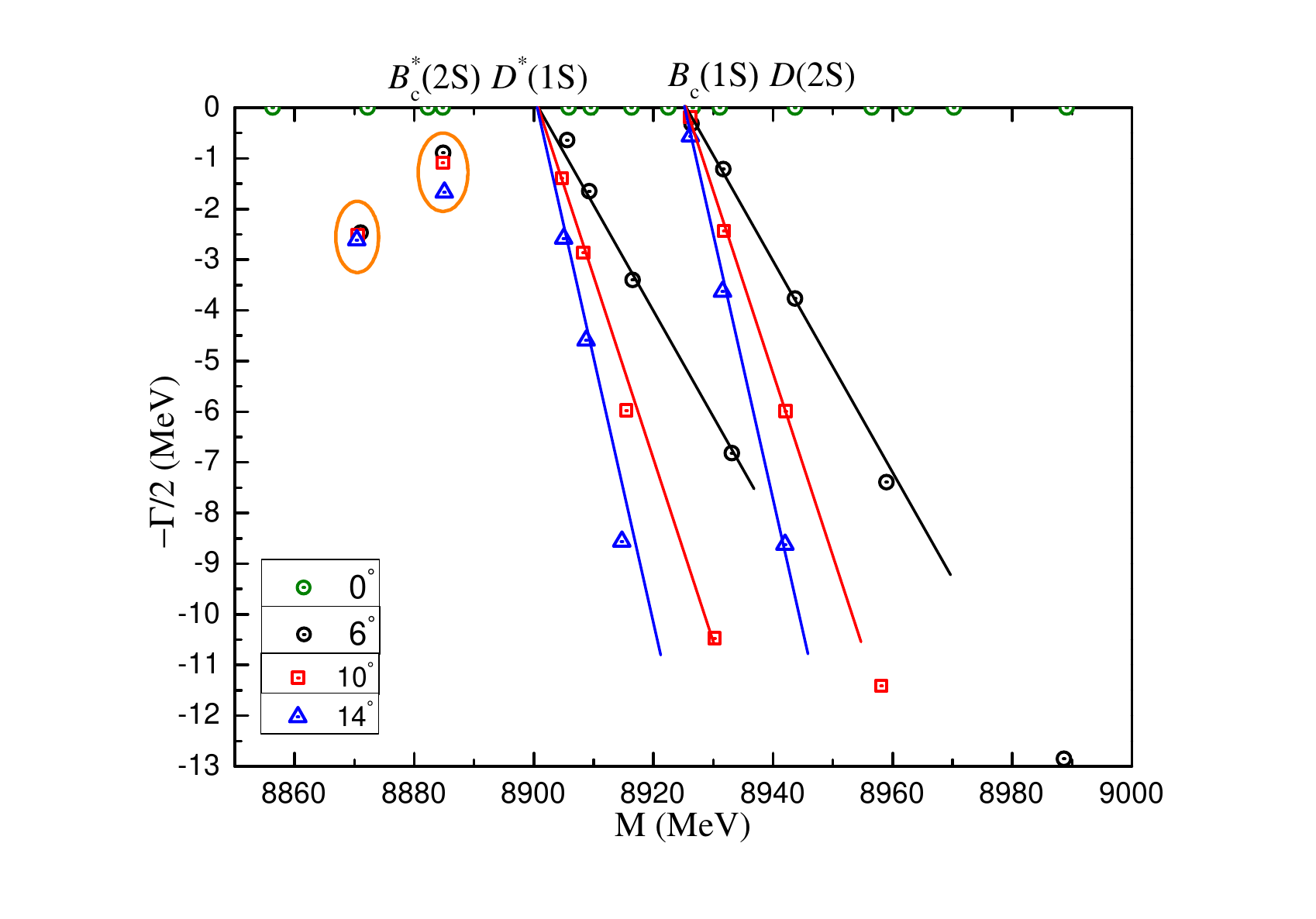}
\caption{\label{PP1} The complete coupled-channels calculation of $\bar{b}c\bar{q}c$ tetraquark system with $I(J^P)=\frac{1}{2}(0^+)$ quantum numbers. Particularly, the bottom panel is an enlarged part of dense energy region from $8.85\,\text{GeV}$ to $9.00\,\text{GeV}$.}
\end{figure}

\begin{table}[!t]
\caption{\label{GresultR1} Compositeness of exotic resonances obtained in a complete coupled-channel calculation in the $\frac{1}{2}(0^+)$ state of $\bar{b}c\bar{q}c$ tetraquark. Particularly, the first column is resonance poles labeled by $M-i\Gamma$, unit in MeV; the second one is the magnetic moment of resonance, unit in $\mu_N$; the distance between any two quarks or quark-antiquark, unit in fm; and the component of resonant state ($S$: dimeson structure in color-singlet channel; $H$: dimeson structure in hidden-color channel; $Di$: diquark-antiquark configuration; $K$: K-type configuration). Herein, two sets of results on a resonance component are listed. Particularly, set I is results on components, that only diagonal elements are employed, and results, that both diagonal and off-diagonal elements are considered, are listed in set II.}
\begin{ruledtabular}
\begin{tabular}{rccc}
Resonance       & \multicolumn{3}{c}{Structure} \\[2ex]
$8870-i5.0$   & \multicolumn{3}{c}{$\mu=0$} \\
  & \multicolumn{3}{c}{$r_{c \bar{b}}:0.71$;\,\,\,\,\,$r_{\bar{b}\bar{q}}:1.02$;\,\,\,\,\,$r_{c\bar{q}}:1.06$;\,\,\,\,\,$r_{cc}:0.94$} \\
$Set$ I: & \multicolumn{3}{c}{$S$: 15.6\%;\, $H$: 3.5\%;\, $Di$: 4.9\%;\, $K$: 76.0\%}\\
$Set$ II: & \multicolumn{3}{c}{$S$: 12.9\%;\, $H$: 5.5\%;\, $Di$: 5.1\%;\, $K$: 76.5\%}\\[2ex]
%%%%%%%%%%
$8885-i2.2$   & \multicolumn{3}{c}{$\mu=0$} \\
  & \multicolumn{3}{c}{$r_{c \bar{b}}:0.73$;\,\,\,\,\,$r_{\bar{b}\bar{q}}:1.12$;\,\,\,\,\,$r_{c\bar{q}}:1.07$;\,\,\,\,\,$r_{cc}:1.03$} \\
$Set$ I: & \multicolumn{3}{c}{$S$: 3.9\%;\, $H$: 7.4\%;\, $Di$: 3.2\%;\, $K$: 85.5\%}\\
$Set$ II: & \multicolumn{3}{c}{$S$: 4.4\%;\, $H$: 7.8\%;\, $Di$: 3.7\%;\, $K$: 84.1\%}\\
\end{tabular}
\end{ruledtabular}
\end{table}

{\bf The $\bm{I(J^P)=\frac{1}{2}(0^+)}$ sector:} A total of 22 channels are considered in this case, as listed in Table~\ref{GresultCC1}. In particular, two meson-meson channels, corresponding to the $B_c D$ and $B_c^* D^*$ configurations, exhibit a scattering nature, with their lowest masses coinciding with the theoretical non-interacting thresholds of 8173 MeV and 8347 MeV, respectively. The hidden-color counterparts of $B_c D$ and $B_c^* D^*$ are found at approximately $8.55$ GeV. The two diquark-antidiquark structures, $(cc)(\bar{b}\bar{q})$ and $(cc)^*(\bar{b}\bar{q})^*$, lie around $8.6$ GeV. Additionally, the 16 K-type configurations are broadly distributed within the $(8.48-8.96)$ GeV range. Notably, no bound state is identified in any individual channel calculation. This conclusion remains valid for both partially and fully coupled-channel scenarios.

The last column of Table~\ref{GresultCC1} presents the results of partially coupled-channel calculations among the eight structural configurations. The lowest mass, $8.17$ GeV, corresponds exactly to the theoretical threshold of the non-interacting $B_c D$ channel. The exotic configurations, including hidden-color, diquark–antidiquark, and K-type arrangements, are all located near $8.5$ GeV. When all 22 channels are considered in a full real-range coupled-channel analysis, the scattering nature of the $B_c D$ state remains unchanged.

Proceeding further, a fully coupled-channel calculation using the CSM reveals two stable resonance poles, as shown in Fig.~\ref{PP1}. In the top panel, six continuum states are clearly identified, including the ground states $B_c D$ and $B_c^* D^*$, along with the radial excitations $B_c^{(*)}(2S) D^{(*)}(1S)$ and $B_c^{(*)}(1S) D^{(*)}(2S)$. The lower panel, which magnifies the energy region from 8.85 to 9.00 GeV, highlights two resonances, encircled for clarity. These resonances possess complex energies of $M-i\Gamma = 8870 - i\,5.0$ MeV and $8885 - i\,2.2$ MeV, respectively.

The internal structure of these two resonances is analyzed by evaluating their root-mean-square (RMS) radii, magnetic moments, and dominant components. Results are summarized in Table~\ref{GresultR1}. Four common features emerge: (i) both resonances are narrow, with widths below 5.0 MeV; (ii) they exhibit compact $\bar{b}c\bar{q}c$ tetraquark structures, with spatial extensions between 0.7 and 1.1 fm; (iii) their magnetic moments are vanishing, suggesting non-magnetic character ($\mu = 0$); and (iv) their dominant structural component, exceeding 76\% in each case, is the K-type configuration. Contributions from off-diagonal terms are found to be minimal by comparison. Finally, both resonances are expected to predominantly decay into the $B_c D$ channel, making them promising candidates for future confirmation in high-energy experiments.

%%%%%%%%%%%%%%%%%%%%%%%%%%%%%%%%%%%%%%%%

\begin{table}[!t]
\caption{\label{GresultCC2} Lowest-lying $\bar{b}c\bar{q}c$ tetraquark states with $I(J^P)=\frac{1}{2}(1^+)$ calculated within the real range formulation of the constituent quark model. Results are similarly organized as those in Table~\ref{GresultCC1} (unit: MeV).}
\begin{ruledtabular}
\begin{tabular}{lcccc}
~~Channel   & Index & $\chi_J^{\sigma_i}$;~$\chi_j^c$ & $M$ & Mixed~~ \\
        &   &$[i; ~j]$ &  \\[2ex]
$(B_c D^*)^1 (8282)$   & 1  & [1;~1]  & $8293$ & \\
$(B^*_c D)^1$  & 2  & [2;~1]   & $8227$ &  \\
$(B^*_c D^*)^1$  & 3  & [3;~1]   & $8347$ & $8227$  \\[2ex]
$(B_c D^*)^8$       & 4  & [1;~2]   & $8582$ &  \\
$(B^*_c D)^8$      & 5  & [2;~2]  & $8588$ & \\
$(B^*_c D^*)^8$  & 6  & [3;~2]   & $8566$ & $8545$ \\[2ex]
$(cc)^*(\bar{b}\bar{q})^*$  & 7  & [6;~3]  & $8633$ & \\
$(cc)^*(\bar{b}\bar{q})$    & 8  & [5;~4]   & $8594$ &  \\
$(cc)(\bar{b}\bar{q})^*$  & 9  & [4;~3]   & $8627$ & $8582$  \\[2ex]
$K_1$      & 10  & [7;~5]   & $8583$ &  \\
      & 11 & [8;~5]  & $8549$ & \\
      & 12  & [9;~5]   & $8574$ & \\
      & 13   & [7;~6]  & $8557$ & \\
      & 14   & [8;~6]  & $8607$ & \\
      & 15   & [9;~6]  & $8551$ & $8466$ \\[2ex]
$K_2$      & 16  & [10;~7]   & $8560$ & \\
                 & 17  & [11;~7]   & $8528$ &  \\
                 & 18  & [12;~7]   & $8517$ & \\
                 & 19  & [10;~8]   & $8558$ & \\
                 & 20  & [11;~8]   & $8570$ & \\
                 & 21  & [12;~8]   & $8576$ & $8488$ \\[2ex]
$K_3$      & 22  & [13;~10]   & $8610$ & \\
                 & 23  & [14;~10]   & $8628$ & \\
                 & 24  & [15;~9]    & $8585$ & $8568$ \\[2ex]
$K_4$      & 25  & [16;~11]   & $8614$ & \\
                 & 26  & [17;~11]   & $8613$ & \\
                 & 27  & [18;~11]   & $9003$ & \\
                 & 28  & [16;~12]   & $8644$ & \\
                 & 29  & [17;~12]   & $8643$ & \\
                 & 30  & [18;~12]   & $8628$ & $8524$ \\[2ex]
$K_5$      & 31  & [19;~14]   & $8641$ & \\
                 & 32  & [20;~14]   & $8624$ & \\
                 & 33  & [21;~13]   & $8561$ & $8548$ \\[2ex]  
\multicolumn{4}{c}{Complete coupled-channels:} & $8227$
\end{tabular}
\end{ruledtabular}
\end{table}

\begin{figure}[!t]
\includegraphics[clip, trim={3.0cm 1.7cm 3.0cm 0.8cm}, width=0.45\textwidth]{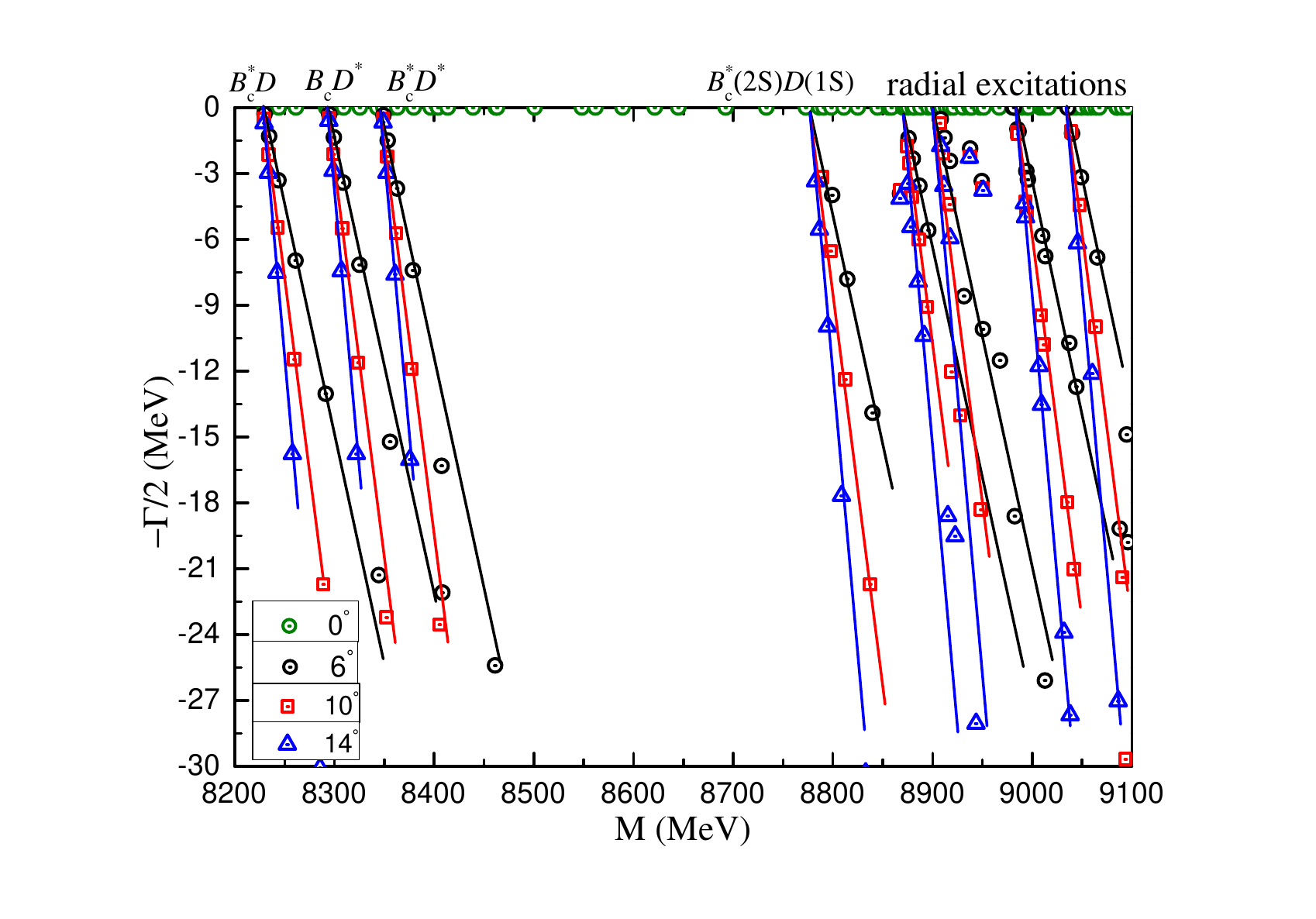} \\[1ex]
\includegraphics[clip, trim={3.0cm 1.7cm 3.0cm 0.8cm}, width=0.45\textwidth]{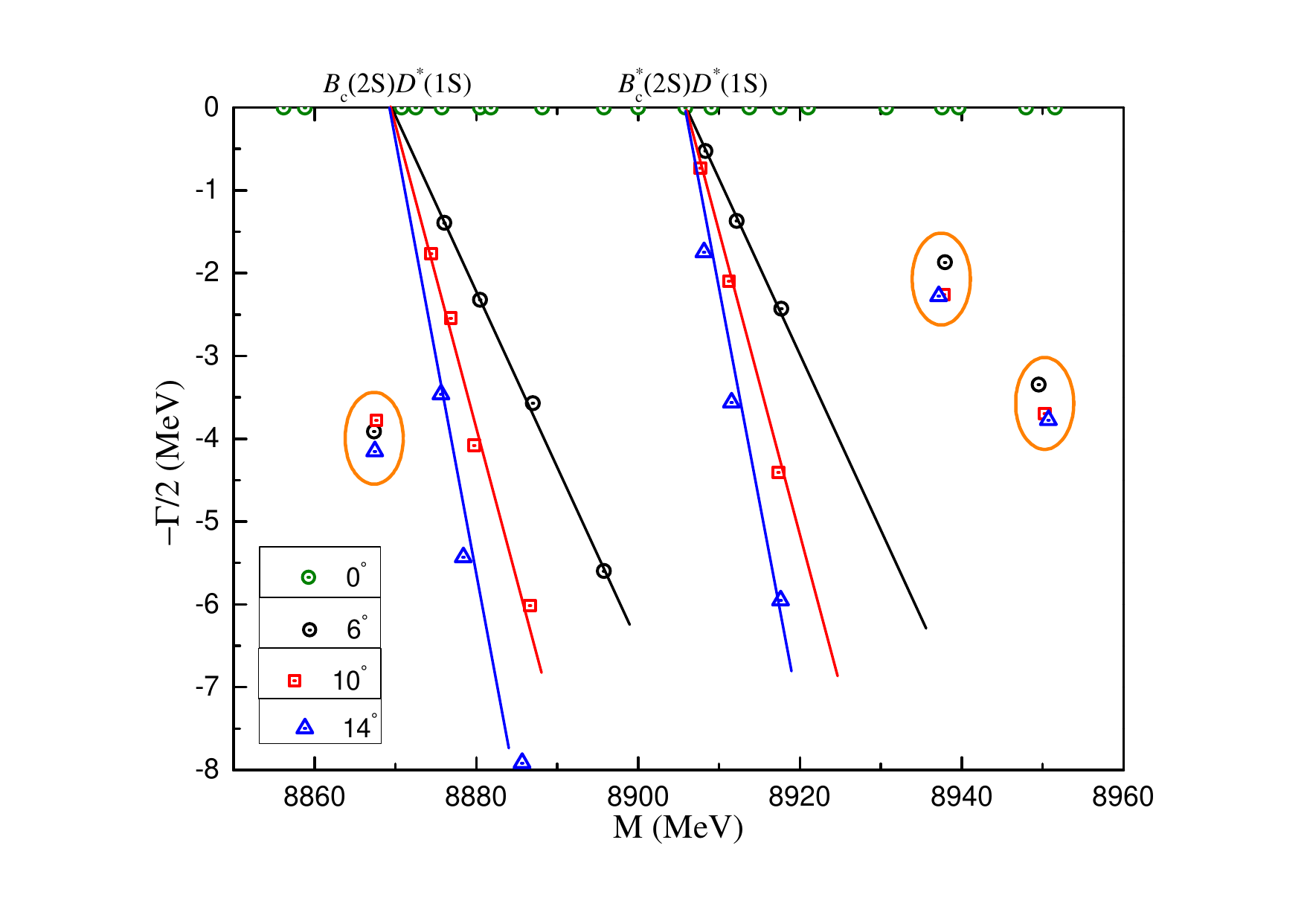}
\caption{\label{PP2} The complete coupled-channels calculation of $\bar{c}c\bar{d}c$ tetraquark system with $I(J^P)=\frac{1}{2}(1^+)$ quantum numbers. Particularly, the bottom panel is enlarged parts of dense energy region from $8.85\,\text{GeV}$ to $8.96\,\text{GeV}$.}
\end{figure}

\begin{table}[!t]
\caption{\label{GresultR2} Compositeness of exotic resonance obtained in a complete coupled-channel calculation in the $\frac{1}{2}(1^+)$ state of $\bar{b}c\bar{q}c$ tetraquark. Results are similarly organized as those in Table~\ref{GresultR1}. Particularly, the magnetic moment of $\bar{b}c\bar{u}c$ and $\bar{b}c\bar{d}c$ tetraquark is $\mu_1$ and $\mu_2$, respectively.}
\begin{ruledtabular}
\begin{tabular}{rccc}
Resonance       & \multicolumn{3}{c}{Structure} \\[2ex]
$8867-i7.8$   & \multicolumn{3}{c}{$\mu_1=-1.42$,\,\,\,$\mu_2=1.20$} \\
  & \multicolumn{3}{c}{$r_{c \bar{b}}:0.82$;\,\,\,\,\,$r_{\bar{b}\bar{q}}:1.12$;\,\,\,\,\,$r_{c\bar{q}}:1.04$;\,\,\,\,\,$r_{cc}:1.02$} \\
$Set$ I: & \multicolumn{3}{c}{$S$: 10.3\%;\, $H$: 5.6\%;\, $Di$: 3.7\%;\, $K$: 80.4\%}\\
$Set$ II: & \multicolumn{3}{c}{$S$: 16.3\%;\, $H$: 10.9\%;\, $Di$: 5.1\%;\, $K$: 67.7\%}\\[2ex]
%%%%%%%%%%
$8937-i4.5$   & \multicolumn{3}{c}{$\mu_1=-1.45$,\,\,\,$\mu_2=1.08$} \\
  & \multicolumn{3}{c}{$r_{c \bar{b}}:0.79$;\,\,\,\,\,$r_{\bar{b}\bar{q}}:1.34$;\,\,\,\,\,$r_{c\bar{q}}:1.27$;\,\,\,\,\,$r_{cc}:1.09$} \\
$Set$ I: & \multicolumn{3}{c}{$S$: 4.4\%;\, $H$: 0.7\%;\, $Di$: 2.3\%;\, $K$: 92.6\%}\\
$Set$ II: & \multicolumn{3}{c}{$S$: 3.4\%;\, $H$: 1.1\%;\, $Di$: 4.7\%;\, $K$: 90.8\%}\\[2ex]
%%%%%%%%%%
$8950-i7.4$   & \multicolumn{3}{c}{$\mu_1=-1.30$,\,\,\,$\mu_2=1.15$} \\
  & \multicolumn{3}{c}{$r_{c \bar{b}}:0.75$;\,\,\,\,\,$r_{\bar{b}\bar{q}}:1.23$;\,\,\,\,\,$r_{c\bar{q}}:1.12$;\,\,\,\,\,$r_{cc}:1.05$} \\
$Set$ I: & \multicolumn{3}{c}{$S$: 3.6\%;\, $H$: 1.1\%;\, $Di$: 2.9\%;\, $K$: 92.4\%}\\
$Set$ II: & \multicolumn{3}{c}{$S$: 4.5\%;\, $H$: 2.0\%;\, $Di$: 2.7\%;\, $K$: 90.8\%}\\
\end{tabular}
\end{ruledtabular}
\end{table}

{\bf The $\bm{I(J^P)=\frac{1}{2}(1^+)}$ sector:} A total of 33 channels are examined for this case, and the results from the real-range computations are summarized in Table~\ref{GresultCC2}. First, in the individual channel analyses, the masses of the three meson-meson configurations: $B_c D^*$, $B_c^* D$, and $B_c^* D^*$, are found to be 8293 MeV, 8227 MeV, and 8347 MeV, respectively. These values correspond precisely to the theoretical thresholds of the associated non-interacting meson pairs. In addition, the hidden-color and diquark-antidiquark channels are clustered around 8.58 GeV and 8.60 GeV, respectively. The remaining 24 K-type channels mostly lie within the range $(8.52-8.64)$ GeV range, except for one outlier, the $K_4$ channel, located at 9.00 GeV.

In the subsequent coupled-channel calculations, carried out separately for the meson-meson, diquark-antidiquark, and K-type configurations, the lowest mass within the meson-meson sector remains at 8227 MeV, corresponding to the theoretical threshold of the $B_c^* D$ state. For both the hidden-color and diquark-antidiquark arrangements, the lowest mass lies around 8.55 GeV. The $K_1$ and $K_2$ configurations yield minimum masses of approximately 8.47 GeV, while the remaining K-type configurations are centered near 8.55 GeV. In the final step of the real-range analysis, all 33 channels are included in a fully coupled calculation; nevertheless, no bound state is observed.

In contrast, when a fully coupled-channel analysis is performed using the CSM, three narrow resonances emerge. The corresponding complex energy eigenvalues are illustrated in Fig.~\ref{PP2}. The top panel displays the ground and radially excited states of the $B_c D^*$, $B_c^* D$, and $B_c^* D^*$ systems, which retain their scattering nature. In the enlarged view of the energy region between 8.85 and 8.96 GeV (bottom panel of Fig.~\ref{PP2}), in addition to the scattering states $B_c(2S) D^*(1S)$ and $B_c^*(2S) D^*(1S)$, three resonance poles are identified at $M-i\Gamma = 8867 - i\,7.8$ MeV, $8937 - i\,4.5$ MeV, and $8950 - i\,7.4$ MeV, respectively.

The internal structure and electromagnetic properties of these resonances are presented in Table~\ref{GresultR2}. The lowest-lying resonance at 8.87 GeV is found to be a compact tetraquark state with a root-mean-square (RMS) radius of less than 1.1 fm. Although off-diagonal contributions lead to approximately 12\% modification, the dominant K-type component remains above 67\%. The golden decay modes for this resonance are identified as $B_c D^*$ and $B_c^* D^*$. In contrast, the other two higher-mass resonances exhibit more diffuse structures with sizes around 1.3 fm. These states also feature K-type components exceeding 90\%, and differences between the two computational sets (with and without off-diagonal elements) are relatively minor. These resonances, located near 8.94 GeV, are expected to be observed in the $B_c^* D^*$ decay channel in future experiments. Lastly, the magnetic moments of the first resonance are $-1.42\,\mu_N$ for the $\bar{b}c\bar{u}c$ configuration and $1.20\,\mu_N$ for $\bar{b}c\bar{d}c$. The corresponding values for the second resonance are $-1.45\,\mu_N$ and $1.08\,\mu_N$, and for the third, $-1.30\,\mu_N$ and $1.15\,\mu_N$, respectively.

%%%%%%%%%%%%%%%%%%%%%%%%%%%%%%%%%%%%%%%%

\begin{table}[!t]
\caption{\label{GresultCC3} Lowest-lying $\bar{b}c\bar{q}c$ tetraquark states with $I(J^P)=\frac{1}{2}(2^+)$ calculated within the real range formulation of the constituent quark model. Results are similarly organized as those in Table~\ref{GresultCC1} (unit: MeV).}
\begin{ruledtabular}
\begin{tabular}{lcccc}
~~Channel   & Index & $\chi_J^{\sigma_i}$;~$\chi_j^c$ & $M$ & Mixed~~ \\
        &   &$[i; ~j]$ &  \\[2ex]
$(B^*_c D^*)^1$  & 1  & [1;~1]   & $8347$ &  \\[2ex]
$(B^*_c D^*)^8$  & 2  & [1;~3]   & $8584$ &  \\[2ex]
$(cc)^*(\bar{b}\bar{q})^*$  & 3  & [1;~7]   & $8649$ & \\[2ex]
$K_1$  & 4  & [1;~8]   & $8581$ & \\
            & 5  & [1;~10]   & $8619$ & $8527$ \\[2ex]
$K_2$  & 6  & [1;~11]   & $8583$ & \\
             & 7  & [1;~12]   & $8582$ & $8568$ \\[2ex]
$K_3$  & 8  & [1;~14]   & $8639$ & \\[2ex]
$K_4$  & 9  & [1;~8]   & $9013$ & \\
             & 10  & [1;~10]   & $8651$ & $8600$ \\[2ex]
$K_5$  & 11  & [1;~11]   & $8652$ & \\[2ex]
\multicolumn{4}{c}{Complete coupled-channels:} & $8347$
\end{tabular}
\end{ruledtabular}
\end{table}

\begin{figure}[!t]
\includegraphics[clip, trim={3.0cm 1.7cm 3.0cm 0.8cm}, width=0.45\textwidth]{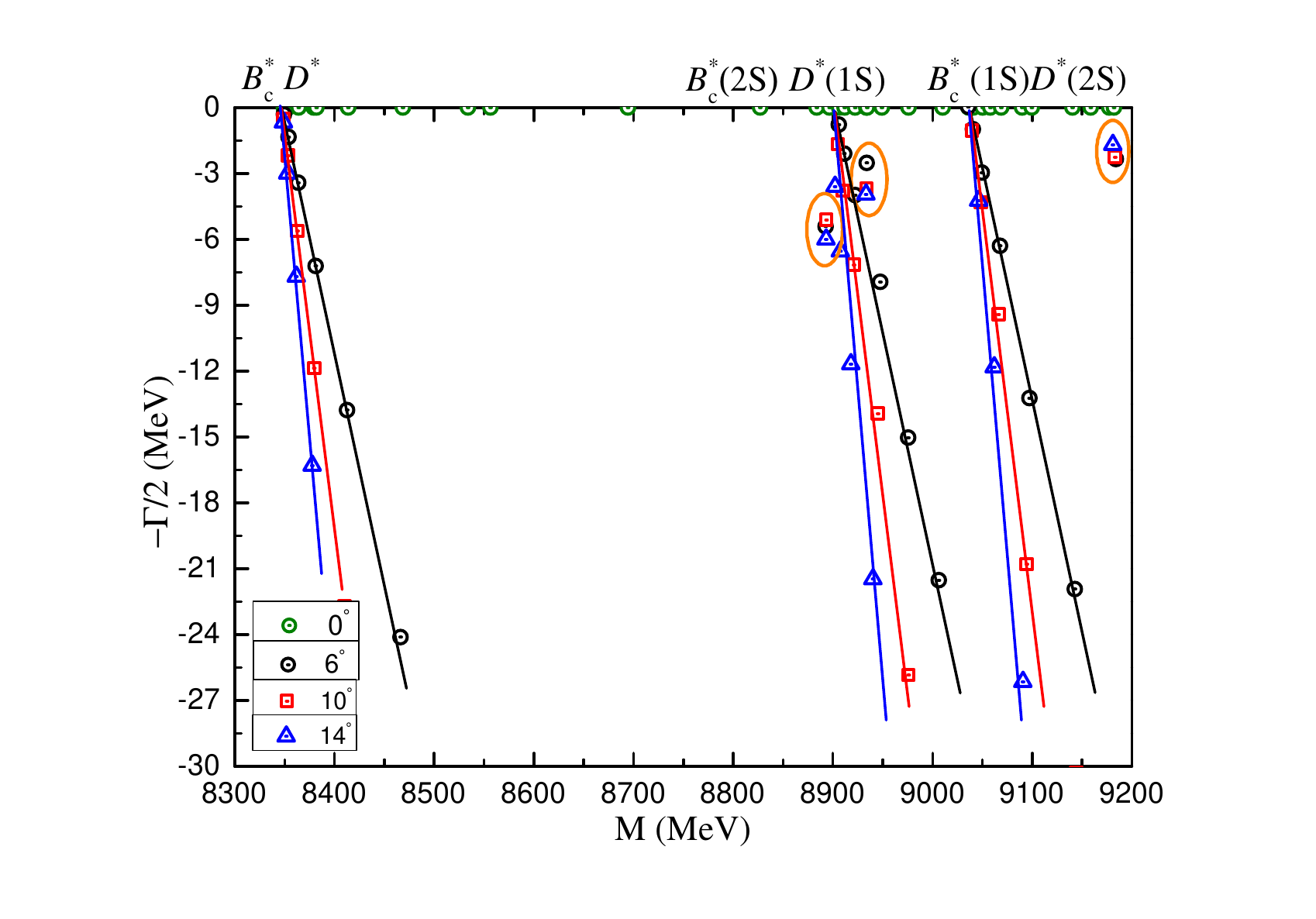}
\caption{\label{PP3} The complete coupled-channels calculation of $\bar{b}c\bar{q}c$ tetraquark system with $I(J^P)=\frac{1}{2}(2^+)$ quantum numbers.}
\end{figure}

\begin{table}[!t]
\caption{\label{GresultR3} Compositeness of exotic resonances obtained in a complete coupled-channel calculation in the $\frac{1}{2}(2^+)$ state of $\bar{b}c\bar{q}c$ tetraquark. Results are similarly organized as those in Table~\ref{GresultR1}.}
\begin{ruledtabular}
\begin{tabular}{rccc}
Resonance       & \multicolumn{3}{c}{Structure} \\[2ex]
$8893-i10.2$   & \multicolumn{3}{c}{$\mu_1=-1.22$,\,\,\,$\mu_2=1.78$} \\
  & \multicolumn{3}{c}{$r_{c \bar{b}}:0.80$;\,\,\,\,\,$r_{\bar{b}\bar{q}}:0.97$;\,\,\,\,\,$r_{c\bar{q}}:0.90$;\,\,\,\,\,$r_{cc}:0.75$} \\
$Set$ I: & \multicolumn{3}{c}{$S$: 5.5\%;\, $H$: 7.7\%;\, $Di$: 0.3\%;\, $K$: 86.5\%}\\
$Set$ II: & \multicolumn{3}{c}{$S$: 1.8\%;\, $H$: 1.3\%;\, $Di$: 0.6\%;\, $K$: 96.3\%}\\[2ex]
%%%%%%%%%%
$8934-i7.4$   & \multicolumn{3}{c}{$\mu_1=-1.22$,\,\,\,$\mu_2=1.78$} \\
  & \multicolumn{3}{c}{$r_{c \bar{b}}:0.87$;\,\,\,\,\,$r_{\bar{b}\bar{q}}:1.36$;\,\,\,\,\,$r_{c\bar{q}}:1.27$;\,\,\,\,\,$r_{cc}:1.22$} \\
$Set$ I: & \multicolumn{3}{c}{$S$: 11.3\%;\, $H$: 1.3\%;\, $Di$: 1.7\%;\, $K$: 85.7\%}\\
$Set$ II: & \multicolumn{3}{c}{$S$: 18.4\%;\, $H$: 1.1\%;\, $Di$: 3.5\%;\, $K$: 77.0\%}\\[2ex]
%%%%%%%%%%
$9183-i4.5$   & \multicolumn{3}{c}{$\mu_1=-1.22$,\,\,\,$\mu_2=1.78$} \\
  & \multicolumn{3}{c}{$r_{c \bar{b}}:0.97$;\,\,\,\,\,$r_{\bar{b}\bar{q}}:0.91$;\,\,\,\,\,$r_{c\bar{q}}:1.00$;\,\,\,\,\,$r_{cc}:0.64$} \\
$Set$ I: & \multicolumn{3}{c}{$S$: 9.9\%;\, $H$: 6.6\%;\, $Di$: 0.8\%;\, $K$: 82.7\%}\\
$Set$ II: & \multicolumn{3}{c}{$S$: 24.4\%;\, $H$: 12.2\%;\, $Di$: 0.4\%;\, $K$: 63.0\%}\\
\end{tabular}
\end{ruledtabular}
\end{table}

{\bf The $\bm{I(J^P)=\frac{1}{2}(2^+)}$ state:} A total of 11 channels are examined for the highest spin state, with the results of the real-range calculations summarized in Table~\ref{GresultCC3}. Among these, the sole meson-meson channel, $B_c^* D$, remains unbound in both single-channel and coupled-channel analyses. Its lowest mass corresponds exactly to the theoretical threshold of 8347 MeV. The hidden-color configuration appears at 8584 MeV, while the diquark-antidiquark structure $(cc)^*(\bar{b}\bar{q})^*$ lies higher, at 8649 MeV. The computed masses of the eight K-type channels are generally clustered around 8.6 GeV, with the exception of the $K_4$ state, which is located at 9.01 GeV. After performing a coupled-channel calculation involving the $K_1$, $K_2$, and $K_4$ structures, the lowest mass obtained is approximately 8.53 GeV.

Figure~\ref{PP3} displays the results of a fully coupled-channel analysis using CSM, where the rotation angle is varied from $0^\circ$ to $14^\circ$. Three scattering states ($B_c^*(1S) D^*(1S)$, $B_c^*(2S) D^*(1S)$, and $B_c^*(1S) D^*(2S)$) can be clearly distinguished in the energy range of 8.3 to 9.2 GeV. In addition, three resonance poles are identified with complex energies of $M-i\Gamma = 8893 - i\,10.2$ MeV, $8934 - i\,7.4$ MeV, and $9183 - i\,4.5$ MeV, respectively.

The internal properties of these three resonances are summarized in Table~\ref{GresultR3}. The lowest and highest resonances are found to be compact $\bar{b}c\bar{q}c$ tetraquark states with root-mean-square radii of less than 1.00 fm. In contrast, the intermediate resonance exhibits a more loosely bound structure, with a spatial extent of approximately 1.3 fm. All three resonances share identical magnetic moments: $-1.22\,\mu_N$ for the $\bar{b}c\bar{u}c$ configuration and $1.78\,\mu_N$ for the $\bar{b}c\bar{d}c$ state. Despite differences of (10-20)\% between the results obtained from the two computational schemes (with and without off-diagonal elements), the K-type structure remains the dominant component for all three resonances, contributing more than 63\%. Finally, these resonances are expected to be identified through the two-body strong decay channel $B_c^* D^*$.

%%%%%%%%%%%%%%%%%%%%%%%%%%%%%%%%%%%%%%%%

\begin{table}[!t]
\caption{\label{GresultCC4} Lowest-lying $\bar{b}c\bar{s}c$ tetraquark states with $I(J^P)=0(0^+)$ calculated within the real range formulation of the constituent quark model. Results are similarly organized as those in Table~\ref{GresultCC1} (unit: MeV).}
\begin{ruledtabular}
\begin{tabular}{lcccc}
~~Channel   & Index & $\chi_J^{\sigma_i}$;~$\chi_j^c$ & $M$ & Mixed~~ \\
        &   &$[i; ~j]$ &  \\[2ex]
$(B_c D_s)^1 (8243)$          & 1  & [1;~1]  & $8265$ & \\
$(B^*_c D^*_s)^1$  & 2  & [2;~1]   & $8445$ & $8265$  \\[2ex]
$(B_c D_s)^8$          & 3  & [1;~2]  & $8693$ & \\
$(B^*_c D^*_s)^8$       & 4  & [2;~2]   & $8658$ & $8638$  \\[2ex]
$(cc)(\bar{b}\bar{s})$      & 5     & [3;~4]  & $8700$ & \\
$(cc)^*(\bar{b}\bar{s})^*$  & 6  & [4;~3]   & $8723$ & $8682$ \\[2ex]
$K_1$  & 7  & [5;~5]   & $8651$ & \\
            & 8  & [5;~6]   & $8693$ & \\
            & 9  & [6;~5]   & $8695$ & \\
            & 10  & [6;~6]   & $8636$ & $8546$ \\[2ex]
$K_2$  & 11  & [7;~7]   & $8665$ & \\
             & 12  & [7;~8]   & $8651$ & \\
             & 13  & [8;~7]   & $8604$ & \\
             & 14  & [8;~8]   & $8694$ & $8589$ \\[2ex]
$K_3$  & 15  & [9;~10]   & $8711$ & \\
             & 16  & [10;~9]   & $8701$ & $8676$ \\[2ex]
$K_4$  & 17  & [11;~12]   & $8723$ & \\
             & 18  & [12;~12]   & $9193$ & \\
             & 19  & [11;~11]   & $9099$ & \\
             & 20  & [12;~11]   & $8693$ & $8631$ \\[2ex]
$K_5$  & 21  & [13;~14]   & $8722$ & \\
             & 22  & [14;~13]   & $8678$ & $8658$ \\[2ex]
\multicolumn{4}{c}{Complete coupled-channels:} & $8265$
\end{tabular}
\end{ruledtabular}
\end{table}

\begin{table}[!t]
\caption{\label{GresultR4} Compositeness of the exotic resonances obtained in a complete coupled-channel calculation in the $0(0^+)$ state of $\bar{b}c\bar{s}c$ tetraquark. Results are similarly organized as those in Table~\ref{GresultR1}.}
\begin{ruledtabular}
\begin{tabular}{rccc}
Resonance       & \multicolumn{3}{c}{Structure} \\[2ex]
$8965-i2.6$   & \multicolumn{3}{c}{$\mu=0$} \\
  & \multicolumn{3}{c}{$r_{c \bar{b}}:0.73$;\,\,\,\,\,$r_{\bar{b}\bar{s}}:0.88$;\,\,\,\,\,$r_{c\bar{s}}:0.94$;\,\,\,\,\,$r_{cc}:0.98$} \\
$Set$ I: & \multicolumn{3}{c}{$S$: 20.4\%;\, $H$: 1.4\%;\, $Di$: 1.3\%;\, $K$: 76.9\%}\\
$Set$ II: & \multicolumn{3}{c}{$S$: 27.7\%;\, $H$: 3.5\%;\, $Di$: 16.7\%;\, $K$: 52.1\%}\\
\end{tabular}
\end{ruledtabular}
\end{table}

%%%%%%%%%%%%%%%%%%%%%%%%%%%%%%%%%%%%%%%%

\subsection{The $\mathbf{\bar{b}c\bar{s}c}$ tetraquarks}

{\bf The $\bm{I(J^P)=0(0^+)}$ sector:} Table~\ref{GresultCC4} lists the 22 channels considered in this study, encompassing meson-meson configurations (both color-singlet and hidden-color), diquark-antidiquark structures, and K-type arrangements. The lowest among them is the $B_c D_s$ scattering channel, with a theoretical mass of 8265 MeV. Another meson-meson channel, $B_c^* D_s^*$, lies at 8445 MeV. All remaining channels, which involve exotic color configurations, are located above 8.60 GeV. Specifically, the two hidden-color channels are found around 8.69 GeV, the two diquark-antidiquark configurations near 8.71 GeV, and the remaining 16 K-type channels generally span the $(8.60-9.19)$ GeV energy range. Upon performing various coupled-channel calculations, the scattering character of the lowest $B_c D_s$ channel remains unaffected. Additionally, the lowest coupled masses among the seven exotic structures cluster around 8.6 GeV.

To search for possible resonant states of the $\bar{b}c\bar{s}c$ tetraquark system, a fully coupled-channel calculation was conducted using CSM. The resulting complex eigenenergies are shown in Fig.~\ref{PP4}. In the top panel, seven continuum states are clearly identified, including the ground states of $B_c D_s$ and $B_c^* D_s^*$, as well as radial excitations such as $B_c^{(*)}(2S) D_s^{(*)}(1S)$, $B_c^{(*)}(1S) D_s^{(*)}(2S)$ and $B_c(3S) D_s(1S)$. Moreover, a single stable resonance pole is circled in the bottom panel within the $(8.96-9.04)$ GeV energy window. The corresponding complex energy of this resonance is $M-i\Gamma = 8965 - i\,2.6$ MeV.

Some properties of this resonance, including quark distances, magnetic moment, and component contributions, are summarized in Table~\ref{GresultR4}. Notably, the magnetic moment remains zero, consistent with the $\bar{b}c\bar{q}c$ resonances in the $\frac{1}{2}(0^+)$ state. The root-mean-square radius lies between 0.73 and 0.98 fm, indicating a compact tetraquark configuration. The dominant component is a K-type structure, contributing over 52\% to the state, though a significant portion, exceeding 18\%, also comes from the color-singlet $B_c^* D_s^*$ channel.

%%%%%%%%%%%%%%%%%%%%%%%%%%%%%%%%%%%%%%%%

\begin{table}[!t]
\caption{\label{GresultCC5} Lowest-lying $\bar{b}c\bar{s}c$ tetraquark states with $I(J^P)=0(1^+)$ calculated within the real range formulation of the constituent quark model. Results are similarly organized as those in Table~\ref{GresultCC1} (unit: MeV).}
\begin{ruledtabular}
\begin{tabular}{lcccc}
~~Channel   & Index & $\chi_J^{\sigma_i}$;~$\chi_j^c$ & $M$ & Mixed~~ \\
        &   &$[i; ~j]$ &  \\[2ex]
$(B_c D^*_s)^1 (8387)$   & 1  & [1;~1]  & $8391$ & \\
$(B^*_c D_s)^1$  & 2  & [2;~1]   & $8319$ &  \\
$(B^*_c D^*_s)^1$  & 3  & [3;~1]   & $8445$ & $8319$  \\[2ex]
$(B_c D^*_s)^8$       & 4  & [1;~2]   & $8684$ &  \\
$(B^*_c D_s)^8$      & 5  & [2;~2]  & $8689$ & \\
$(B^*_c D^*_s)^8$  & 6  & [3;~2]   & $8670$ & $8654$ \\[2ex]
$(cc)^*(\bar{b}\bar{s})^*$  & 7  & [6;~3]  & $8730$ & \\
$(cc)^*(\bar{b}\bar{s})$    & 8  & [5;~4]   & $8696$ &  \\
$(cc)(\bar{b}\bar{s})^*$  & 9  & [4;~3]   & $8723$ & $8686$  \\[2ex]
$K_1$      & 10  & [7;~5]   & $8692$ &  \\
      & 11 & [8;~5]  & $8661$ & \\
      & 12  & [9;~5]   & $8683$ & \\
      & 13   & [7;~6]  & $8673$ & \\
      & 14   & [8;~6]  & $8714$ & \\
      & 15   & [9;~6]  & $8660$ & $8578$ \\[2ex]
$K_2$      & 16  & [10;~7]   & $8673$ & \\
                 & 17  & [11;~7]   & $8648$ &  \\
                 & 18  & [12;~7]   & $8631$ & \\
                 & 19  & [10;~8]   & $8668$ & \\
                 & 20  & [11;~8]   & $8679$ & \\
                 & 21  & [12;~8]   & $8684$ & $8608$ \\[2ex]
$K_3$      & 22  & [13;~10]   & $8710$ & \\
                 & 23  & [14;~10]   & $8724$ & \\
                 & 24  & [15;~9]    & $8696$ & $8680$ \\[2ex]
$K_4$      & 25  & [16;~11]   & $8716$ & \\
                 & 26  & [17;~11]   & $8715$ & \\
                 & 27  & [18;~11]   & $9129$ & \\
                 & 28  & [16;~12]   & $8740$ & \\
                 & 29  & [17;~12]   & $8739$ & \\
                 & 30  & [18;~12]   & $8724$ & $8641$ \\[2ex]
$K_5$      & 31  & [19;~14]   & $8736$ & \\
                 & 32  & [20;~14]   & $8723$ & \\
                 & 33  & [21;~13]   & $8672$ & $8660$ \\[2ex]  
\multicolumn{4}{c}{Complete coupled-channels:} & $8319$
\end{tabular}
\end{ruledtabular}
\end{table}

\begin{figure}[!t]
\includegraphics[clip, trim={3.0cm 1.7cm 3.0cm 0.8cm}, width=0.45\textwidth]{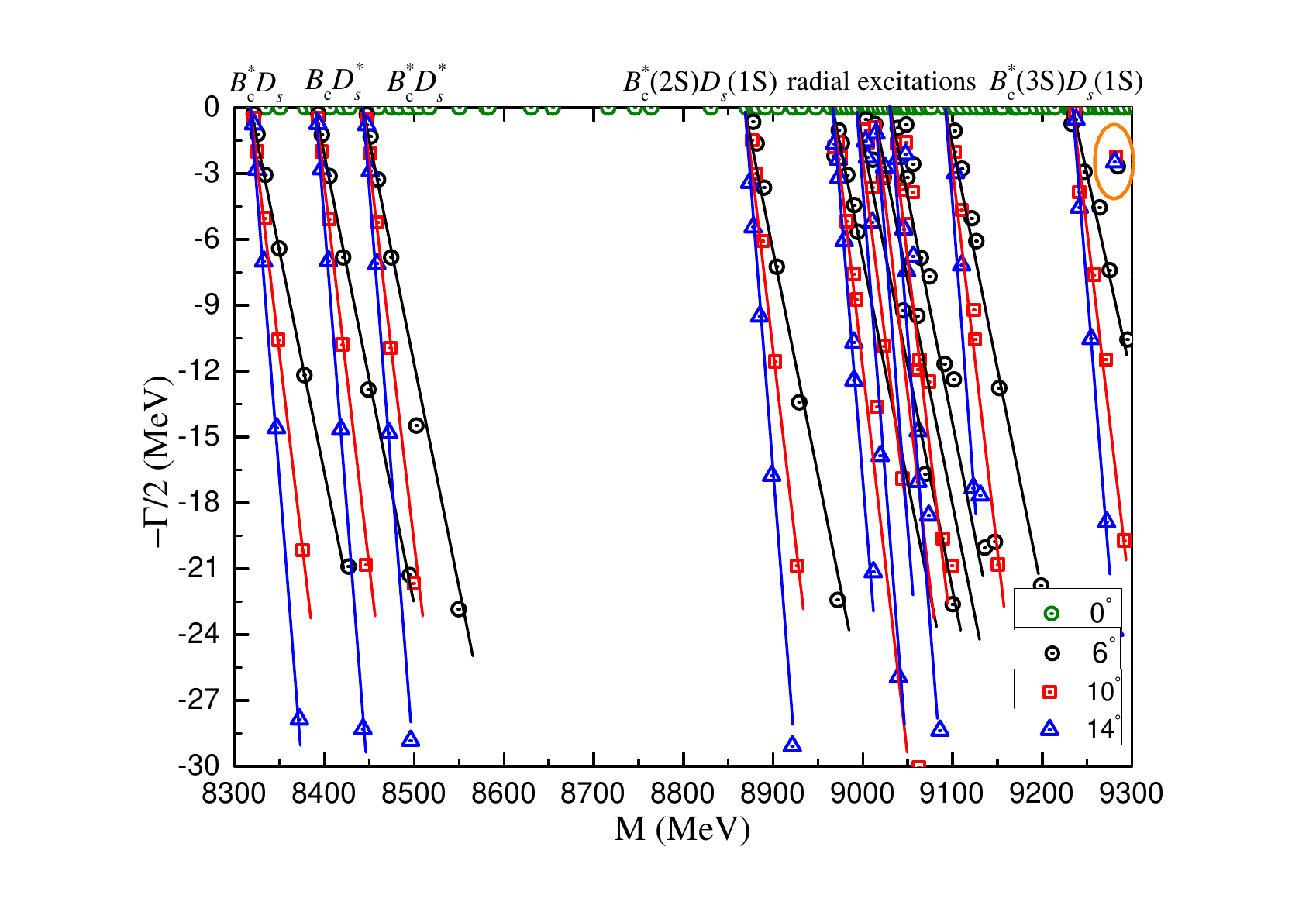} \\[1ex]
\includegraphics[clip, trim={3.0cm 1.7cm 3.0cm 0.8cm}, width=0.45\textwidth]{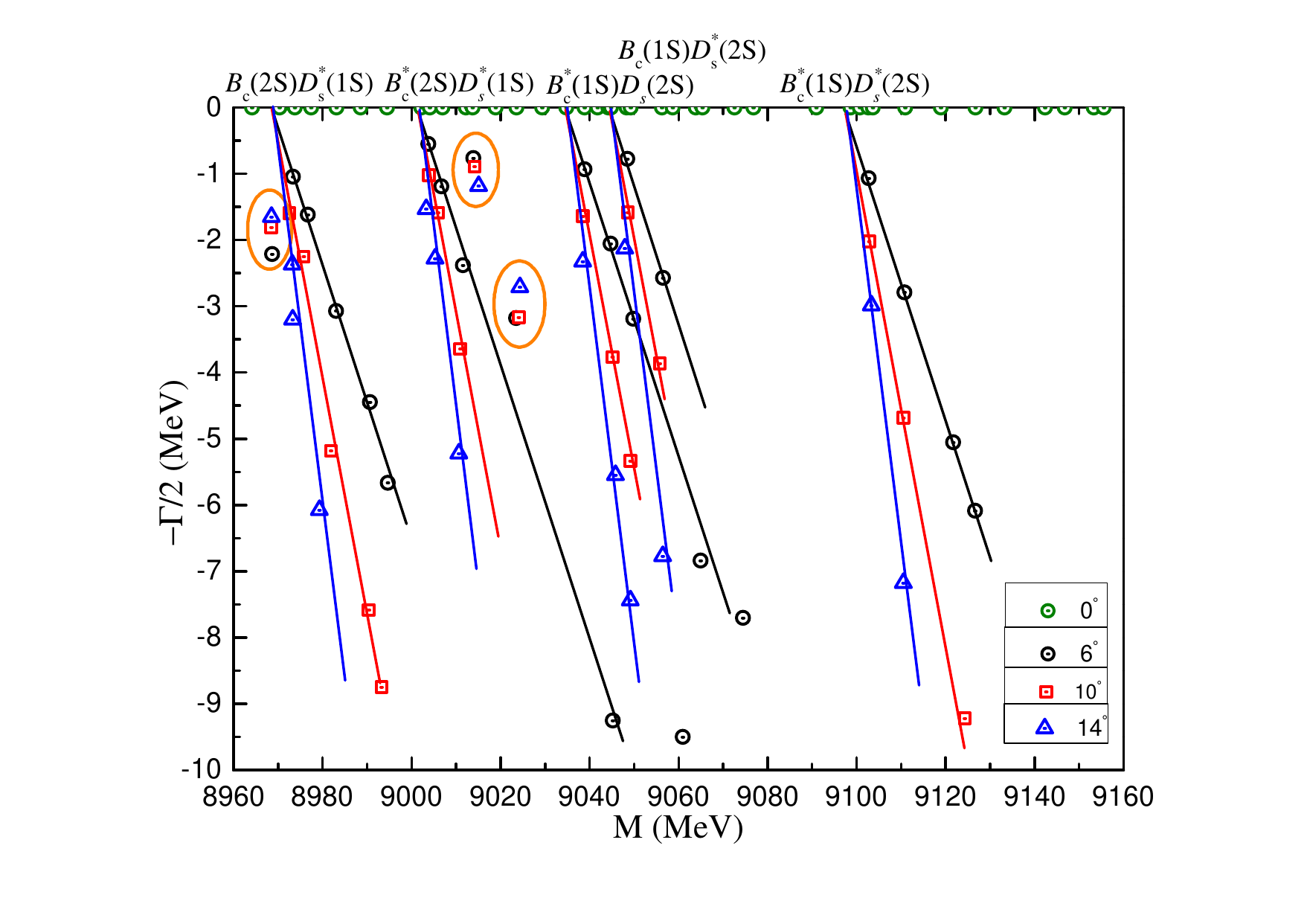}
\caption{\label{PP5} The complete coupled-channels calculation of $\bar{b}c\bar{s}c$ tetraquark system with $I(J^P)=0(1^+)$ quantum numbers. Particularly, the bottom panel is enlarged parts of dense energy region from $8.96\,\text{GeV}$ to $9.16\,\text{GeV}$.}
\end{figure}

\begin{table}[!t]
\caption{\label{GresultR5} Compositeness of exotic resonances obtained in a complete coupled-channel calculation in the $0(1^+)$ state of $\bar{b}c\bar{s}c$ tetraquark. Results are similarly organized as those in Table~\ref{GresultR1}.}
\begin{ruledtabular}
\begin{tabular}{rccc}
Resonance       & \multicolumn{3}{c}{Structure} \\[2ex]
$8968-i3.6$   & \multicolumn{3}{c}{$\mu=0.86$} \\
  & \multicolumn{3}{c}{$r_{c \bar{b}}:0.74$;\,\,\,\,\,$r_{\bar{b}\bar{s}}:1.02$;\,\,\,\,\,$r_{c\bar{s}}:0.99$;\,\,\,\,\,$r_{cc}:0.98$} \\
$Set$ I: & \multicolumn{3}{c}{$S$: 5.1\%;\, $H$: 6.9\%;\, $Di$: 8.7\%;\, $K$: 79.3\%}\\
$Set$ II: & \multicolumn{3}{c}{$S$: 7.7\%;\, $H$: 9.4\%;\, $Di$: 7.3\%;\, $K$: 75.6\%}\\[2ex]
%%%%%%%%%%
$9014-i1.8$   & \multicolumn{3}{c}{$\mu=0.53$} \\
  & \multicolumn{3}{c}{$r_{c \bar{b}}:0.73$;\,\,\,\,\,$r_{\bar{b}\bar{s}}:0.91$;\,\,\,\,\,$r_{c\bar{s}}:0.90$;\,\,\,\,\,$r_{cc}:0.98$} \\
$Set$ I: & \multicolumn{3}{c}{$S$: 4.2\%;\, $H$: 1.2\%;\, $Di$: 6.3\%;\, $K$: 88.3\%}\\
$Set$ II: & \multicolumn{3}{c}{$S$: 4.6\%;\, $H$: 1.4\%;\, $Di$: 7.3\%;\, $K$: 86.7\%}\\[2ex]
%%%%%%%%%%
$9024-i6.4$   & \multicolumn{3}{c}{$\mu=0.74$} \\
  & \multicolumn{3}{c}{$r_{c \bar{b}}:0.80$;\,\,\,\,\,$r_{\bar{b}\bar{s}}:1.13$;\,\,\,\,\,$r_{c\bar{s}}:0.91$;\,\,\,\,\,$r_{cc}:1.10$} \\
$Set$ I: & \multicolumn{3}{c}{$S$: 13.6\%;\, $H$: 3.2\%;\, $Di$: 12.8\%;\, $K$: 70.4\%}\\
$Set$ II: & \multicolumn{3}{c}{$S$: 10.3\%;\, $H$: 4.0\%;\, $Di$: 9.6\%;\, $K$: 76.1\%}\\[2ex]
%%%%%%%%%%
$9283-i4.4$   & \multicolumn{3}{c}{$\mu=1.04$} \\
  & \multicolumn{3}{c}{$r_{c \bar{b}}:0.92$;\,\,\,\,\,$r_{\bar{b}\bar{s}}:0.77$;\,\,\,\,\,$r_{c\bar{s}}:0.87$;\,\,\,\,\,$r_{cc}:0.58$} \\
$Set$ I: & \multicolumn{3}{c}{$S$: 7.3\%;\, $H$: 0.8\%;\, $Di$: 7.9\%;\, $K$: 84.0\%}\\
$Set$ II: & \multicolumn{3}{c}{$S$: 4.9\%;\, $H$: 2.7\%;\, $Di$: 6.2\%;\, $K$: 86.2\%}
\end{tabular}
\end{ruledtabular}
\end{table}

{\bf The $\bm{I(J^P)=0(1^+)}$ sector:} A total of 33 channels contribute to this case, and their masses, as obtained in the real-range investigation, are listed in Table~\ref{GresultCC5}. Several key observations can be drawn from these results: First, the theoretical threshold masses of the individual meson-meson channels $B_c D_s^*$, $B_c^* D_s$, and $B_c^* D_s^*$ are 8391 MeV, 8319 MeV, and 8445 MeV, respectively. Additionally, in the single-channel calculations, the hidden-color and diquark–antidiquark configurations are found to lie around 8.68 GeV and 8.72 GeV, respectively. The 24 K-type channels span a broader energy range between 8.63 GeV and 9.12 GeV. Second, no bound state is identified in this sector. In the coupled-channel calculations, the lowest mass remains at 8319 MeV, corresponding exactly to the theoretical threshold of the $B_c^* D_s$ channel. The lowest coupled-channel masses for the other configurations are approximately 8.6 GeV.

When a complete coupled-channel calculation is performed using the complex-range method, four resonances are identified. In the top panel of Fig.~\ref{PP5}, both the ground and radially excited states of the $B_c D_s^*$, $B_c^* D_s$, and $B_c^* D_s^*$ channels are clearly depicted. One resonance pole is especially distinct and is highlighted with a circle; its complex energy is $M-i\Gamma = 9283 - i\,4.4$ MeV. An enlarged view, covering the $(8.96-9.16)$ GeV energy region, is presented in the bottom panel. This includes the scattering states $B_c(2S) D_s^*(1S)$, $B_c^*(2S) D_s^*(1S)$, $B_c^*(1S) D_s(2S)$, $B_c(1S) D_s^*(2S)$, and $B_c^*(1S) D_s^*(2S)$. Among these complex energy points, three additional resonance poles are identified at $8968 - i\,3.6$ MeV, $9014 - i\,1.8$ MeV and $9024 - i\,6.4$ MeV.

The magnetic moments of these four isoscalar resonances with a $\bar{b}c\bar{s}c$ quark composition are given in Table~\ref{GresultR5}, with values of $0.86\,\mu_N$, $0.53\,\mu_N$, $0.74\,\mu_N$, and $1.04\,\mu_N$, respectively. All of these states exhibit compact tetraquark structures, with sizes less than 1.1 fm. A comparison between component sets I and II shows minimal modifications due to off-diagonal elements, which generally contributes less than 6\%; and, in all cases, the dominant configuration is the K-type structure. The decay channels $B_c D_s^*$ and $B_c^* D_s^*$ serve as the primary decay modes for these resonances, except for the resonance at 9.01 GeV, for which the dominant decay channel is theoretically predicted to be $B_c^* D_s$.

%%%%%%%%%%%%%%%%%%%%%%%%%%%%%%%%%%%%%

\begin{table}[!t]
\caption{\label{GresultCC6} Lowest-lying $\bar{b}c\bar{s}c$ tetraquark states with $I(J^P)=0(2^+)$ calculated within the real range formulation of the constituent quark model. Results are similarly organized as those in Table~\ref{GresultCC1} (unit: MeV).}
\begin{ruledtabular}
\begin{tabular}{lcccc}
~~Channel   & Index & $\chi_J^{\sigma_i}$;~$\chi_j^c$ & $M$ & Mixed~~ \\
        &   &$[i; ~j]$ &  \\[2ex]
$(B^*_c D^*_s)^1$  & 1  & [1;~1]   & $8445$ &  \\[2ex]
$(B^*_c D^*_s)^8$            & 2  & [1;~3]   & $8685$ &  \\[2ex]
$(cc)^*(\bar{b}\bar{s})^*$  & 3  & [1;~7]   & $8744$ & \\[2ex]
$K_1$  & 4  & [1;~8]   & $8689$ & \\
            & 5  & [1;~10]   & $8724$ & $8629$ \\[2ex]
$K_2$  & 6  & [1;~11]   & $8693$ & \\
             & 7  & [1;~12]   & $8689$ & $8678$ \\[2ex]
$K_3$  & 8  & [1;~14]   & $8735$ & \\[2ex]
$K_4$  & 9  & [1;~8]   & $9138$ & \\
             & 10  & [1;~10]   & $8745$ & $8710$ \\[2ex]
$K_5$  & 11  & [1;~11]   & $8747$ & \\[2ex]
\multicolumn{4}{c}{Complete coupled-channels:} & $8445$
\end{tabular}
\end{ruledtabular}
\end{table}

\begin{figure}[!t]
\includegraphics[clip, trim={3.0cm 1.7cm 3.0cm 0.8cm}, width=0.45\textwidth]{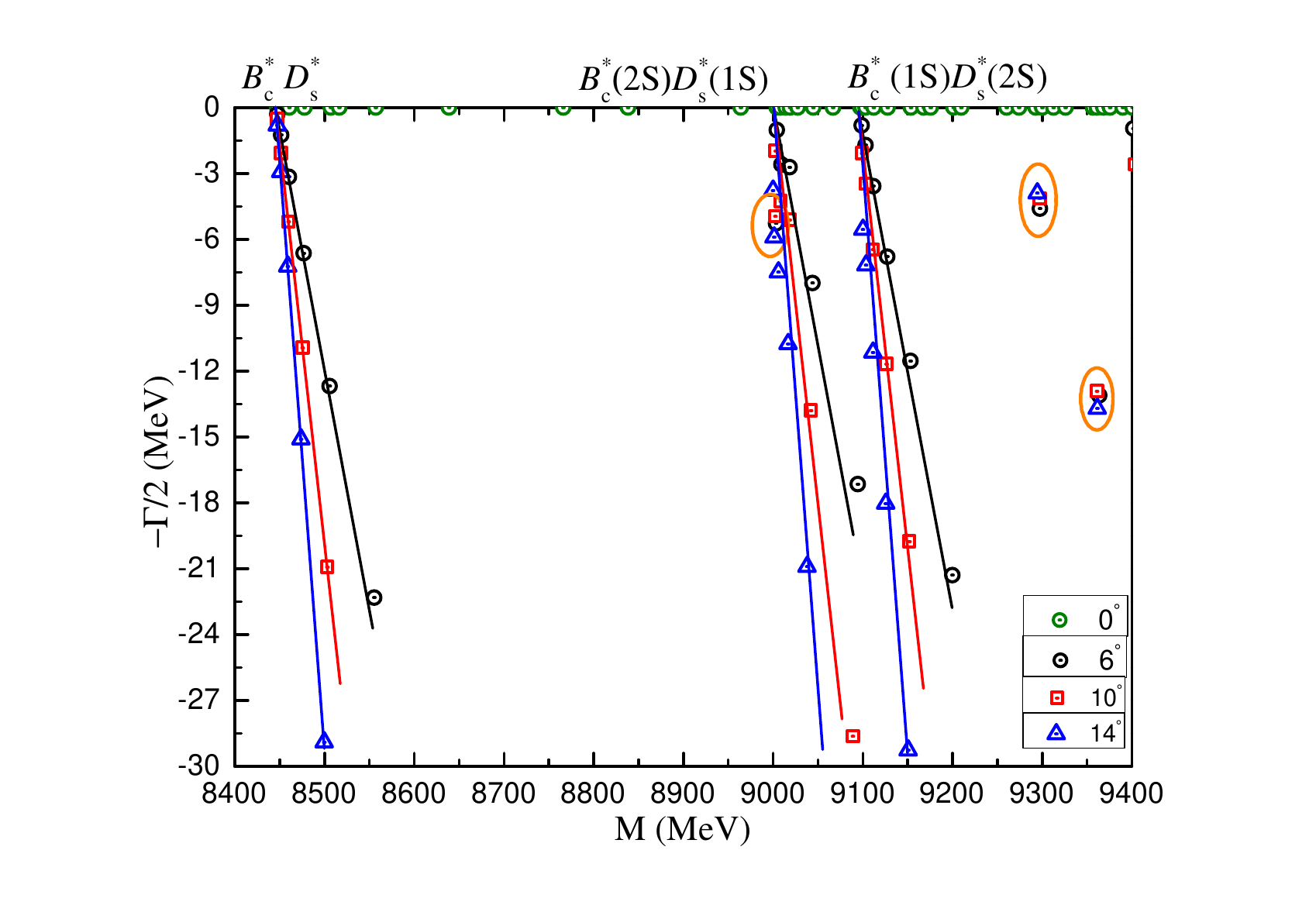}
\caption{\label{PP6} The complete coupled-channels calculation of $\bar{b}c\bar{s}c$ tetraquark system with $I(J^P)=0(2^+)$ quantum numbers.}
\end{figure}

\begin{table}[!t]
\caption{\label{GresultR6} Compositeness of exotic resonances obtained in a complete coupled-channel calculation in the $0(2^+)$ state of $\bar{b}c\bar{s}c$ tetraquark. Results are similarly organized as those in Table~\ref{GresultR1}.}
\begin{ruledtabular}
\begin{tabular}{rccc}
Resonance       & \multicolumn{3}{c}{Structure} \\[2ex]
$9004-i10.6$   & \multicolumn{3}{c}{$\mu=1.34$} \\
  & \multicolumn{3}{c}{$r_{c \bar{b}}:0.77$;\,\,\,\,\,$r_{\bar{b}\bar{s}}:0.88$;\,\,\,\,\,$r_{c\bar{s}}:0.80$;\,\,\,\,\,$r_{cc}:0.80$} \\
$Set$ I: & \multicolumn{3}{c}{$S$: 1.5\%;\, $H$: 0.1\%;\, $Di$: 3.4\%;\, $K$: 95.0\%}\\
$Set$ II: & \multicolumn{3}{c}{$S$: 0.6\%;\, $H$: 0.1\%;\, $Di$: 11.8\%;\, $K$: 87.5\%}\\[2ex]
%%%%%%%%%%
$9298-i9.2$   & \multicolumn{3}{c}{$\mu=1.34$} \\
  & \multicolumn{3}{c}{$r_{c \bar{b}}:0.92$;\,\,\,\,\,$r_{\bar{b}\bar{s}}:0.73$;\,\,\,\,\,$r_{c\bar{s}}:0.88$;\,\,\,\,\,$r_{cc}:0.55$} \\
$Set$ I: & \multicolumn{3}{c}{$S$: 1.6\%;\, $H$: 0.2\%;\, $Di$: 18.5\%;\, $K$: 79.7\%}\\
$Set$ II: & \multicolumn{3}{c}{$S$: 2.0\%;\, $H$: 0.3\%;\, $Di$: 16.6\%;\, $K$: 81.1\%}\\[2ex]
%%%%%%%%%%
$9364-i26.2$   & \multicolumn{3}{c}{$\mu=1.34$} \\
  & \multicolumn{3}{c}{$r_{c \bar{b}}:0.59$;\,\,\,\,\,$r_{\bar{b}\bar{s}}:0.68$;\,\,\,\,\,$r_{c\bar{s}}:0.77$;\,\,\,\,\,$r_{cc}:0.54$} \\
$Set$ I: & \multicolumn{3}{c}{$S$: 2.0\%;\, $H$: 0.6\%;\, $Di$: 13.5\%;\, $K$: 83.9\%}\\
$Set$ II: & \multicolumn{3}{c}{$S$: 2.6\%;\, $H$: 1.7\%;\, $Di$: 16.7\%;\, $K$: 79.0\%}
\end{tabular}
\end{ruledtabular}
\end{table}

{\bf The $\bm{I(J^P)=0(2^+)}$ sector:} The results of real-range calculations for the highest spin state of the $\bar{b}c\bar{s}c$ tetraquark are presented in Table~\ref{GresultCC6}. Among the 11 channels considered, the lowest corresponds to the color-singlet $B_c^* D_s^*$ configuration, with its mass equal to the theoretical threshold of 8445 MeV. The calculated lowest masses of the hidden-color channels $(B_c^* D_s^*)^8$ and $(cc)^*(\bar{b}\bar{s})^*$ are approximately 8.68 GeV and 8.74 GeV, respectively. The eight K-type arrangements span the energy range from 8.68 GeV to 9.13 GeV, with the lowest masses in the coupled-channel calculations for each K-type structure around 8.67 GeV. Notably, no bound state emerges even when performing a full coupled-channels analysis.

Figure~\ref{PP6} shows that, within the energy interval of 8.4 to 9.4 GeV, the ground state of $B_c^* D_s^*$ and the first radial excitations $B_c^*(2S) D_s^*(1S)$ and $B_c^*(1S) D_s^*(2S)$ are clearly identified. Additionally, three resonances are found near the threshold lines associated with these radial excitations. Their complex energies are $M-i\Gamma = 9004 - i\,10.6$ MeV, $9298 - i\,9.2$ MeV, and $9364 - i\,26.2$ MeV, respectively.

A detailed characterization of these resonances is summarized in Table~\ref{GresultR6}. Remarkably, all three states share the same magnetic moment value of $\mu = 1.34\,\mu_N$. They exhibit compact $\bar{b}c\bar{s}c$ tetraquark configurations, with sizes ranging from 0.6 to 0.9 fm. The K-type components contribute more than 80\% to these exotic states, while the effects of the non-orthogonal basis remain minor (less than 7\%) in this sector. Consequently, these three narrow resonances are promising candidates for experimental confirmation via two-body strong decay channels involving $B_c^* D_s^*$.

%%%%%%%%%%%%%%%%%%%%%%%%%%%%%%%%%%%%%%%%%%%%%%%%%%%%%%%%%%

\subsection{The $\mathbf{\bar{c}b\bar{q}b}$ tetraquarks}

\begin{table}[!t]
\caption{\label{GresultCC7} Lowest-lying $\bar{c}b\bar{q}b$ tetraquark states with $I(J^P)=\frac{1}{2}(0^+)$ calculated within the real range formulation of the constituent quark model. Results are similarly organized as those in Table~\ref{GresultCC1} (unit: MeV).}
\begin{ruledtabular}
\begin{tabular}{lcccc}
~~Channel   & Index & $\chi_J^{\sigma_i}$;~$\chi_j^c$ & $M$ & Mixed~~ \\
        &   &$[i; ~j]$ &  \\[2ex]
$(B_c B)^1 (11555)$          & 1  & [1;~1]  & $11554$ & \\
$(B^*_c B^*)^1$  & 2  & [2;~1]   & $11649$ & $11554$  \\[2ex]
$(B_c B)^8$          & 3  & [1;~2]  & $11879$ & \\
$(B^*_c B^*)^8$       & 4  & [2;~2]   & $11909$ & $11844$  \\[2ex]
$(bb)(\bar{c}\bar{q})$      & 5     & [3;~4]  & $11851$ & \\
$(bb)^*(\bar{c}\bar{q})^*$  & 6  & [4;~3]   & $11838$ & $11827$ \\[2ex]
$K_1$  & 7  & [5;~5]   & $11861$ & \\
            & 8  & [5;~6]   & $11843$ & \\
            & 9  & [6;~5]   & $11848$ & \\
            & 10  & [6;~6]   & $11797$ & $11719$ \\[2ex]
$K_2$  & 11  & [7;~7]   & $11773$ & \\
             & 12  & [7;~8]   & $11878$ & \\
             & 13  & [8;~7]   & $11726$ & \\
             & 14  & [8;~8]   & $11862$ & $11711$ \\[2ex]
$K_3$  & 15  & [9;~10]   & $11838$ & \\
             & 16  & [10;~9]   & $11837$ & $11818$ \\[2ex]
$K_4$  & 17  & [11;~12]   & $11839$ & \\
             & 18  & [12;~12]   & $12254$ & \\
             & 19  & [11;~11]   & $12237$ & \\
             & 20  & [12;~11]   & $11857$ & $11793$ \\[2ex]
$K_5$  & 21  & [13;~14]   & $11844$ & \\
             & 22  & [14;~13]   & $11833$ & $11818$ \\[2ex]
\multicolumn{4}{c}{Complete coupled-channels:} & $11554$
\end{tabular}
\end{ruledtabular}
\end{table}

\begin{figure}[!t]
\includegraphics[clip, trim={3.0cm 1.7cm 3.0cm 0.8cm}, width=0.45\textwidth]{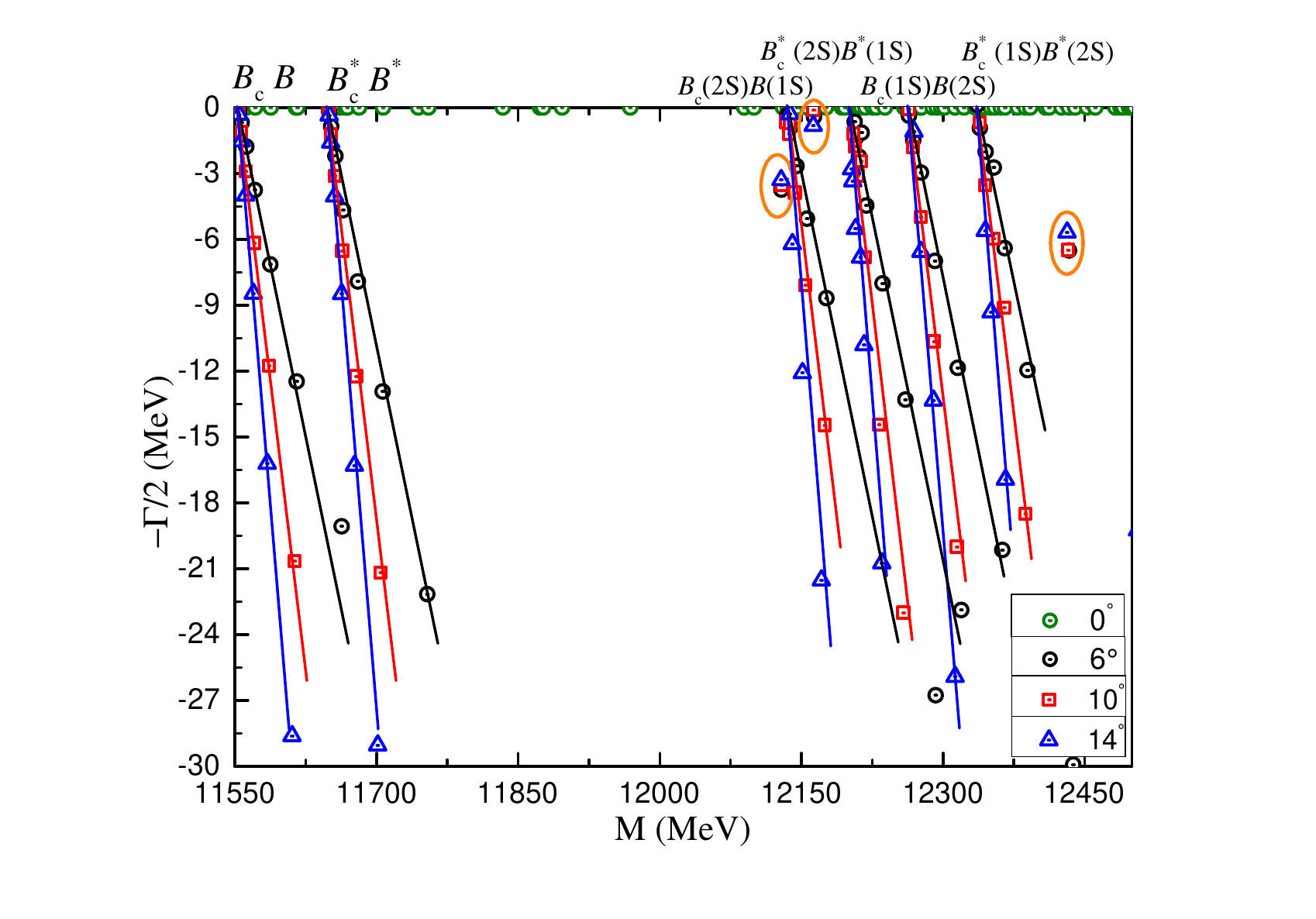}
\caption{\label{PP7} The complete coupled-channels calculation of $\bar{c}b\bar{q}b$ tetraquark system with $I(J^P)=\frac{1}{2}(0^+)$ quantum numbers.}
\end{figure}

\begin{table}[!t]
\caption{\label{GresultR7} Compositeness of exotic resonances obtained in a complete coupled-channel calculation in the $\frac{1}{2}(0^+)$ state of $\bar{c}b\bar{q}b$ tetraquark. Results are similarly organized as those in Table~\ref{GresultR1}.}
\begin{ruledtabular}
\begin{tabular}{rccc}
Resonance       & \multicolumn{3}{c}{Structure} \\[2ex]
$12129-i7.5$   & \multicolumn{3}{c}{$\mu=0$} \\
  & \multicolumn{3}{c}{$r_{b \bar{c}}:0.85$;\,\,\,\,\,$r_{\bar{c}\bar{q}}:1.10$;\,\,\,\,\,$r_{b\bar{q}}:0.87$;\,\,\,\,\,$r_{bb}:0.81$} \\
$Set$ I: & \multicolumn{3}{c}{$S$: 6.8\%;\, $H$: 17.6\%;\, $Di$: 2.9\%;\, $K$: 72.7\%}\\
$Set$ II: & \multicolumn{3}{c}{$S$: 1.4\%;\, $H$: 10.3\%;\, $Di$: 4.5\%;\, $K$: 83.8\%}\\[2ex]
%%%%%%%%%%
$12163-i0.8$   & \multicolumn{3}{c}{$\mu=0$} \\
  & \multicolumn{3}{c}{$r_{b \bar{c}}:0.67$;\,\,\,\,\,$r_{\bar{c}\bar{q}}:0.92$;\,\,\,\,\,$r_{b\bar{q}}:0.81$;\,\,\,\,\,$r_{bb}:0.69$} \\
$Set$ I: & \multicolumn{3}{c}{$S$: 9.8\%;\, $H$: 9.5\%;\, $Di$: 4.3\%;\, $K$: 76.4\%}\\
$Set$ II: & \multicolumn{3}{c}{$S$: 8.1\%;\, $H$: 11.0\%;\, $Di$: 9.1\%;\, $K$: 71.8\%}\\[2ex]
%%%%%%%%%%
$12433-i13.0$   & \multicolumn{3}{c}{$\mu=0$} \\
  & \multicolumn{3}{c}{$r_{b \bar{c}}:1.03$;\,\,\,\,\,$r_{\bar{c}\bar{q}}:0.94$;\,\,\,\,\,$r_{b\bar{q}}:0.91$;\,\,\,\,\,$r_{bb}:0.39$} \\
$Set$ I: & \multicolumn{3}{c}{$S$: 4.5\%;\, $H$: 3.3\%;\, $Di$: 1.9\%;\, $K$: 90.3\%}\\
$Set$ II: & \multicolumn{3}{c}{$S$: 5.2\%;\, $H$: 3.7\%;\, $Di$: 9.7\%;\, $K$: 81.4\%}
\end{tabular}
\end{ruledtabular}
\end{table}

{\bf The $\bm{I(J^P)=\frac{1}{2}(0^+)}$ sector:} single-channel, partially coupled-channel, and fully coupled-channel analyses are performed, with the results summarized in Table~\ref{GresultCC7}. Specifically, the lowest masses of the color-singlet $B_c B$ and $B_c^* B^*$ channels are 11.55 GeV and 11.65 GeV, respectively. Their corresponding hidden-color channels lie at 11.88 GeV and 11.91 GeV. The masses of the two diquark-antidiquark configurations are approximately 11.84 GeV. The 16 K-type channels span a mass range from 11.72 GeV to 12.25 GeV. The theoretical results indicate that the system is unbound in the lowest $B_c B$ channel, as the masses obtained across all coupling schemes remain above threshold, 11.55 GeV.

Nevertheless, a complex-range analysis incorporating all 22 channels reveals three narrow resonances. Figure~\ref{PP7} displays the distribution of the complex eigenenergies obtained. Six scattering states, corresponding to the ground states of $B_c B$ and $B_c^* B^*$, as well as the first radial excitations $B_c(2S) B(1S)$, $B_c^*(2S) B^*(1S)$, $B_c(1S) B(2S)$, and $B_c^*(1S) B^*(2S)$, are clearly identified in the complex energy plane. Additionally, three resonance poles are distinctly circled, with complex energies $M-i\Gamma = 12129 - i\,7.5$ MeV, $12163 - i\,0.8$ MeV, and $12433 - i\,13.0$ MeV, respectively.

As shown in Table~\ref{GresultR7}, all resonances possess zero magnetic moment and correspond to compact $\bar{c}b\bar{q}b$ tetraquark configurations with sizes smaller than 1.1 fm. A common characteristic of these exotic states is that over 71\% of their components arise from K-type channels, and the effects of off-diagonal elements are very small. The $B_c B$ channel serves as the golden decay one for the first and third resonances, whereas the resonance at 12.16 GeV can be identified experimentally in both the $B_c B$ and $B_c^* B^*$ channels.

%%%%%%%%%%%%%%%%%%%%%%%%%%%%%%%%%%%%%%%

\begin{table}[!t]
\caption{\label{GresultCC8} Lowest-lying $\bar{c}b\bar{q}b$ tetraquark states with $I(J^P)=\frac{1}{2}(1^+)$ calculated within the real range formulation of the constituent quark model. Results are similarly organized as those in Table~\ref{GresultCC1} (unit: MeV).}
\begin{ruledtabular}
\begin{tabular}{lcccc}
~~Channel   & Index & $\chi_J^{\sigma_i}$;~$\chi_j^c$ & $M$ & Mixed~~ \\
        &   &$[i; ~j]$ &  \\[2ex]
$(B_c B^*)^1 (11600)$   & 1  & [1;~1]  & $11595$ & \\
$(B^*_c B)^1$  & 2  & [2;~1]   & $11609$ &  \\
$(B^*_c B^*)^1$  & 3  & [3;~1]   & $11649$ & $11595$  \\[2ex]
$(B_c B^*)^8$       & 4  & [1;~2]   & $11867$ &  \\
$(B^*_c B)^8$      & 5  & [2;~2]  & $11870$ & \\
$(B^*_c B^*)^8$  & 6  & [3;~2]   & $11869$ & $11830$ \\[2ex]
$(bb)^*(\bar{c}\bar{q})^*$  & 7  & [6;~3]  & $11842$ & \\
$(bb)^*(\bar{c}\bar{q})$    & 8  & [5;~4]   & $11837$ &  \\
$(bb)(\bar{c}\bar{q})^*$  & 9  & [4;~3]   & $11806$ & $11803$  \\[2ex]
$K_1$      & 10  & [7;~5]   & $11874$ &  \\
      & 11 & [8;~5]  & $11798$ & \\
      & 12  & [9;~5]   & $11833$ & \\
      & 13   & [7;~6]  & $11828$ & \\
      & 14   & [8;~6]  & $11835$ & \\
      & 15   & [9;~6]  & $11803$ & $11723$ \\[2ex]
$K_2$      & 16  & [10;~7]   & $11772$ & \\
                 & 17  & [11;~7]   & $11755$ &  \\
                 & 18  & [12;~7]   & $11737$ & \\
                 & 19  & [10;~8]   & $11819$ & \\
                 & 20  & [11;~8]   & $11886$ & \\
                 & 21  & [12;~8]   & $11848$ & $11715$ \\[2ex]
$K_3$      & 22  & [13;~10]   & $11824$ & \\
                 & 23  & [14;~10]   & $11837$ & \\
                 & 24  & [15;~9]    & $11820$ & $11807$ \\[2ex]
$K_4$      & 25  & [16;~11]   & $11883$ & \\
                 & 26  & [17;~11]   & $11917$ & \\
                 & 27  & [18;~11]   & $12266$ & \\
                 & 28  & [16;~12]   & $11845$ & \\
                 & 29  & [17;~12]   & $11844$ & \\
                 & 30  & [18;~12]   & $11807$ & $11779$ \\[2ex]
$K_5$      & 31  & [19;~14]   & $11836$ & \\
                 & 32  & [20;~14]   & $11840$ & \\
                 & 33  & [21;~13]   & $11817$ & $11808$ \\[2ex]  
\multicolumn{4}{c}{Complete coupled-channels:} & $11595$
\end{tabular}
\end{ruledtabular}
\end{table}

\begin{figure}[!t]
\includegraphics[clip, trim={3.0cm 1.7cm 3.0cm 0.7cm}, width=0.45\textwidth]{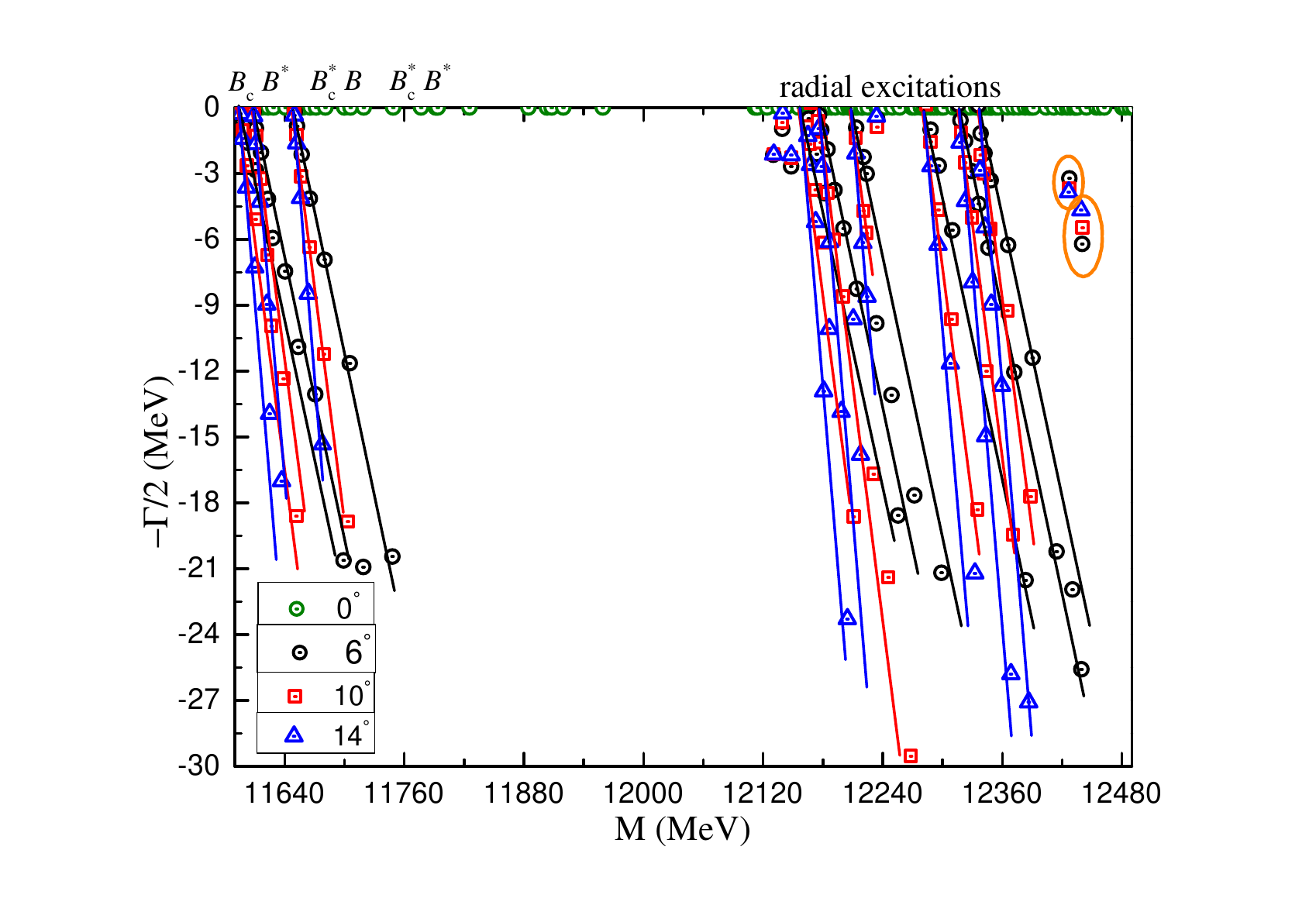} \\[1ex]
\includegraphics[clip, trim={3.0cm 1.7cm 3.0cm 0.7cm}, width=0.45\textwidth]{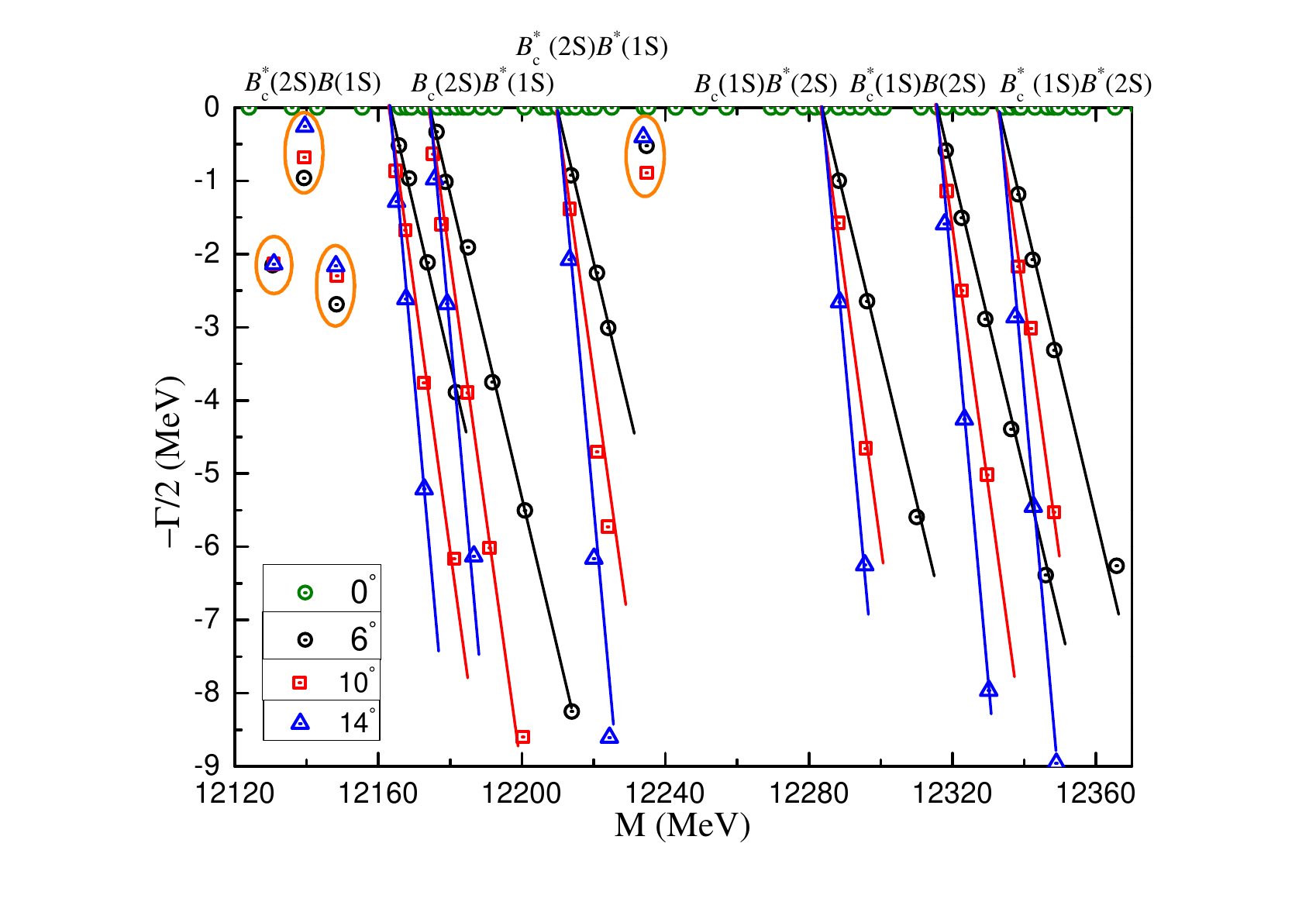}
\caption{\label{PP8} The complete coupled-channels calculation of $\bar{c}b\bar{q}b$ tetraquark system with $I(J^P)=\frac{1}{2}(1^+)$ quantum numbers. Particularly, the bottom panel is enlarged parts of dense energy region from $12.12\,\text{GeV}$ to $12.38\,\text{GeV}$.}
\end{figure}

\begin{table}[!t]
\caption{\label{GresultR8} Compositeness of exotic resonances obtained in a complete coupled-channel calculation in the $\frac{1}{2}(1^+)$ state of $\bar{c}b\bar{q}b$ tetraquark. Results are similarly organized as those in Table~\ref{GresultR1}. Particularly, the magnetic moment of $\bar{c}b\bar{u}b$ and $\bar{c}b\bar{d}b$ tetraquark is $\mu_1$ and $\mu_2$, respectively.}
\begin{ruledtabular}
\begin{tabular}{rccc}
Resonance       & \multicolumn{3}{c}{Structure} \\[2ex]
$12130-i4.3$   & \multicolumn{3}{c}{$\mu_1=-0.56$,\,\,\,$\mu_2=0.17$} \\
  & \multicolumn{3}{c}{$r_{b \bar{c}}:0.74$;\,\,\,\,\,$r_{\bar{c}\bar{q}}:0.89$;\,\,\,\,\,$r_{b\bar{q}}:0.84$;\,\,\,\,\,$r_{bb}:0.51$} \\
$Set$ I: & \multicolumn{3}{c}{$S$: 18.7\%;\, $H$: 15.8\%;\, $Di$: 9.3\%;\, $K$: 56.2\%}\\
$Set$ II: & \multicolumn{3}{c}{$S$: 17.7\%;\, $H$: 16.6\%;\, $Di$: 9.7\%;\, $K$: 56.0\%}\\[2ex]
%%%%%%%%%%
$12139-i1.9$   & \multicolumn{3}{c}{$\mu_1=-1.32$,\,\,\,$\mu_2=0.34$} \\
  & \multicolumn{3}{c}{$r_{b \bar{c}}:0.73$;\,\,\,\,\,$r_{\bar{c}\bar{q}}:1.14$;\,\,\,\,\,$r_{b\bar{q}}:1.03$;\,\,\,\,\,$r_{bb}:0.82$} \\
$Set$ I: & \multicolumn{3}{c}{$S$: 42.3\%;\, $H$: 4.6\%;\, $Di$: 1.3\%;\, $K$: 51.8\%}\\
$Set$ II: & \multicolumn{3}{c}{$S$: 13.0\%;\, $H$: 7.6\%;\, $Di$: 2.1\%;\, $K$: 77.3\%}\\[2ex]
%%%%%%%%%%
$12148-i5.4$   & \multicolumn{3}{c}{$\mu_1=-1.61$,\,\,\,$\mu_2=0.57$} \\
  & \multicolumn{3}{c}{$r_{b \bar{c}}:0.78$;\,\,\,\,\,$r_{\bar{c}\bar{q}}:1.01$;\,\,\,\,\,$r_{b\bar{q}}:0.88$;\,\,\,\,\,$r_{bb}:0.66$} \\
$Set$ I: & \multicolumn{3}{c}{$S$: 28.0\%;\, $H$: 0.8\%;\, $Di$: 4.8\%;\, $K$: 66.4\%}\\
$Set$ II: & \multicolumn{3}{c}{$S$: 6.9\%;\, $H$: 3.9\%;\, $Di$: 2.9\%;\, $K$: 86.3\%}\\[2ex]
%%%%%%%%%%
$12234-i1.1$   & \multicolumn{3}{c}{$\mu_1=-1.80$,\,\,\,$\mu_2=0.80$} \\
  & \multicolumn{3}{c}{$r_{b \bar{c}}:0.63$;\,\,\,\,\,$r_{\bar{c}\bar{q}}:0.92$;\,\,\,\,\,$r_{b\bar{q}}:0.83$;\,\,\,\,\,$r_{bb}:0.67$} \\
$Set$ I: & \multicolumn{3}{c}{$S$: 24.9\%;\, $H$: 1.2\%;\, $Di$: 1.3\%;\, $K$: 72.6\%}\\
$Set$ II: & \multicolumn{3}{c}{$S$: 26.9\%;\, $H$: 1.9\%;\, $Di$: 2.0\%;\, $K$: 69.2\%}\\[2ex]
%%%%%%%%%%
$12427-i6.5$   & \multicolumn{3}{c}{$\mu_1=-1.03$,\,\,\,$\mu_2=0.44$} \\
  & \multicolumn{3}{c}{$r_{b \bar{c}}:1.01$;\,\,\,\,\,$r_{\bar{c}\bar{q}}:0.87$;\,\,\,\,\,$r_{b\bar{q}}:0.95$;\,\,\,\,\,$r_{bb}:0.38$} \\
$Set$ I: & \multicolumn{3}{c}{$S$: 7.7\%;\, $H$: 7.0\%;\, $Di$: 5.1\%;\, $K$: 80.2\%}\\
$Set$ II: & \multicolumn{3}{c}{$S$: 9.9\%;\, $H$: 10.0\%;\, $Di$: 8.1\%;\, $K$: 72.0\%}\\[2ex]
%%%%%%%%%%
$12440-i12.4$   & \multicolumn{3}{c}{$\mu_1=-1.08$,\,\,\,$\mu_2=0.12$} \\
  & \multicolumn{3}{c}{$r_{b \bar{c}}:1.03$;\,\,\,\,\,$r_{\bar{c}\bar{q}}:0.91$;\,\,\,\,\,$r_{b\bar{q}}:0.92$;\,\,\,\,\,$r_{bb}:0.39$} \\
$Set$ I: & \multicolumn{3}{c}{$S$: 6.2\%;\, $H$: 2.1\%;\, $Di$: 0.8\%;\, $K$: 90.9\%}\\
$Set$ II: & \multicolumn{3}{c}{$S$: 16.4\%;\, $H$: 8.3\%;\, $Di$: 11.5\%;\, $K$: 63.8\%}
\end{tabular}
\end{ruledtabular}
\end{table}

{\bf The $\bm{I(J^P)=\frac{1}{2}(1^+)}$ sector:} 33 channels are considered in this case, and the corresponding calculated results are presented in Table~\ref{GresultCC8}. Specifically, three meson-meson color-singlet channels: $B_c B^*$, $B_c^* B$, and $B_c^* B^*$, are located at 11.59, 11.61, and 11.65 GeV, respectively. Their hidden-color counterparts are found at 11.86, 11.87, and 11.87 GeV. The masses of the three diquark-antidiquark configurations are 11.84, 11.83, and 11.81 GeV, respectively. The 24 K-type channels span the $(11.74-12.26)$ GeV mass range. No bound state is found; the lowest state remains the $B_c B^*$ scattering channel, both in single-channel and coupled-channel analyses. Additionally, coupled-channel results for hidden-color, diquark-antidiquark, and the $K_3$- and $K_5$-type structures yield masses around 11.81 GeV, while the $K_1$-, $K_2$-, and $K_4$-type configurations are located near 11.72 GeV.

Figure~\ref{PP8} shows the results of a fully coupled-channel calculation performed using CSM. In the top panel, the scattering nature of $B_c B^*$, $B_c^* B$, and $B_c^* B^*$ channels, including both ground and radially excited states, is depicted. Notably, two nearby resonance poles are identified, with complex energies $M-i\Gamma = 12427 - i\,6.5$ MeV and $12440 - i\,12.4$ MeV. Furthermore, we show in the bottom panel an enlarged view of the energy region between 12.12 and 12.38 GeV, identifying four additional resonances. The corresponding resonance poles are located at $12130 - i\,4.3$ MeV, $12139 - i\,1.9$ MeV, $12148 - i\,5.4$ MeV, and $12234 - i\,1.1$ MeV, respectively.

Some properties of these six resonances are summarized in Table~\ref{GresultR8}. All of them are identified as compact $\bar{c}b\bar{q}b$ tetraquark configurations, with spatial extensions of less than 1.0 fm. While off-diagonal effects contribute moderately (around (20-30)\%) the overall structure of all resonances remains consistent with K-type dominance, with contributions exceeding 50\%. The magnetic moments of the $\bar{c}b\bar{u}b$ resonances are calculated to be $-0.56\,\mu_N$, $-1.32\,\mu_N$, $-1.61\,\mu_N$, $-1.80\,\mu_N$, $-1.03\,\mu_N$, and $-1.08\,\mu_N$, respectively. For the corresponding $\bar{c}b\bar{d}b$ states, the values are $0.17\,\mu_N$, $0.34\,\mu_N$, $0.57\,\mu_N$, $0.80\,\mu_N$, $0.44\,\mu_N$, and $0.12\,\mu_N$. Theoretically, the $B_c B^*$ channel is the dominant decay mode for the first three resonances, while the remaining three states can potentially be observed through both $B_c B^*$ and $B_c^* B$ final decay channels.

%%%%%%%%%%%%%%%%%%%%%%%%%%%%%%%%%%%%%

\begin{table}[!t]
\caption{\label{GresultCC9} Lowest-lying $\bar{c}b\bar{q}b$ tetraquark states with $I(J^P)=\frac{1}{2}(2^+)$ calculated within the real range formulation of the constituent quark model. Results are similarly organized as those in Table~\ref{GresultCC1} (unit: MeV).}
\begin{ruledtabular}
\begin{tabular}{lcccc}
~~Channel   & Index & $\chi_J^{\sigma_i}$;~$\chi_j^c$ & $M$ & Mixed~~ \\
        &   &$[i; ~j]$ &  \\[2ex]
$(B^*_c B^*)^1$   & 1  & [1;~1]   & $11649$ &  \\[2ex]
$(B^*_c B^*)^8$   & 2  & [1;~3]   & $11860$ &  \\[2ex]
$(bb)^*(\bar{c}\bar{q})^*$  & 3  & [1;~7]   & $11851$ & \\[2ex]
$K_1$  & 4  & [1;~8]   & $11834$ & \\
            & 5  & [1;~10]   & $11848$ & $11778$ \\[2ex]
$K_2$  & 6  & [1;~11]   & $11797$ & \\
             & 7  & [1;~12]   & $11847$ & $11790$ \\[2ex]
$K_3$  & 8  & [1;~14]   & $11853$ & \\[2ex]
$K_4$  & 9  & [1;~8]   & $12264$ & \\
             & 10  & [1;~10]   & $11852$ & $11830$ \\[2ex]
$K_5$  & 11  & [1;~11]   & $11859$ & \\[2ex]
\multicolumn{4}{c}{Complete coupled-channels:} & $11649$
\end{tabular}
\end{ruledtabular}
\end{table}

\begin{table}[!t]
\caption{\label{GresultR9} Compositeness of exotic resonances obtained in a complete coupled-channel calculation in the $\frac{1}{2}(2^+)$ state of $\bar{c}b\bar{q}b$ tetraquark. Results are similarly organized as those in Table~\ref{GresultR1}.}
\begin{ruledtabular}
\begin{tabular}{rccc}
Resonance       & \multicolumn{3}{c}{Structure} \\[2ex]
$12164-i5.4$   & \multicolumn{3}{c}{$\mu_1=-2.48$,\,\,\,$\mu_2=0.52$} \\
  & \multicolumn{3}{c}{$r_{b \bar{c}}:0.76$;\,\,\,\,\,$r_{\bar{c}\bar{q}}:0.89$;\,\,\,\,\,$r_{b\bar{q}}:0.81$;\,\,\,\,\,$r_{bb}:0.43$} \\
$Set$ I: & \multicolumn{3}{c}{$S$: 3.9\%;\, $H$: 7.3\%;\, $Di$: 0.1\%;\, $K$: 88.7\%}\\
$Set$ II: & \multicolumn{3}{c}{$S$: 8.6\%;\, $H$: 15.1\%;\, $Di$: 0.1\%;\, $K$: 76.2\%}\\[2ex]
%%%%%%%%%%
$12455-i11.2$   & \multicolumn{3}{c}{$\mu_1=-2.48$,\,\,\,$\mu_2=0.52$} \\
  & \multicolumn{3}{c}{$r_{b \bar{c}}:1.01$;\,\,\,\,\,$r_{\bar{c}\bar{q}}:0.86$;\,\,\,\,\,$r_{b\bar{q}}:0.93$;\,\,\,\,\,$r_{bb}:0.41$} \\
$Set$ I: & \multicolumn{3}{c}{$S$: 4.7\%;\, $H$: 3.8\%;\, $Di$: 0.1\%;\, $K$: 91.4\%}\\
$Set$ II: & \multicolumn{3}{c}{$S$: 25.5\%;\, $H$: 17.3\%;\, $Di$: 0.1\%;\, $K$: 57.1\%}
\end{tabular}
\end{ruledtabular}
\end{table}

\begin{figure}[!t]
\includegraphics[clip, trim={3.0cm 1.7cm 3.0cm 0.8cm}, width=0.45\textwidth]{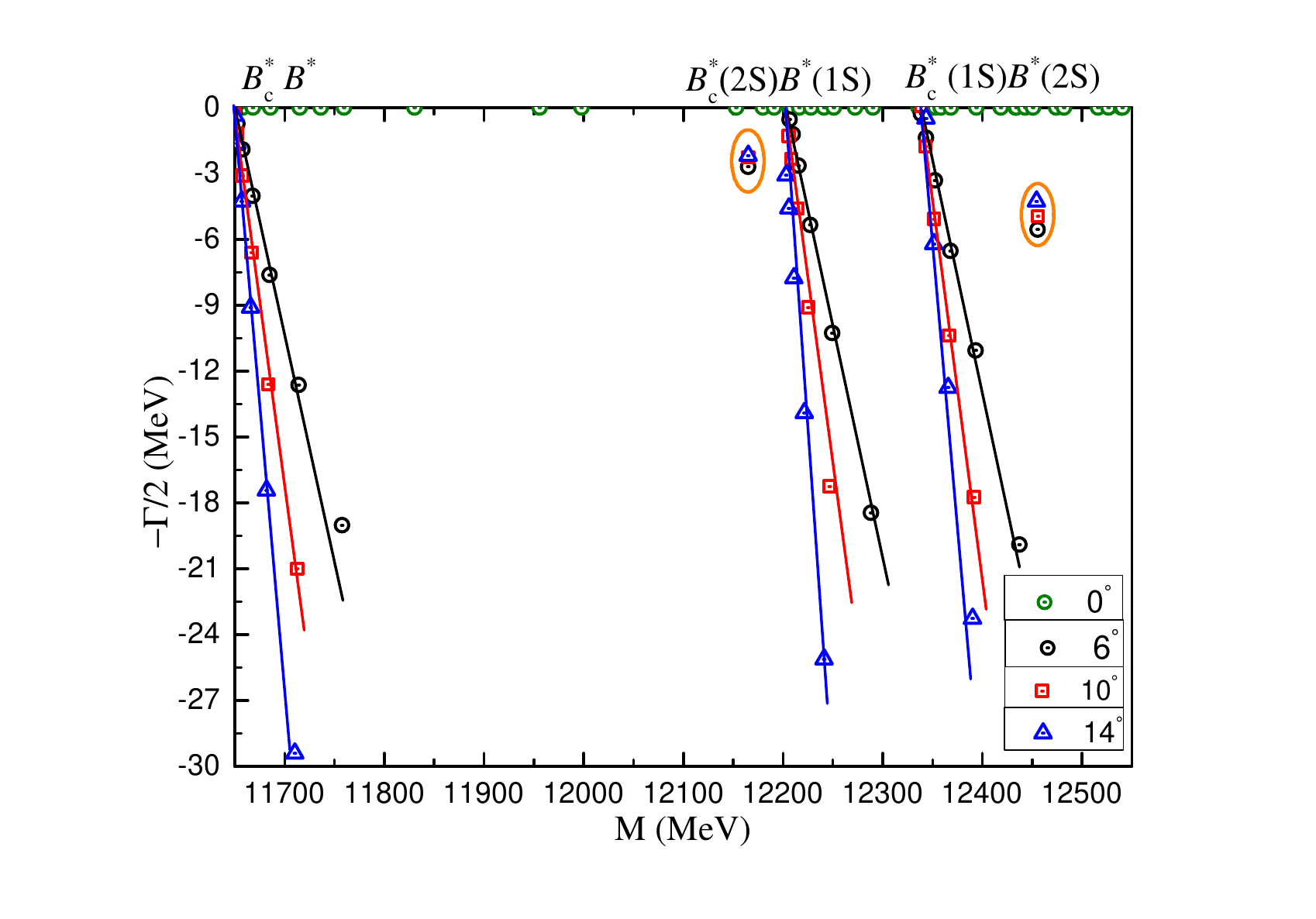}
\caption{\label{PP9} The complete coupled-channels calculation of $\bar{c}b\bar{q}b$ tetraquark system with $I(J^P)=\frac{1}{2}(2^+)$ quantum numbers.}
\end{figure}

{\bf The $\bm{I(J^P)=\frac{1}{2}(2^+)}$ sector:} Only one meson-meson configuration, $B_c^* B^*$, is considered, both in the color-singlet and hidden-color channels, along with a single $(bb)^*(\bar{c}\bar{q})^*$ configuration and eight K-type arrangements. The real-range calculation results are summarized in Table~\ref{GresultCC9}, and the conclusions are consistent with those found in other cases. First, the lowest state, corresponding to the color-singlet $B_c^* B^*$ channel, is of scattering nature, with a computed mass of 11.65 GeV. This result persists in the full coupled-channel analysis. The remaining exotic configurations yield masses around 11.85 GeV, with the exception of two K-type channels: $K_2$, with a mass of 11.80 GeV, and $K_4$, at 12.26 GeV. In the coupled-channel computations involving the $K_1$, $K_2$, and $K_4$ structures, the lowest mass in each arrangement is consistently around 11.80 GeV.

Further analysis using CSM is presented in Fig.~\ref{PP9}, which shows the results of a fully coupled-channel calculation. In the energy range from 11.65 to 12.55 GeV, the ground state of $B_c^* B^*$, along with its first radial excitations: $B_c^*(2S) B^*(1S)$ and $B_c^*(1S) B^*(2S)$, are clearly identified. Additionally, two narrow resonances are observed, with complex energies located at $M-i\Gamma = 12164 - i\,5.4$ MeV and $12455 - i\,11.2$ MeV, respectively.

Some properties of these exotic hadrons are detailed in Table~\ref{GresultR9}. Both resonances exhibit a compact tetraquark structure, with spatial sizes below 1.0 fm. The magnetic moments of the two $\bar{c}b\bar{u}b$ resonances are notably large, each having $\mu = -2.48\,\mu_N$, while the corresponding $\bar{c}b\bar{d}b$ states have a value of $\mu = 0.52\,\mu_N$. K-type components dominate the wave functions of these resonances, contributing more than 57\%. Notably, the higher resonance at 12.46 GeV exhibits strong mixing among color-singlet, hidden-color, and K-type channels.

%%%%%%%%%%%%%%%%%%%%%%%%%%%%%%%%%%%%

\subsection{The $\mathbf{\bar{c}b\bar{s}b}$ tetraquarks}

\begin{figure}[!t]
\includegraphics[clip, trim={3.0cm 1.7cm 3.0cm 0.8cm}, width=0.45\textwidth]{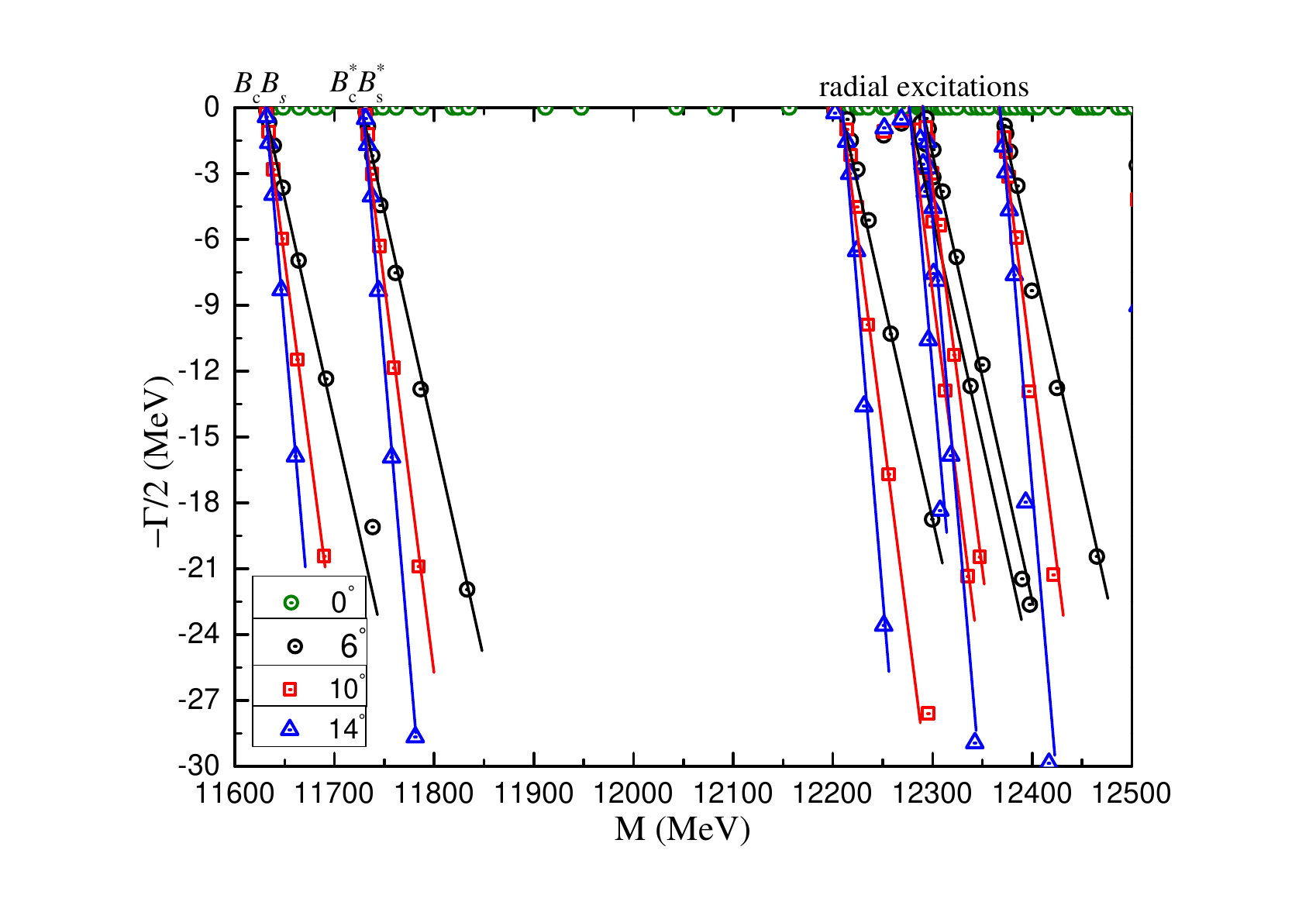} \\[1ex]
\includegraphics[clip, trim={3.0cm 1.7cm 3.0cm 0.8cm}, width=0.45\textwidth]{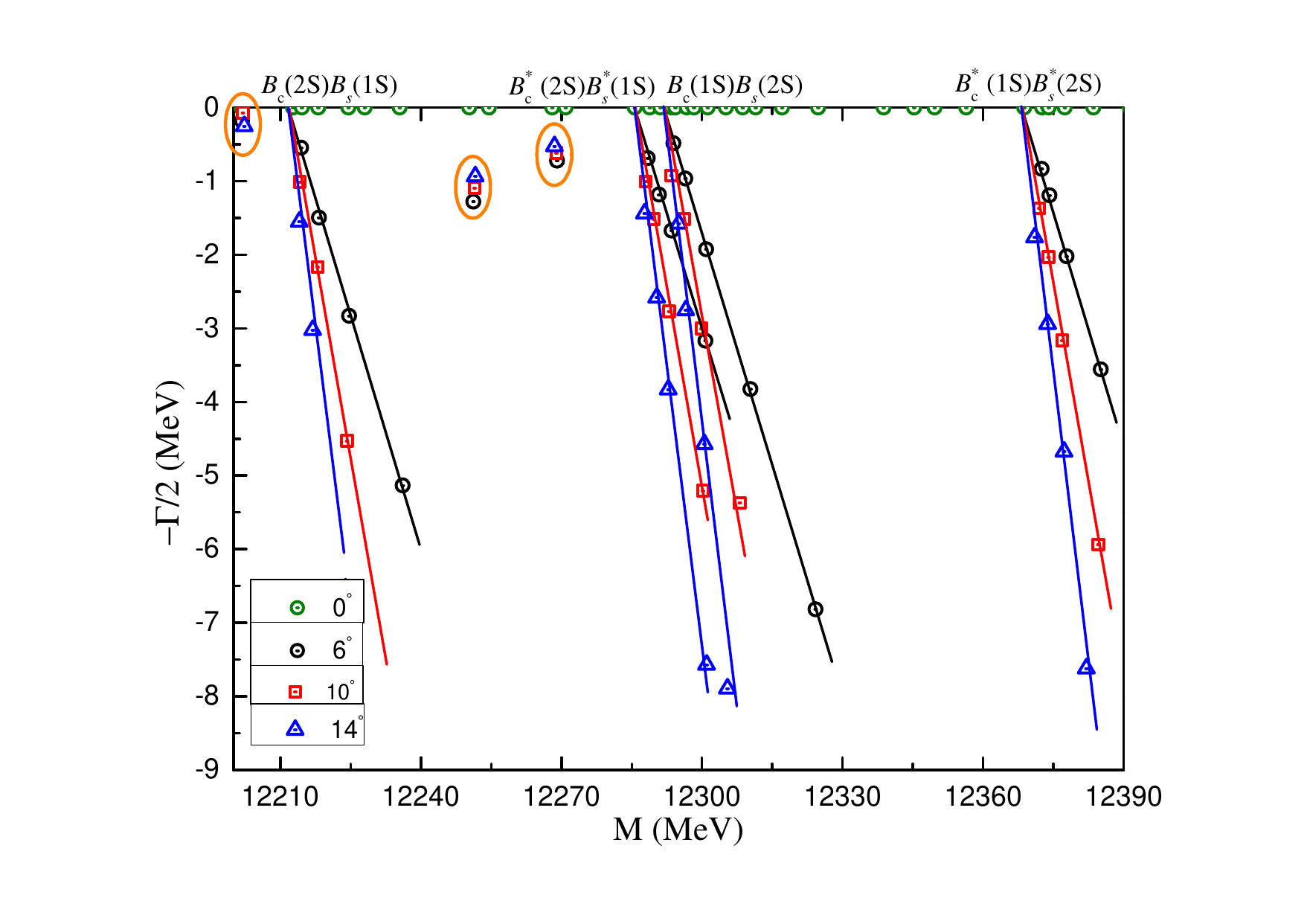}
\caption{\label{PP10} The complete coupled-channels calculation of $\bar{c}b\bar{s}b$ tetraquark system with $I(J^P)=0(0^+)$ quantum numbers. Particularly, the bottom panel is enlarged parts of dense energy region from $12.19\,\text{GeV}$ to $12.39\,\text{GeV}$.}
\end{figure}

\begin{table}[!t]
\caption{\label{GresultCC10} Lowest-lying $\bar{c}b\bar{s}b$ tetraquark states with $I(J^P)=0(0^+)$ calculated within the real range formulation of the constituent quark model. Results are similarly organized as those in Table~\ref{GresultCC1} (unit: MeV).}
\begin{ruledtabular}
\begin{tabular}{lcccc}
~~Channel   & Index & $\chi_J^{\sigma_i}$;~$\chi_j^c$ & $M$ & Mixed~~ \\
        &   &$[i; ~j]$ &  \\[2ex]
$(B_c B_s)^1 (11642)$          & 1  & [1;~1]  & $11631$ & \\
$(B^*_c B^*_s)^1$  & 2  & [2;~1]   & $11731$ & $11631$  \\[2ex]
$(B_c B_s)^8$          & 3  & [1;~2]  & $11970$ & \\
$(B^*_c B^*_s)^8$       & 4  & [2;~2]   & $11995$ & $11937$  \\[2ex]
$(bb)(\bar{c}\bar{s})$      & 5     & [3;~4]  & $11932$ & \\
$(bb)^*(\bar{c}\bar{s})^*$  & 6  & [4;~3]   & $11931$ & $11915$ \\[2ex]
$K_1$  & 7  & [5;~5]   & $11944$ & \\
            & 8  & [5;~6]   & $11945$ & \\
            & 9  & [6;~5]   & $11940$ & \\
            & 10  & [6;~6]   & $11897$ & $11812$ \\[2ex]
$K_2$  & 11  & [7;~7]   & $11872$ & \\
             & 12  & [7;~8]   & $11965$ & \\
             & 13  & [8;~7]   & $11828$ & \\
             & 14  & [8;~8]   & $11957$ & $11813$ \\[2ex]
$K_3$  & 15  & [9;~10]   & $11935$ & \\
             & 16  & [10;~9]   & $11926$ & $11911$ \\[2ex]
$K_4$  & 17  & [11;~12]   & $11931$ & \\
             & 18  & [12;~12]   & $12356$ & \\
             & 19  & [11;~11]   & $12341$ & \\
             & 20  & [12;~11]   & $11938$ & $11898$ \\[2ex]
$K_5$  & 21  & [13;~14]   & $11940$ & \\
             & 22  & [14;~13]   & $11922$ & $11910$ \\[2ex]
\multicolumn{4}{c}{Complete coupled-channels:} & $11631$
\end{tabular}
\end{ruledtabular}
\end{table}

\begin{table}[!t]
\caption{\label{GresultR10} Compositeness of exotic resonances obtained in a complete coupled-channel calculation in the $0(0^+)$ state of $\bar{c}b\bar{s}b$ tetraquark. Results are similarly organized as those in Table~\ref{GresultR1}.}
\begin{ruledtabular}
\begin{tabular}{rccc}
Resonance       & \multicolumn{3}{c}{Structure} \\[2ex]
$12201-i0.4$   & \multicolumn{3}{c}{$\mu=0$} \\
  & \multicolumn{3}{c}{$r_{b \bar{c}}:0.66$;\,\,\,\,\,$r_{\bar{c}\bar{s}}:1.01$;\,\,\,\,\,$r_{b\bar{s}}:0.92$;\,\,\,\,\,$r_{bb}:0.84$} \\
$Set$ I: & \multicolumn{3}{c}{$S$: 43.0\%;\, $H$: 2.4\%;\, $Di$: 0.2\%;\, $K$: 54.4\%}\\
$Set$ II: & \multicolumn{3}{c}{$S$: 22.7\%;\, $H$: 8.8\%;\, $Di$: 0.8\%;\, $K$: 67.7\%}\\[2ex]
%%%%%%%%%%
$12251-i2.6$   & \multicolumn{3}{c}{$\mu=0$} \\
  & \multicolumn{3}{c}{$r_{b \bar{c}}:0.72$;\,\,\,\,\,$r_{\bar{c}\bar{s}}:1.05$;\,\,\,\,\,$r_{b\bar{s}}:0.89$;\,\,\,\,\,$r_{bb}:0.90$} \\
$Set$ I: & \multicolumn{3}{c}{$S$: 33.8\%;\, $H$: 3.8\%;\, $Di$: 4.0\%;\, $K$: 58.4\%}\\
$Set$ II: & \multicolumn{3}{c}{$S$: 24.2\%;\, $H$: 5.3\%;\, $Di$: 5.0\%;\, $K$: 65.5\%}\\[2ex]
%%%%%%%%%%
$12269-i1.4$   & \multicolumn{3}{c}{$\mu=0$} \\
  & \multicolumn{3}{c}{$r_{b \bar{c}}:0.73$;\,\,\,\,\,$r_{\bar{c}\bar{s}}:0.97$;\,\,\,\,\,$r_{b\bar{s}}:0.81$;\,\,\,\,\,$r_{bb}:0.76$} \\
$Set$ I: & \multicolumn{3}{c}{$S$: 16.3\%;\, $H$: 4.1\%;\, $Di$: 0.7\%;\, $K$: 78.9\%}\\
$Set$ II: & \multicolumn{3}{c}{$S$: 14.8\%;\, $H$: 4.9\%;\, $Di$: 0.9\%;\, $K$: 79.4\%}\\
\end{tabular}
\end{ruledtabular}
\end{table}

{\bf The $\bm{I(J^P)=0(0^+)}$ sector:} 22 channels, listed in Table~\ref{GresultCC10}, are initially investigated through real-range calculations. The lowest channel, $B_c B_s$, is clearly a scattering state, with a theoretical threshold at 11.63 GeV. Another meson-meson configuration, $B_c^* B_s^*$, lies at 11.73 GeV. Their corresponding hidden-color counterparts are found at 11.97 and 11.99 GeV, respectively. The two diquark-antidiquark configurations, $(bb)(\bar{c}\bar{s})$ and $(bb)^*(\bar{c}\bar{s})^*$, have masses around 11.93 GeV. Additionally, the 16 K-type configurations span an energy range from 11.83 to 12.35 GeV. Upon performing coupled-channel calculations within each of the seven exotic color structures, the lowest mass obtained is approximately 11.91 GeV for the hidden-color, diquark-antidiquark, $K_3$, $K_4$, and $K_5$ configurations, and around 11.81 GeV for the $K_1$ and $K_2$ arrangements. In the fully coupled-channel analysis, the results remain consistent with those for the $B_c B_s$ system, indicating no bound state formation.

Figure~\ref{PP10} presents the outcomes of a complex-range study for the fully coupled-channel case. In the top panel ($11.6-12.5$ GeV energy region), the ground states of $B_c B_s$ and $B_c^* B_s^*$, along with their radial excitations, are clearly illustrated. The bottom panel zooms into the energy region between 12.19 and 12.39 GeV, revealing four more scattering states: $B_c(2S)B_s(1S)$, $B_c^*(2S)B_s^*(1S)$, $B_c(1S)B_s(2S)$, and $B_c^*(1S)B_s^*(2S)$. Within this region, three narrow resonances are identified at complex energies $M-i\Gamma = 12201 - i\,0.4$ MeV, $12251 - i\,2.6$ MeV, and $12269 - i\,1.4$ MeV.

Table~\ref{GresultR10} summarizes the properties of these resonances, including size, magnetic moment, and dominant components. All three states exhibit a compact $\bar{c}b\bar{s}b$ tetraquark configuration, with spatial sizes below 1.0 fm. Each has a magnetic moment $\mu = 0$. The first two resonances display strong mixing between the color-singlet and K-type configurations, although the K-type remains dominant with contributions exceeding 54\%. In contrast, the third resonance shows weak coupling effects, and its dominant component ($\sim 79\%$) is of K-type nature. The golden decay channel for the first resonance at 12.20 GeV is expected to be $B_c B_s$, while the remaining two higher-mass resonances are likely to be confirmed through the $B_c^* B_s^*$ decay mode.

%%%%%%%%%%%%%%%%%%%%%%%%%%%%%%%%%%%%%%%%%%%%%%%%%%%%%%%%%%%%%%%%%

\begin{figure}[!t]
\includegraphics[clip, trim={3.0cm 1.7cm 3.0cm 0.8cm}, width=0.45\textwidth]{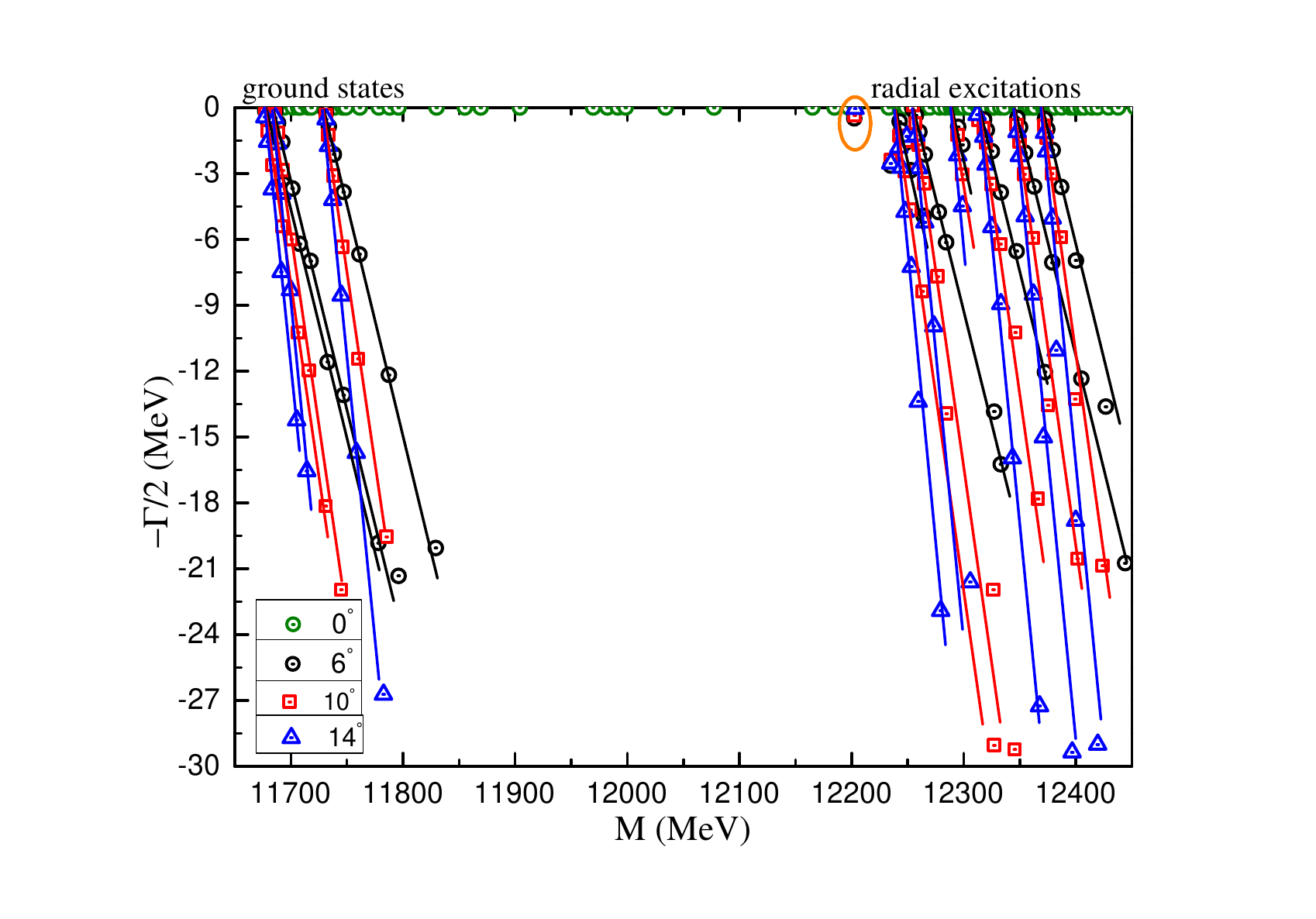} \\[1ex]
\includegraphics[clip, trim={3.0cm 1.7cm 3.0cm 0.8cm}, width=0.45\textwidth]{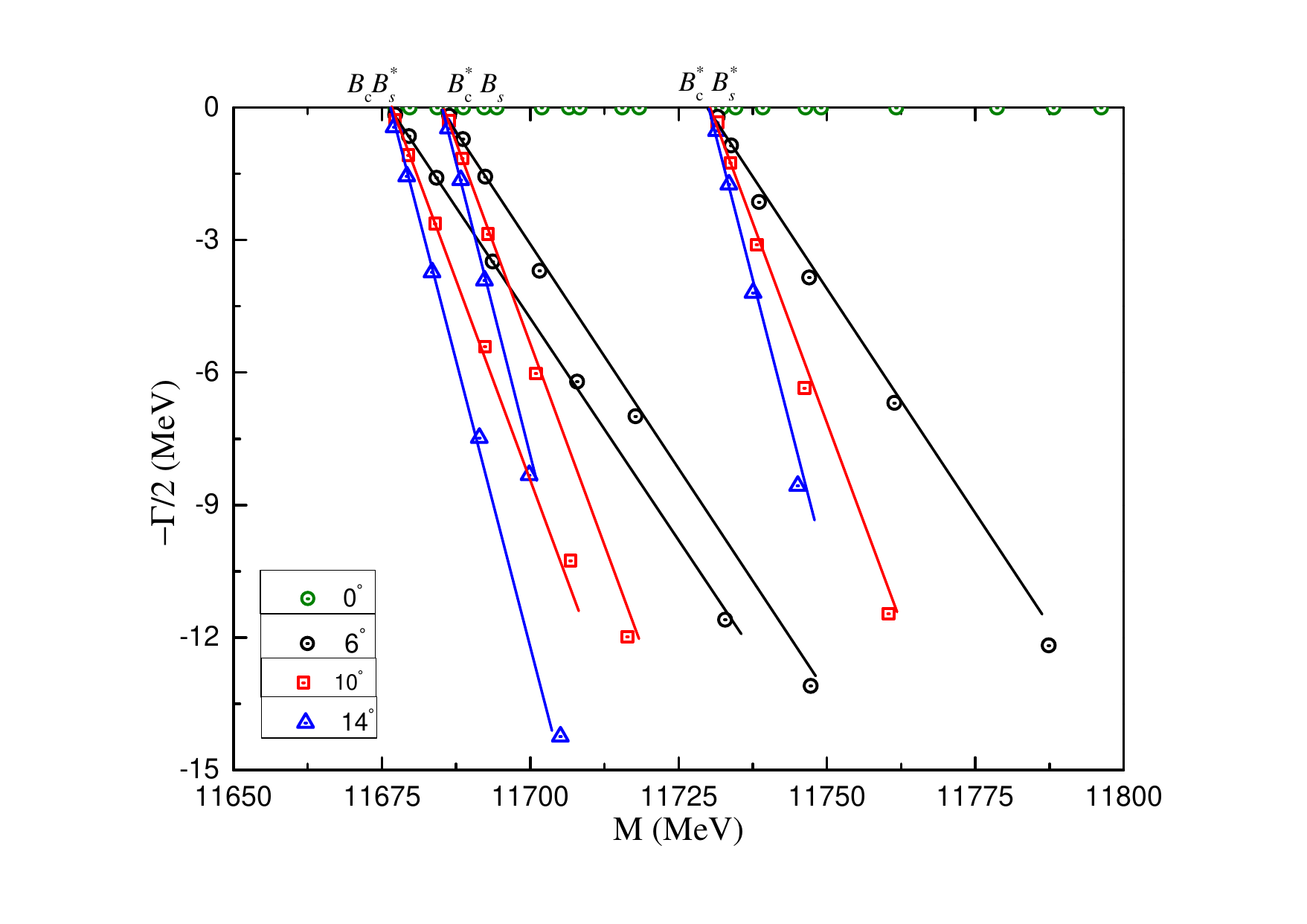} \\[1ex]
\includegraphics[clip, trim={3.0cm 1.7cm 3.0cm 0.8cm}, width=0.45\textwidth]{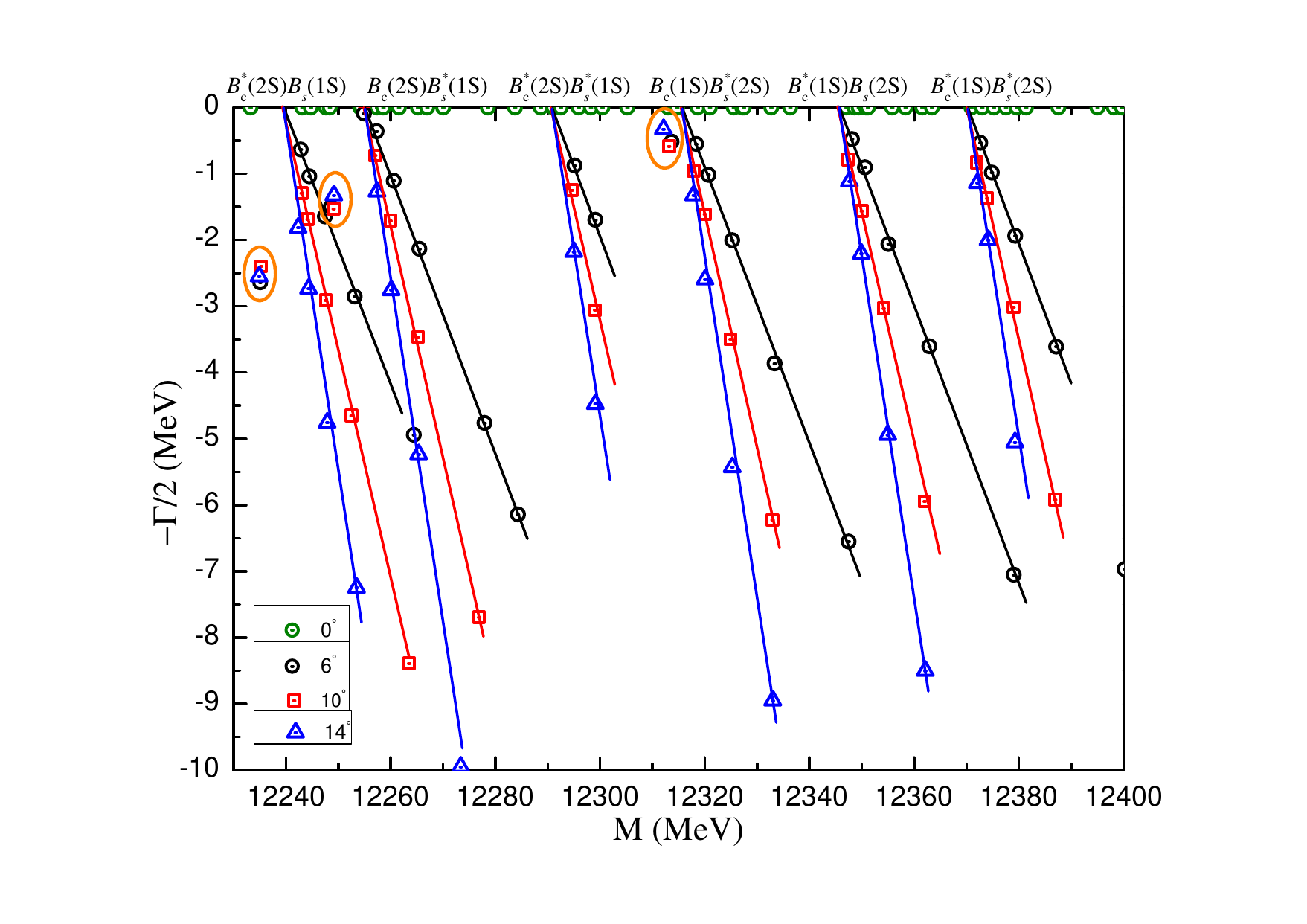}
\caption{\label{PP11} The complete coupled-channels calculation of $\bar{c}b\bar{s}b$ tetraquark system with $I(J^P)=0(1^+)$ quantum numbers. Particularly, the middle panel is enlarged parts of dense energy region from $11.65\,\text{GeV}$ to $11.80\,\text{GeV}$, and the bottom one is enlarged parts of dense energy region from $12.23\,\text{GeV}$ to $12.40\,\text{GeV}$.}
\end{figure}

\begin{table}[!t]
\caption{\label{GresultCC11} Lowest-lying $\bar{c}b\bar{s}b$ tetraquark states with $I(J^P)=0(1^+)$ calculated within the real range formulation of the constituent quark model. Results are similarly organized as those in Table~\ref{GresultCC1} (unit: MeV).}
\begin{ruledtabular}
\begin{tabular}{lcccc}
~~Channel   & Index & $\chi_J^{\sigma_i}$;~$\chi_j^c$ & $M$ & Mixed~~ \\
        &   &$[i; ~j]$ &  \\[2ex]
$(B_c B^*_s)^1 (11690)$   & 1  & [1;~1]  & $11676$ & \\
$(B^*_c B_s)^1$  & 2  & [2;~1]   & $11685$ &  \\
$(B^*_c B^*_s)^1$  & 3  & [3;~1]   & $11731$ & $11676$  \\[2ex]
$(B_c B^*_s)^8$       & 4  & [1;~2]   & $11960$ &  \\
$(B^*_c B_s)^8$      & 5  & [2;~2]  & $11961$ & \\
$(B^*_c B^*_s)^8$  & 6  & [3;~2]   & $11962$ & $11928$ \\[2ex]
$(bb)^*(\bar{c}\bar{s})^*$  & 7  & [6;~3]  & $11935$ & \\
$(bb)^*(\bar{c}\bar{s})$    & 8  & [5;~4]   & $11920$ &  \\
$(bb)(\bar{c}\bar{s})^*$  & 9  & [4;~3]   & $11901$ & $11897$  \\[2ex]
$K_1$      & 10  & [7;~5]   & $11959$ &  \\
      & 11 & [8;~5]  & $11896$ & \\
      & 12  & [9;~5]   & $11927$ & \\
      & 13   & [7;~6]  & $11931$ & \\
      & 14   & [8;~6]  & $11935$ & \\
      & 15   & [9;~6]  & $11902$ & $11819$ \\[2ex]
$K_2$      & 16  & [10;~7]   & $11875$ & \\
                 & 17  & [11;~7]   & $11856$ &  \\
                 & 18  & [12;~7]   & $11838$ & \\
                 & 19  & [10;~8]   & $11920$ & \\
                 & 20  & [11;~8]   & $11974$ & \\
                 & 21  & [12;~8]   & $11945$ & $11816$ \\[2ex]
$K_3$      & 22  & [13;~10]   & $11924$ & \\
                 & 23  & [14;~10]   & $11934$ & \\
                 & 24  & [15;~9]    & $11912$ & $11903$ \\[2ex]
$K_4$      & 25  & [16;~11]   & $11967$ & \\
                 & 26  & [17;~11]   & $11999$ & \\
                 & 27  & [18;~11]   & $12366$ & \\
                 & 28  & [16;~12]   & $11938$ & \\
                 & 29  & [17;~12]   & $11937$ & \\
                 & 30  & [18;~12]   & $11902$ & $11884$ \\[2ex]
$K_5$      & 31  & [19;~14]   & $11934$ & \\
                 & 32  & [20;~14]   & $11937$ & \\
                 & 33  & [21;~13]   & $11909$ & $11902$ \\[2ex]  
\multicolumn{4}{c}{Complete coupled-channels:} & $11676$
\end{tabular}
\end{ruledtabular}
\end{table}

\begin{table}[!t]
\caption{\label{GresultR11} Compositeness of exotic resonances obtained in a complete coupled-channel calculation in the $0(1^+)$ state of $\bar{c}b\bar{s}b$ tetraquark. Results are similarly organized as those in Table~\ref{GresultR1}.}
\begin{ruledtabular}
\begin{tabular}{rccc}
Resonance       & \multicolumn{3}{c}{Structure} \\[2ex]
$12202-i1.0$   & \multicolumn{3}{c}{$\mu=0.51$} \\
  & \multicolumn{3}{c}{$r_{b \bar{c}}:0.66$;\,\,\,\,\,$r_{\bar{c}\bar{s}}:1.02$;\,\,\,\,\,$r_{b\bar{s}}:0.93$;\,\,\,\,\,$r_{bb}:0.84$} \\
$Set$ I: & \multicolumn{3}{c}{$S$: 51.0\%;\, $H$: 0.9\%;\, $Di$: 0.4\%;\, $K$: 47.7\%}\\
$Set$ II: & \multicolumn{3}{c}{$S$: 2.9\%;\, $H$: 3.4\%;\, $Di$: 1.4\%;\, $K$: 92.3\%}\\[2ex]
%%%%%%%%%%
$12235-i5.3$   & \multicolumn{3}{c}{$\mu=0.59$} \\
  & \multicolumn{3}{c}{$r_{b \bar{c}}:0.79$;\,\,\,\,\,$r_{\bar{c}\bar{s}}:0.93$;\,\,\,\,\,$r_{b\bar{s}}:0.77$;\,\,\,\,\,$r_{bb}:0.69$} \\
$Set$ I: & \multicolumn{3}{c}{$S$: 20.8\%;\, $H$: 7.5\%;\, $Di$: 4.4\%;\, $K$: 67.3\%}\\
$Set$ II: & \multicolumn{3}{c}{$S$: 25.3\%;\, $H$: 9.2\%;\, $Di$: 2.1\%;\, $K$: 63.4\%}\\[2ex]
%%%%%%%%%%
$12248-i3.0$   & \multicolumn{3}{c}{$\mu=0.30$} \\
  & \multicolumn{3}{c}{$r_{b \bar{c}}:0.70$;\,\,\,\,\,$r_{\bar{c}\bar{s}}:0.96$;\,\,\,\,\,$r_{b\bar{s}}:0.85$;\,\,\,\,\,$r_{bb}:0.79$} \\
$Set$ I: & \multicolumn{3}{c}{$S$: 29.9\%;\, $H$: 6.5\%;\, $Di$: 0.8\%;\, $K$: 62.8\%}\\
$Set$ II: & \multicolumn{3}{c}{$S$: 28.2\%;\, $H$: 7.5\%;\, $Di$: 0.9\%;\, $K$: 63.4\%}\\[2ex]
%%%%%%%%%%
$12313-i1.0$   & \multicolumn{3}{c}{$\mu=0.45$} \\
   & \multicolumn{3}{c}{$r_{b \bar{c}}:0.67$;\,\,\,\,\,$r_{\bar{c}\bar{s}}:0.92$;\,\,\,\,\,$r_{b\bar{s}}:0.78$;\,\,\,\,\,$r_{bb}:0.75$} \\
$Set$ I: & \multicolumn{3}{c}{$S$: 24.2\%;\, $H$: 1.3\%;\, $Di$: 3.0\%;\, $K$: 71.5\%}\\
$Set$ II: & \multicolumn{3}{c}{$S$: 23.6\%;\, $H$: 1.7\%;\, $Di$: 4.5\%;\, $K$: 70.2\%}\\
\end{tabular}
\end{ruledtabular}
\end{table}

{\bf The $\bm{I(J^P)=0(1^+)}$ sector:} A total of 33 channels are considered for this case, with the corresponding results presented in Table~\ref{GresultCC11}. Notably, the masses of the three color-singlet configurations: $B_c B_s^*$, $B_c^* B_s$, and $B_c^* B_s^*$, are 11.67 GeV, 11.68 GeV, and 11.73 GeV, respectively. Their corresponding hidden-color channels are nearly degenerate, each having a mass around 11.96 GeV. The three diquark-antidiquark configurations exhibit masses of 11.93 GeV, 11.92 GeV, and 11.90 GeV, respectively. The remaining 24 K-type configurations span an energy interval from approximately 11.84 to 12.36 GeV. When coupled-channel calculations are performed within each structure, the lowest masses are consistently found around 11.90 GeV for all exotic configurations, except for the $K_1$ and $K_2$ cases, where the lowest masses are both approximately 11.82 GeV. Across all computational scenarios, the lowest-lying channel, $B_c B_s^*$, remains a scattering state.

Figure~\ref{PP11} illustrates the results from a complete coupled-channel calculation using the CSM, where four resonance poles are identified. In the top panel, a remarkably narrow resonance with a complex energy of $M-i\Gamma = 12202 - i\,1.0$ MeV is found embedded among the continuum states of $B_c(1S) B_s^*(1S)$, $B_c^*(1S) B_s(1S)$, $B_c^*(1S) B_s^*(1S)$, and their radial excitations. The middle panel clearly depicts the scattering nature of the ground-state channels $B_c B_s^*$, $B_c^* B_s$, and $B_c^* B_s^*$. The bottom panel focuses on the energy window between 12.23 GeV and 12.40 GeV, where six radial excitations, namely $B_c^*(2S) B_s(1S)$, $B_c(2S) B_s^*(1S)$, $B_c^*(2S) B_s^*(1S)$, $B_c(1S) B_s^*(2S)$, $B_c^*(1S) B_s(2S)$, and $B_c^*(1S) B_s^*(2S)$, are well resolved. In this region, three additional narrow resonance poles are observed at $12235 - i\,5.3$ MeV, $12248 - i\,3.0$ MeV, and $12313 - i\,1.0$ MeV, respectively.

As shown in Table~\ref{GresultR11}, all resonances are identified as compact $\bar{c}b\bar{s}b$ tetraquark configurations with sizes smaller than 1.0 fm. Their magnetic moments are $0.51\,\mu_N$, $0.59\,\mu_N$, $0.30\,\mu_N$, and $0.45\,\mu_N$, respectively. A strong coupling between the color-singlet (S) and K-type (K) components is observed, particularly for the three resonances at 12.23, 12.24, and 12.31 GeV, where the S and K contributions are approximately 25\% and 64\%, respectively. For the lowest resonance at 12.20 GeV, a significant mixing between S and K structures is evident; nonetheless, the K-type configuration remains the dominant component. The golden decay channel for the lowest resonance at 12.20 GeV is suggested to be $B_c B_s^*$. The resonances at 12.23 GeV and 12.24 GeV are expected to be confirmed through both $B_c^* B_s$ and $B_c^* B_s^*$ final states. The highest resonance, at 12.31 GeV, could be observed in either the $B_c B_s^*$ or $B_c^* B_s$ decay channels.

%%%%%%%%%%%%%%%%%%%%%%%%%%%%%%%%%%%%%%%%%%%%%%%%%%%%%%%%%%%%%%%%%

\begin{figure}[!t]
\includegraphics[clip, trim={3.0cm 1.7cm 3.0cm 0.8cm}, width=0.45\textwidth]{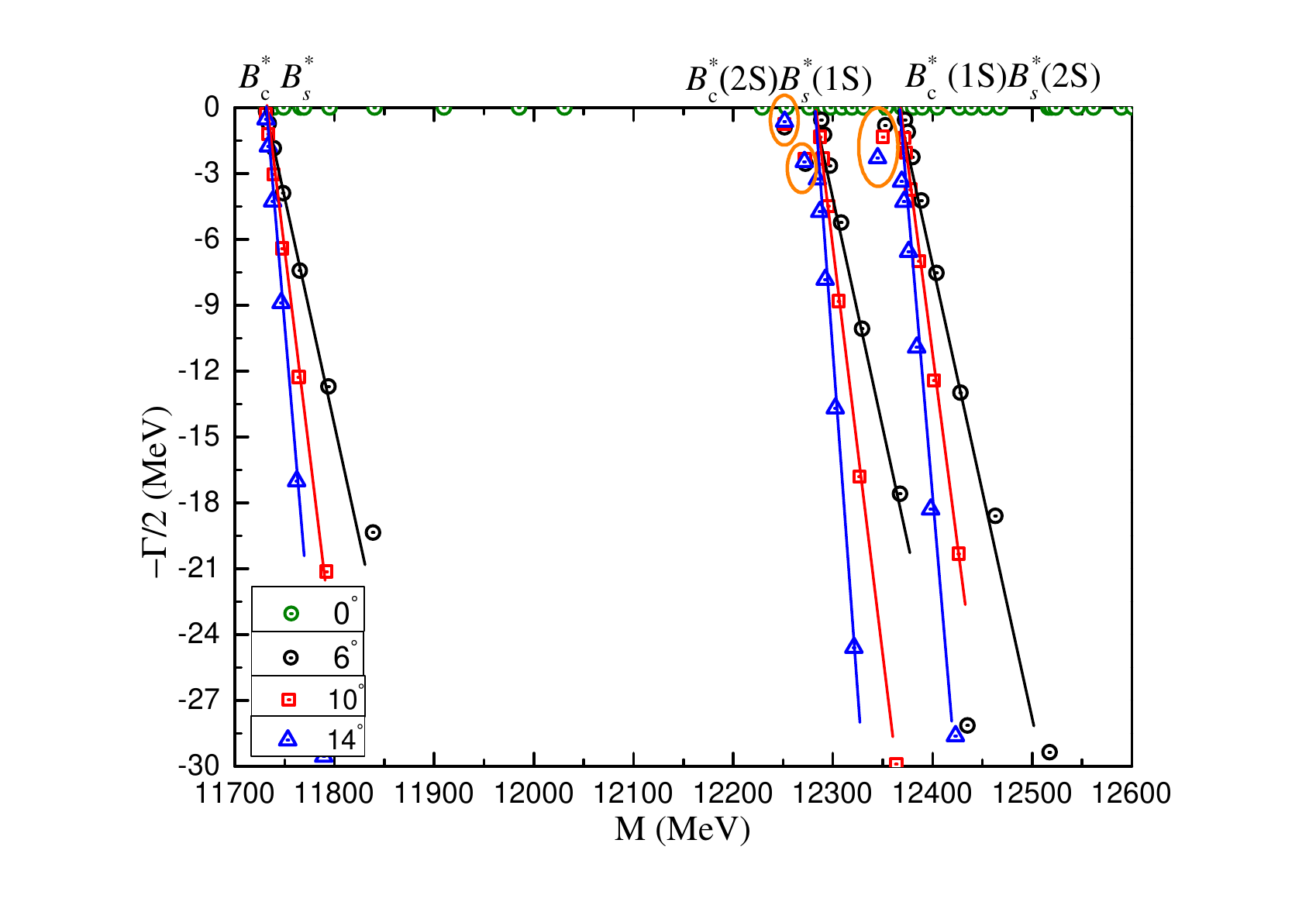}
\caption{\label{PP12} The complete coupled-channels calculation of $\bar{c}b\bar{s}b$ tetraquark system with $I(J^P)=0(2^+)$ quantum numbers.}
\end{figure}

\begin{table}[!t]
\caption{\label{GresultCC12} Lowest-lying $\bar{c}b\bar{s}b$ tetraquark states with $I(J^P)=0(2^+)$ calculated within the real range formulation of the constituent quark model. Results are similarly organized as those in Table~\ref{GresultCC1} (unit: MeV).}
\begin{ruledtabular}
\begin{tabular}{lcccc}
~~Channel   & Index & $\chi_J^{\sigma_i}$;~$\chi_j^c$ & $M$ & Mixed~~ \\
        &   &$[i; ~j]$ &  \\[2ex]
$(B^*_c B^*_s)^1$  & 1  & [1;~1]   & $11731$ &  \\[2ex]
$(B^*_c B^*_s)^8$            & 2  & [1;~3]   & $11953$ &  \\[2ex]
$(bb)^*(\bar{c}\bar{s})^*$  & 3  & [1;~7]   & $11943$ & \\[2ex]
$K_1$  & 4  & [1;~8]   & $11929$ & \\
            & 5  & [1;~10]   & $11944$ & $11871$ \\[2ex]
$K_2$  & 6  & [1;~11]   & $11897$ & \\
             & 7  & [1;~12]   & $11944$ & $11891$ \\[2ex]
$K_3$  & 8  & [1;~14]   & $11949$ & \\[2ex]
$K_4$  & 9  & [1;~8]   & $12365$ & \\
             & 10  & [1;~10]   & $11944$ & $11932$ \\[2ex]
$K_5$  & 11  & [1;~11]   & $11954$ & \\[2ex]
\multicolumn{4}{c}{Complete coupled-channels:} & $11731$
\end{tabular}
\end{ruledtabular}
\end{table}

\begin{table}[!t]
\caption{\label{GresultR12} Compositeness of exotic resonances obtained in a complete coupled-channel calculation in the $0(2^+)$ state of $\bar{c}b\bar{s}b$ tetraquark. Results are similarly organized as those in Table~\ref{GresultR1}.}
\begin{ruledtabular}
\begin{tabular}{rccc}
Resonance       & \multicolumn{3}{c}{Structure} \\[2ex]
$12251-i1.5$   & \multicolumn{3}{c}{$\mu=0.08$} \\
  & \multicolumn{3}{c}{$r_{b \bar{c}}:0.66$;\,\,\,\,\,$r_{\bar{c}\bar{s}}:0.96$;\,\,\,\,\,$r_{b\bar{s}}:0.89$;\,\,\,\,\,$r_{bb}:0.78$} \\
$Set$ I: & \multicolumn{3}{c}{$S$: 0.1\%;\, $H$: 0.1\%;\, $Di$: 0.1\%;\, $K$: 99.7\%}\\
$Set$ II: & \multicolumn{3}{c}{$S$: 22.0\%;\, $H$: 4.6\%;\, $Di$: 1.1\%;\, $K$: 72.3\%}\\[2ex]
%%%%%%%%%%
$12272-i4.7$   & \multicolumn{3}{c}{$\mu=0.08$} \\
  & \multicolumn{3}{c}{$r_{b \bar{c}}:0.77$;\,\,\,\,\,$r_{\bar{c}\bar{s}}:0.87$;\,\,\,\,\,$r_{b\bar{s}}:0.74$;\,\,\,\,\,$r_{bb}:0.59$} \\
$Set$ I: & \multicolumn{3}{c}{$S$: 0.1\%;\, $H$: 0.1\%;\, $Di$: 0.1\%;\, $K$: 99.7\%}\\
$Set$ II: & \multicolumn{3}{c}{$S$: 8.3\%;\, $H$: 7.2\%;\, $Di$: 0.4\%;\, $K$: 84.1\%}\\[2ex]
%%%%%%%%%%
$12350-i2.7$   & \multicolumn{3}{c}{$\mu=0.08$} \\
  & \multicolumn{3}{c}{$r_{b \bar{c}}:0.83$;\,\,\,\,\,$r_{\bar{c}\bar{s}}:1.20$;\,\,\,\,\,$r_{b\bar{s}}:1.04$;\,\,\,\,\,$r_{bb}:1.08$} \\
$Set$ I: & \multicolumn{3}{c}{$S$: 0.3\%;\, $H$: 0.1\%;\, $Di$: 0.1\%;\, $K$: 99.5\%}\\
$Set$ II: & \multicolumn{3}{c}{$S$: 24.5\%;\, $H$: 4.7\%;\, $Di$: 1.2\%;\, $K$: 69.6\%}\\
\end{tabular}
\end{ruledtabular}
\end{table}

{\bf The $\bm{I(J^P)=0(2^+)}$ sector:} 11 channels, listed in Table~\ref{GresultCC12}, are investigated in this sector. These include one $B_c^* B_s^*$ configuration in both color-singlet and hidden-color channels, one $(bb)^*(\bar{c}\bar{s})^*$ diquark-antidiquark configuration, and eight K-type configurations. In the real-range analysis, no bound state is found in this sector. The lowest configuration corresponds to a scattering state of $B_c^* B_s^*$, with a theoretical mass of 11.73 GeV. Both the hidden-color and diquark-antidiquark channels exhibit masses around 11.95 GeV, while the K-type configurations lie within the energy range of 11.89 to 12.36 GeV.

In the complete coupled-channel calculation using CSM, three resonance states are identified, as shown in Fig.~\ref{PP12}. Their complex energies are found to be $12251 - i\,1.5$ MeV, $12272 - i\,4.7$ MeV, and $12350 - i\,2.7$ MeV, respectively. Also, within this analysis, the scattering states $B_c^*(1S) B_s^*(1S)$, $B_c^*(2S) B_s^*(1S)$, and $B_c^*(1S) B_s^*(2S)$ are clearly displayed.

As displayed in Table~\ref{GresultR12}, all three resonant states exhibit very small values of magnetic moments, with $\mu \approx 0.08\,\mu_N$. The first two resonances are compact $\bar{c}b\bar{s}b$ tetraquark structures, with spatial sizes less than 0.96 fm. In contrast, the highest-mass resonance appears to be a more loosely bound state, with an estimated size of approximately 1.20 fm. Lastly, a strong coupling between the color-singlet and K-type channels is observed across all three resonances. Nevertheless, the K-type configurations remain dominant, contributing over 70\% to the composition of each state.

%%%%%%%%%%%%%%%%%%%%%%%%%%%%%%%%%%%%%%%%%%%%%%%%%%%%%%%%%%%%%%%%%

\begin{table}[!t]
\caption{\label{GresultCCT} Summary of resonance structures found in the triply heavy tetraquark systems. The first column shows the isospin, total spin and parity of each singularity. The second column refers to the theoretical resonance with notation: $E=M-i\Gamma$ (unit: MeV). Size ($r$, unit: fm) and magnetic moment ($\mu$, unit: $\mu_N$) of resonance are presented in the last column. Particularly, the magnetic moment of states with $I_m=\frac{1}{2}$ and $-\frac{1}{2}$ are assigned with $\mu_{-\frac{1}{2}}(\mu_{\frac{1}{2}})$, respectivey.}
\begin{ruledtabular}
\begin{tabular}{lcc}
~ $I(J^P)$ & Theoretical resonance   & Structure~~ \\
                   & $E=M-i\Gamma$   & $r,\,\mu$ \\
\hline
\multicolumn{3}{c}{$\bar{b}c\bar{q}c$ tetraquarks}\\
~~$\frac{1}{2}(0^+)$  & $8870-i5.0$  & $0.71\sim1.06$, $0$~~  \\
                                    & $8885-i2.2$   & $0.73\sim1.12$, $0$~~  \\[2ex]
~~$\frac{1}{2}(1^+)$   & $8867-i7.8$   & $0.82\sim1.12$, $-1.42(1.20)$~~ \\
                                    & $8937-i4.5$   & $0.79\sim1.34$, $-1.45(1.08)$~~  \\
                                    & $8950-i7.4$   & $0.75\sim1.23$, $-1.30(1.15)$~~  \\[2ex]
~~$\frac{1}{2}(2^+)$   & $8893-i10.2$   & $0.75\sim0.97$, $-1.22(1.78)$~~ \\
                                     & $8934-i7.4$   & $0.87\sim1.36$, $-1.22(1.78)$~~ \\
                                     & $9183-i4.5$   & $0.64\sim1.00$, $-1.22(1.78)$~~ \\[2ex]
~~$0(0^+)$    & $8965-i2.6$   & $0.73\sim0.98$, $0$~~ \\[2ex]
~~$0(1^+)$     & $8968-i3.6$  & $0.74\sim1.02$, $0.86$~~  \\
                       & $9014-i1.8$   & $0.73\sim0.98$, $0.53$~~  \\
                       & $9024-i6.4$   & $0.80\sim1.13$, $0.74$~~  \\
                       & $9283-i4.4$   & $0.58\sim0.92$, $1.04$~~  \\[2ex]
~~$0(2^+)$    & $9004-i10.6$   & $0.77\sim0.88$, $1.34$~~ \\
                       & $9298-i9.2$   & $0.55\sim0.92$, $1.34$~~ \\
                       & $9364-i26.2$   & $0.54\sim0.77$, $1.34$~~ \\
\hline
\multicolumn{3}{c}{$\bar{c}b\bar{q}b$ tetraquarks}\\
~~$\frac{1}{2}(0^+)$  & $12129-i7.5$  & $0.81\sim1.10$, $0$~~  \\
                                    & $12163-i0.8$   & $0.67\sim0.92$, $0$~~  \\
                                    & $12433-i13.0$   & $0.39\sim1.03$, $0$~~  \\[2ex]
~~$\frac{1}{2}(1^+)$   & $12130-i4.3$  & $0.51\sim0.89$, $-0.56(0.17)$~~  \\
                                    & $12139-i1.9$   & $0.73\sim1.14$, $-1.32(0.34)$~~  \\
                                    & $12148-i5.4$   & $0.66\sim1.01$, $-1.61(0.57)$~~  \\
                                    & $12234-i1.1$   & $0.63\sim0.92$, $-1.80(0.80)$~~  \\
                                    & $12427-i6.5$   & $0.38\sim1.01$, $-1.03(0.44)$~~  \\
                                    & $12440-i12.4$   & $0.39\sim1.03$, $-1.08(0.12)$~~  \\[2ex]
~~$\frac{1}{2}(2^+)$   & $12164-i5.4$   & $0.43\sim0.89$, $-2.48(0.52)$~~ \\
                                     & $12455-i11.2$   & $0.41\sim1.01$, $-2.48(0.52)$~~ \\[2ex]
~~$0(0^+)$    & $12201-i0.4$  & $0.66\sim1.01$, $0$~~  \\
                       & $12251-i2.6$   & $0.72\sim1.05$, $0$~~  \\
                       & $12269-i1.4$   & $0.73\sim0.97$, $0$~~  \\[2ex]
~~$0(1^+)$     & $12202-i1.0$  & $0.66\sim1.02$, $0.51$~~  \\
                       & $12235-i5.3$   & $0.69\sim0.93$, $0.59$~~  \\
                       & $12248-i3.0$   & $0.70\sim0.96$, $0.30$~~  \\
                       & $12313-i1.0$   & $0.67\sim0.92$, $0.45$~~  \\[2ex]
~~$0(2^+)$    & $12251-i1.5$   & $0.66\sim0.96$, $0.08$~~ \\
                       & $12272-i4.7$   & $0.59\sim0.87$, $0.08$~~ \\
                       & $12350-i2.7$   & $0.83\sim1.20$, $0.08$~~
\end{tabular}
\end{ruledtabular}
\end{table}

\section{Summary}
\label{sec:summary}

Within the framework of a constituent quark model, triply heavy tetraquark systems of the forms $\bar{b}c\bar{q}c$ and $\bar{c}b\bar{q}b$ ($q = u,\,d,\,s$) with spin-parity quantum numbers $J^P = 0^+$, $1^+$, and $2^+$, and isospin $I = 0$ or $\frac{1}{2}$, are systematically studied. Special attention is given to fully S-wave tetraquark configurations, including color-singlet and hidden-color meson-meson structures, diquark-antidiquark arrangements with allowed color triplet-antitriplet and sextet-antisextet combinations, as well as five K-type configurations. To address the four-body bound- and resonant states, we solve the Schrödinger equation using the Gaussian Expansion Method (GEM) combined with the Complex Scaling Method (CSM).

Multiple narrow resonances are identified across the different isospin-spin-parity channels when fully coupled-channel calculations using CSM are performed. These are summarized in Table~\ref{GresultCCT}, which includes the mass, $M$, width, $\Gamma$, size, $r$, and magnetic moment, $\mu$, of the found resonances, organized by their quantum numbers $I(J^P)$. For a clearer representation of the magnetic moments in isodoublet $\bar{Q}Q\bar{q}Q$ ($q = u,\,d$) states, the third component of isospin, $I_m$, is indicated using the notation $\mu_{-\frac{1}{2}}(\mu_{\frac{1}{2}})$.

The following key conclusions can be drawn:
\begin{itemize}
\item Resonance states with $\bar{b}c\bar{q}c$ content appear in the mass range of $(8.87-9.36)$ GeV, while those of $\bar{c}b\bar{q}b$ lie between $12.13$ and $12.45$ GeV.

\item The vast majority of the identified resonances are compact $\bar{b}c\bar{q}c$ and $\bar{c}b\bar{q}b$ tetraquark configurations with spatial sizes smaller than 1.0 fm.

\item The K-type configurations constitute the dominant components of the obtained resonance states, while some cases exhibit moderate mixing with other quark configurations such as meson-meson ones.

\item The magnetic moment $\mu$ of the $0^+$ resonance state is zero. For the $1^+$ and $2^+$ states with $I_m = -\frac{1}{2}$, $\mu$ ranges from $-1.03$ to $-1.61\,\mu_N$, except for a $\bar{c}b\bar{u}b$ state in the $1^+$ channel, which has a smaller magnetic moment of $-0.56\,\mu_N$. Two $\bar{c}b\bar{u}b$ resonances in the $2^+$ channel exhibit larger magnetic moments, reaching $-2.48\,\mu_N$.

\item For the $I_m = \frac{1}{2}$ and $0$ components, $\mu$ ranges from 0.53 to 1.78 $\mu_N$ for $\bar{b}c\bar{q}c$ ($q = d,\,s$) states in the $1^+$ and $2^+$ sectors. In contrast, $\mu$ is smaller for the corresponding $\bar{c}b\bar{q}b$ ($q = d,\,s$) resonances, falling in the range 0.08-0.80 $\mu_N$.
\end{itemize}

All of these exotic resonances involving heavy flavors are expected to be accessible in future high-energy experiments.

%%%%%%%%%%%%%%%%%%%%%%%%%%%%%%%%%%%%%%%%%%%%%%%%%%%%%%%%%%%%%%%%%

\begin{acknowledgments}
Work partially financed by National Natural Science Foundation of China under Grant Nos. 12305093, 11535005 and 11775118; Ministerio Espa\~nol de Ciencia e Innovaci\'on under grant No. PID2022-140440NB-C22; Junta de Andaluc\'ia under contract No. FQM-370 as well as PCI+D+i under the title: ``Tecnolog\'\i as avanzadas para la exploraci\'on del universo y sus componentes" (Code AST22-0001).
\end{acknowledgments}

%%%%%%%%%%%%%%%%%%%%%%%%%%%%%%%%%%%%%%%%%%%%%%%%%%%%%%%%%%%%%%%%%

\bibliography{THTII}

\end{document}